\documentclass[fleqn,usenatbib]{mnras}

\usepackage[T1]{fontenc}
\usepackage{ae,aecompl}
\usepackage{layouts}

\usepackage{graphicx}	
\usepackage{graphbox}
\usepackage{adjustbox}
\usepackage{mathtools}
\usepackage{amsmath}
\usepackage{pdflscape}	
\usepackage{xcolor}
\usepackage[normalem]{ulem}
\usepackage{siunitx}
\usepackage{accents}

\usepackage[labelformat=parens,labelsep=none,subrefformat=parens,]{subcaption}
\newcommand{\hi}{H\,{\textsc{i}}}
\newcommand{\radms}{rad\,m$^{-2}$}

\graphicspath{{./}{figures/}}
\newcommand{\subfigimg}[3][,]{%
  \setkeys{Gin,subfigpos}{pos,font,vsep,hsep,#1}
  \setbox1=\hbox{\includegraphics{#3}}
  \ifnum\pdfstrcmp{\sfp@pos}{ul}=0
    \leavevmode\rlap{\usebox1}
    \rlap{\hspace*{\sfp@hsep}\raisebox{\dimexpr\ht1-\sfp@vsep}{\sfp@font{#2}}}
    \phantom{\usebox1}
  \else\ifnum\pdfstrcmp{\sfp@pos}{ur}=0
    \leavevmode\usebox1
    \llap{\raisebox{\dimexpr\ht1-\sfp@vsep}{\sfp@font{#2}}\hspace*{\sfp@hsep}}
  \else\ifnum\pdfstrcmp{\sfp@pos}{lr}=0
    \leavevmode\usebox1
    \llap{\raisebox{\sfp@vsep}{\sfp@font{#2}}\hspace*{\sfp@hsep}}
  \else
    \leavevmode\rlap{\usebox1}
    \rlap{\hspace*{\sfp@hseplen}\raisebox{\sfp@vsep}{\sfp@font{#2}}}
    \phantom{\usebox1}
  \fi\fi\fi
}
\usepackage{tikz}
\usetikzlibrary{arrows,decorations.pathmorphing}
\newcommand{\eye}[4]
{   \draw[rotate around={#4:(#2,#3)}] (#2,#3) -- ++(-.5*55:#1) (#2,#3) -- ++(.5*55:#1);
    \draw (#2,#3) ++(#4+55:.75*#1) arc (#4+55:#4-55:.75*#1);
    \draw[fill=gray] (#2,#3) ++(#4+55/3:.75*#1) arc (#4+180-55:#4+180+55:.28*#1);
    \draw[fill=black] (#2,#3) ++(#4+55/3:.75*#1) arc (#4+55/3:#4-55/3:.75*#1);
}

\newcommand{\dish}[3]
{   \draw[black] (#2,#3) arc (45:-45:#1);
    \draw[black] (#2,#3) -- ++(45+180:#1) -- +(-45:#1);
    \draw[red] (#2,#3) -- ++(0+180:#1);
}
\usepackage{tikzscale}
\usepackage[utf8]{inputenc}
\usepackage{pgf}
\usepackage{pgfplots}
\DeclareUnicodeCharacter{2212}{−}
\usepgfplotslibrary{groupplots,dateplot}
\usetikzlibrary{patterns,shapes.arrows}
\pgfplotsset{compat=newest}

\title[GMIMS: Brightest Southern polarized region at 75\,cm]{The Global Magneto-Ionic Medium Survey (GMIMS): The brightest polarized region in the Southern sky at 75\,cm and its implications for Radio Loop II}

\author[A. J. M. Thomson et al.]{
\newauthor Alec J. M. Thomson,$^{1}$\thanks{E-mail: alec.thomson@csiro.au} 
T. L. Landecker,$^{2}$ 
N. M. McClure-Griffiths,$^{3}$
\newauthor John M. Dickey,$^{4}$
J. L. Campbell,$^{5,6}$
Ettore Carretti,$^{7}$
S. E. Clark,$^{8}$
\newauthor Christoph Federrath,$^{3}$
B. M. Gaensler,$^{5,6}$
J. L. Han$^{9,10,11}$
Marijke Haverkorn,$^{12}$
\newauthor
Alex. S. Hill,$^{13,2}$
S.A. Mao,$^{14}$
Anna Ordog,$^{13,2}$
Luke Pratley,$^{6}$
Wolfgang Reich,$^{14}$
\newauthor
Cameron L. Van Eck,$^{6}$
J. L. West,$^{6}$
M. Wolleben$^{15}$
\\
$^{1}$CSIRO Astronomy \& Space Science, PO Box 1130, Bentley, WA 6102, Australia\\
$^{2}$Dominion Radio Astrophysical Observatory, Herzberg Astronomy and Astrophysics Research Centre, National Research Council \\Canada, PO Box 248, Penticton, B.C. Canada, V2A 6J9\\
$^{3}$Research School of Astronomy and Astrophysics, Australian National University, Canberra, ACT 2611, Australia\\
$^{4}$School of Natural Sciences, Private Bag 37, University of Tasmania, Hobart, TAS, 7001, Australia\\
$^{5}$David A. Dunlap Department of Astronomy \& Astrophysics, University of Toronto, 50 St. George St., Toronto, Ontario, Canada, \\M5S 3H4\\
$^{6}$Dunlap Institute for Astronomy and Astrophysics, University of Toronto, 50 St. George St., Toronto, Ontario, Canada, M5S 3H4\\
$^{7}$INAF - Istituto di Radioastronomia, Via Gobetti 101, 40129 Bologna, Italy\\
$^{8}$Institute for Advanced Study, 1 Einstein Drive, Princeton, NJ, USA\\
$^{9}$National Astronomical Observatories, Chinese Academy of Sciences, 20A Datun Road, Chaoyang District, Beijing 100012, China\\
$^{10}$CAS Key Laboratory of FAST, NAOC, Chinese Academy of Sciences, Beijing 100101, China \\
$^{11}$School of Astronomy, University of Chinese Academy of Sciences, Beijing 100049, China \\
$^{12}$Department of Astrophysics/IMAPP, Radboud University Nijmegen, P.O. Box 9010, 6500 GL Nijmegen, The Netherlands\\
$^{13}$Department of Computer Science, Math, Physics, \& Statistics, The University of British Columbia, Okanagan Campus, Kelowna, \\BC V1V 1V7 Canada\\
$^{14}$Max Planck Institute for Radio Astronomy, Auf dem H{\"u}gel 69, 53121 Bonn, Germany\\
$^{15}$Skaha Remote Sensing Ltd., 3165 Juniper Drive, Naramata, British Columbia V0H 1N0, Canada\\
}

\date{Accepted XXX. Received YYY; in original form ZZZ}

\pubyear{2021}
\usepackage{newtxtext,newtxmath}
\begin{document}
\label{firstpage}
\pagerange{\pageref{firstpage}--\pageref{lastpage}}
\maketitle
\begin{abstract}
Using the Global Magneto-Ionic Medium Survey (GMIMS) Low-Band South (LBS) southern sky polarization survey, covering 300 to 480 MHz at 81\arcmin\ resolution, we reveal the brightest region in the Southern polarized sky at these frequencies. The region, G150$-$50, covers nearly 20\,$\deg^2$, near ${(l,b)}{\approx}{(150\degr,-50\degr)}$. Using GMIMS-LBS and complementary data at higher frequencies ($\sim0.6$--$30$\,GHz), we apply Faraday tomography and Stokes $QU$-fitting techniques. We find that the magnetic field associated with G150$-$50 is both coherent and primarily in the plane of the sky, and indications that the region is associated with Radio Loop II. The Faraday depth spectra across G150$-$50 are broad and contain a large-scale spatial gradient. We model the magnetic field in the region as an expanding shell, and we can reproduce both the observed Faraday rotation and the synchrotron emission in the GMIMS-LBS band. Using $QU$-fitting, we find that the Faraday spectra are produced by several Faraday dispersive sources along the line-of-sight. Alternatively, polarization horizon effects that we cannot model are adding complexity to the high-frequency polarized spectra. The magnetic field structure of Loop II dominates a large fraction of the sky, and studies of the large-scale polarized sky will need to account for this object. Studies of G150$-$50 with high angular resolution could mitigate polarization horizon effects, and clarify the nature of G150$-$50.
\end{abstract}




\section{Introduction}
At low radio frequencies ($\lesssim5$\,GHz), the polarized sky often bears little resemblance to its total intensity counterpart, despite the fact that the dominant emission mechanism, synchrotron radiation, can be highly polarized \citep[up to $\sim70$\,per\,cent,][]{Rybicki1985}. Faraday rotation, strongly wavelength ($\lambda$) dependent, modulates the otherwise smoothly polarized emission. Polarized emission and Faraday rotation are often co-extensive in the interstellar medium (ISM) of the Milky Way. We can extract information on the magneto-ionic medium (MIM) of the Galaxy through analysis of the received linearly polarized radio signal, provided that observations are sensitive to large angular scales, and cover a large area of sky and a broad range of wavelengths~\citep{Wolleben2009}.

Away from the Galactic plane, the radio sky is dominated by large-scale `radio loops'~\citep{Berkhuijsen1971}. \citet{Vidal2015} provide a recent summary of the properties of these loops, including their radio polarization properties. The radio emission from these features is non-thermal \citep{Berkhuijsen1973, Borka2007}, and has been associated with emission from other ISM tracers \citep{Meaburn1965,Heiles1989}. The precise origin of the loops is yet to be determined, but their magnetic field structures have been shown to be consistent with expanding shell \citep[e.g.][]{VanDerLaan} models~\citep{Spoelstra1972, Berkhuijsen1973}. Here we pay particular attention to Loop II~\citep{Large1962}, also known as the Cetus Arc. This diffuse feature can be mapped in total intensity by a circle centred at $l,b\sim(100\degr,-32.5\degr)$ with a radius of $45.5\degr$, extending South from the Galactic plane. The distance to the centre of this object has been estimated~\citep[e.g. by][]{Spoelstra1972,Berkhuijsen1973,Case1998, Borka2007} to be $\sim100\,$pc from the Sun, with a corresponding diameter of $\sim180\,$pc. \citet{Planck2016-diffuseforeground} reported both a detection of Loop II in total intensity at 30\,GHz and coherent magnetic field vectors along the loop.

We can measure the degree of Faraday rotation, the change in polarization angle ($\psi$) at given $\lambda^2$ value through a magneto-ionic medium, by determining the Faraday depth \citep[$\phi$,][]{Burn1966,Brentjens2005,Ferriere2021}:
\begin{equation}
\phi(d) \equiv 0.812\int_{\text{src}}^{\text{obs}}n_e(r)\mathbf{B}(r)\cdot \mathrm{d}\mathbf{r} =  0.812\int_{0}^{d}n_e(r)B_\parallel(r) \mathrm{d}r,
\label{eqn:faradaydepth}
\end{equation}
where $d$ is the distance from the observer in pc, $n_e$ is the thermal electron density in cm$^{-3}$, $\mathbf{B}$ is the magnetic field vector in $\mu$G, and $\mathbf{r}=r\hat{\mathbf{r}}$ is a radial vector, measured in pc, directed away from the observer along the line-of-sight (LOS). $B_\parallel$ is therefore the LOS component of the magnetic field in $\mu$G. When there is a single source of polarized emission behind a purely rotating medium, the Faraday depth is equal to the rotation measure (RM). The RM is defined as the derivative of the polarization angle  with respect to $\lambda^2$:
\begin{equation}
	\text{RM} = \frac{\mathrm{d}\psi(\lambda^2)}{\mathrm{d}\lambda^2} = 0.812\int_{0}^{l}n_e(r)B_\parallel(r) \mathrm{d}r,
\end{equation}
and is equal to the degree of Faraday rotation integrated along the entire line-of-sight (LOS) to the source at distance $l$. The scenario where $\text{RM} = \phi(l)$ is referred to as `Faraday simple'. 

If the given LOS is not Faraday simple, particularly when the rotating and emitting volumes are mixed, the RM does not accurately describe the Faraday rotation. The specific case of mixed emission and rotation is referred to as `Faraday thick'. We refer the reader to \citet{Alger2021} for an in-depth discussion on Faraday complexity. In such `Faraday complex' cases there are two widely used methods of evaluating the Faraday depth structure from polarization observations: the Fourier-transform-like rotation measure synthesis \citep[RM synthesis, ][]{Burn1966,Brentjens2005, Heald2009} and model fitting of the spectral energy distribution of Stokes $Q$ and $U$ with $\lambda^2$; so-called $QU$-fitting. $QU$-fitting has been used to great effect in studies of extra-galactic sources \citep[e.g.][]{Law2011,OSullivan2012,OSullivan2013,Anderson2015,Anderson2016,Kim2016b,OSullivan2018,Pasetto2018,Schnitzeler2019,Ma2019}. Here we apply the technique for the first time to the large-scale Galactic emission. We also note that there are several new analysis methods that have recently been presented, such as non-parametric $QU$-fitting~\citep{Pratley2020}, sparse modelling~\citep{Akiyama2018}, and iterative reconstruction~\citep{Cooray2021}. Polarized emission is Faraday rotated by the complex structure of the MIM along the LOS. Interpretation of these effects is possible with RM synthesis and $QU$ fitting, but requires observations over a broad range of $\lambda^2$ to obtain unambiguous results.

The Global Magneto-Ionic Medium Survey \citep[GMIMS,]{Wolleben2009, Wolleben2019, Wolleben2021} is providing the necessary bandwidth to map Faraday rotation of diffuse emission across the entire sky. Recently, the low-band, Southern component \citep[GMIMS-LBS,][]{Wolleben2019} has been completed, which presents an unprecedented view of the polarized sky. This survey maps diffuse polarized emission from 300 to 480\,MHz with 500\,kHz frequency resolution. This broad bandwidth corresponds to a $\lambda^2$ coverage of 0.39 to 1\,m$^2$, which enables a variety of polarization features to be detected. Additionally, the lower frequencies of this survey allow for high-precision RM synthesis. As RM synthesis has been applied to every LOS observed, this survey allows for the application of `Faraday tomography'~\citep[e.g.][]{VanEck2017,VanEck2019,Thomson2019}. Here we use the term `Faraday tomography' to refer to the mapping of Faraday depths from diffuse emission across the sky~\citep{Ferriere2016}. The angular resolution of GMIMS-LBS is 81\arcmin\ at 300\,MHz, which is relatively coarse. This limits the distance probed along the LOS \citep[i.e. the `polarization horizon'][]{Uyaniker2003} to be within about 500\,pc of the Sun~\citep{Dickey2018}. Despite the potential depolarization that likely affects GMIMS-LBS, there remain regions within these data with high polarized intensity. By investigating these regions, we can reveal the magneto-ionic structure of the local ISM.

In this paper we present results from GMIMS-LBS towards the brightest polarized region in the survey. This region is roughly centred on $l,b\sim(150\degr,-50\degr)$ and covers nearly $20\deg^2$. Throughout we will refer to the region as G150$-$50. In Section~\ref{sec:observations} we briefly describe the GMIMS-LBS observations, and the complementary data we utilize. We provide our results in Section~\ref{sec:results}, first discussing the morphology of G150$-$50 and how it relates to other ISM tracers in Section~\ref{sec:morphology}. In Section~\ref{sec:tomography} we show the results of Faraday tomography towards this region. We propose a simple physical model which can reproduce our observations in Section~\ref{sec:physical-model}. In Section~\ref{sec:qufit} we apply $QU$-fitting techniques to the data. Finally, we discuss our results in Section~\ref{sec:discussion} and provide our conclusions in Section~\ref{sec:conclusion}. We describe our formalisms for Faraday tomography in the Appendix.

\section{Observations}\label{sec:observations}
\subsection{GMIMS-LBS}
GMIMS-LBS is described in detail by \citet{Wolleben2019}, which we will hereafter refer to as the `survey paper'. We summarise the properties of this survey in Table~\ref{tab:survey}. The data from the survey are available in two forms: Stokes $I$, $Q$, and $U$ cubes as function of frequency, and Faraday depth cubes resulting from Faraday tomography. The frequency cubes cover 300 to 480\,MHz between declinations $-90\degr$ and $+20\degr$ with a common angular resolution of $81\arcmin$. For this analysis we regrid both sets of data cubes into HEALPix\footnote{\url{http://healpix.sourceforge.net}} format~\citep{Gorski2005} with a resolution parameter $N_\text{side}$ of 256, corresponding to a pixel size of $\sim13.7\arcmin$. HEALPix is useful to us for two primary reasons. First, each pixel represents an equal area of sky, meaning that statistics computed across pixels are not biased by changes in the sky area corresponding to each pixel. Second, the scheme allows for spherical harmonic decomposition, which in turn allows for efficient rotation and accurate sky-coordinate transformations.

GMIMS-LBS was observed with the Parkes radio telescope / \textit{Murriyang} using the same strategy as S-PASS~\citep{Carretti2019}. The advantages of the technique are discussed at length by both \citet{Carretti2019} and the survey paper, however there are several important properties which we will discuss here. The basketweaving algorithm, used to reconcile the many telescope scans into images at each frequency plane, subtracts the sky minimum value from the total intensity images. This does not affect the Stokes $Q$ and $U$ images \citep[see $\S$4 of][]{Carretti2019}. The published 300--480\,MHz data represent the lower portion of the frequencies observed by GMIMS-LBS, and were highly spatially oversampled; each point on the sky was observed many times, with large gaps in time between each pass. Coupled with the radio-frequency interference (RFI) rejection strategy described in the survey paper, this scanning method means that RFI will not present as spatially coherent structure. Rather, individual channels may be affected in a minor fashion. Instrumental polarization is still present in GMIMS-LBS, however this only affects very bright Stokes $I$ sources, such as emission from the Galactic plane. Here we avoid all such sources.

GMIMS was designed for the implementation of Faraday tomography. The broad bandwidth of GMIMS-LBS provides many unique and powerful properties for probing the Galactic magneto-ionic medium. For GMIMS-LBS the Faraday resolution, maximum Faraday depth, and maximum Faraday scale are $\delta\phi=5.9$\,\radms, $ \phi_{\text{max}}=1700 $\,\radms, and $ \phi_{\text{max-scale}}=8.6 $\,\radms, respectively (see the Appendix for detailed definitions). These values vary slightly across the sky depending on which channels are affected by RFI. The Faraday spectra are deconvolved with \texttt{RM-CLEAN}~\citep{Heald2009}, with a \texttt{CLEAN} cut-off of 60\,mK, the RMS noise in GMIMS-LBS spectra. We note that all GMIMS-LBS data were corrected for ionospheric Faraday rotation (see the survey paper for a detailed description).

\begin{table}
	\centering
	\caption{Summary of the observational parameters of the GMIMS-LBS \citep{Dickey2018, Wolleben2019}. $^a$ -- This range was determined by the high and low signal-to-noise limits. $^b$ -- These values were selected during Faraday tomography. Here we refer to the linearly polarized intensity as $L$.} \label{tab:survey}
	\footnotesize
	\begin{tabular}{llcc}
		\hline
		Survey parameter 					& Symbol  & min. 					& max. 		\\
		\hline
		Declination [$\degr$] 				& $\delta$ & $-90$ 				& $+20$ 	\\
		Beamwidth [$\arcmin$] 		& & 45 					& 81 		\\
		Frequency [MHz] 					& $f$ & 300 				& 480 	\\
		Frequency resolution [MHz] 			& $\delta f$ & \multicolumn{2}{c}{0.5} \\
		Wavelength-squared [m$^{2}$] 		& $\lambda^2$ & 0.4 				& 1 	\\
		$\lambda^2$ bandwidth [m$^{2}$] 	& $\Delta\lambda^2$ & \multicolumn{2}{c}{0.608} \\
		$\lambda^2$ resolution [m$^{2}$] 	& $\delta\lambda^2$ & \multicolumn{2}{c}{$3.32\times10^{-3}$} \\
		$Q$ and $U$ RMS noise [mK] 	& $\sigma_{QU}$ 	& \multicolumn{2}{c}{60} \\
		$L$ RMS noise$^a$ [mK]					& $\sigma_{\text{L}}$ 	& 39 & 60 \\
		Faraday resolution [\radms] 		& $\delta\phi$ & \multicolumn{2}{c}{5.9}	\\
		Max. Faraday depth [\radms] 		& $\phi_{\text{max}}$ & \multicolumn{2}{c}{$1.7\times10^3$} \\
		Faraday max. scale [\radms] 		& $\phi_{\text{max-scale}}$ & \multicolumn{2}{c}{8.6} \\
		$\phi$ range$^b$ [\radms] 				& & $-100$ 				& $+100$ 	\\
		$\phi$ sampling$^b$ [\radms] 			& & \multicolumn{2}{c}{0.5} \\
		\hline
	\end{tabular}
\end{table}

\subsubsection{Additional RFI flagging}
Some RFI remains in the published data. This does not seriously affect Faraday tomography, but does affect $QU$ fitting, where the process will treat RFI as real data and will attempt to fit components to highly discordant numbers. We go beyond the RFI flagging described in the survey paper, and simply remove channels which have more than 3\% of the data points masked (Fig. 9 of the survey paper). We also apply extra flagging to the single-pixel frequency spectra that we analyse, flagging channels in which the polarized intensity differs by more than 5\,mK from neighbouring channels. 

\subsection{Maps at 408 MHz}
We use the absolutely calibrated all-sky 408-MHz map of \citet{Haslam1982} in HEALPix format\footnote{Obtained from \url{https://lambda.gsfc.nasa.gov/} via the MPIfR Survey Sampler \url{https://www3.mpifr-bonn.mpg.de/survey.html}.} to determine the total intensity in directions of interest to us. The GMIMS-LBS intensity scale matches the scale of the absolutely calibrated Haslam survey within 10\% (see the survey paper), but the sky minimum has been removed from the GMIMS-LBS data by the basketweaving process. The spatial resolution of the Haslam data is 51\arcmin, and we convolve the data to match the GMIMS-LBS resolution of 81\arcmin.

We use the polarized intensity data from \citet{Mathewson1965}, gridded onto a HEALpix map with $N_\text{side} = 64$ (pixel size $\sim55\arcmin$), to confirm that the features we observe are not artefacts from RFI or data processing problems. We multiply the data by 0.5 to bring them to the current definition of polarized intensity \citep{Berkhuijsen1975}. These data provide the only information on the polarized Southern sky in our frequency range, apart from the GMIMS-LBS data.

\subsection{Maps at 1.4 GHz}
To complement the low-frequency surveys we use absolutely calibrated all-sky continuum data at 21\,cm. We use the total-intensity data at 1420\,MHz from \citet{Reich1982}, \citet{Reich1986}, and \citet{Reich2001}, regridded into HEALpix format with $N_\text{side} = 256$, and the polarization data at 1410\,MHz from \citet{Wolleben2006} and \citet{Testori2008}\footnote{Obtained from CADE \url{http://cade.irap.omp.eu} in combined HEALPix format.} We have not corrected for the slight frequency difference between the total-intensity and polarization maps, which results in a difference in brightness temperatures of less than $2\%$.

\subsection{Maps at 30 GHz}
For information on the polarized sky free from Faraday rotation, we need maps at high radio frequencies. We use the component-separated synchrotron map from the 2018 \textit{Planck} release \citep[PR3;][]{Planck2018-diffuse}, which avoids contamination from non-synchrotron emission. Stokes $Q$ and $U$ images are available, but synchrotron total intensity is not. We can therefore investigate polarized intensity and angle, but not polarized fraction. We convert the Planck data to the IAU polarization conventions \citep[see e.g.][]{Hamaker1996,Benvenuti2015,deSerego2017} and refer polarization angles to the equatorial coordinate system.

\subsection{Dwingeloo 25 m survey}
The surveys of \citet{Brouw1976} provide absolutely calibrated measurements of the linear polarization from declination $+90\degr$ to $\sim-20\degr$, at 408, 465, 610, 820, and 1411 MHz, albeit with sparse spatial sampling. We use the 610 and 820\,MHz data to bridge the gap between GMIMS-LBS and the 1.4\,GHz data. We employ the maps produced by \citet{Carretti2005} by interpolating tabulated data with a $4\degr$ full width at half maximum (FWHM) Gaussian convolution kernel, and the tabulated data themselves. We estimate the uncertainty in these maps by cross-matching the uncertainty value in the tabulated data with the location of each HEALPix pixel.

\subsection{Extragalactic polarization and rotation measures}
Finally, we obtain polarization measurements of extragalactic sources. To measure the contribution of these background polarized sources we obtain the catalogue of \citet{Taylor2009}\footnote{Compiled in \url{https://github.com/Cameron-Van-Eck/RMTable} v0.1.7}, which was derived from the NRAO VLA Sky Survey (NVSS). Accompanying this, we also use the `Faraday sky' map of \citet{Hutschenreuter2021}, which primarily uses the \citet{Taylor2009} catalogue to estimate the RM through the entire Milky Way across the sky. Conveniently, this map is already produced on the exact HEALPix grid we require.

\section{Results and Analysis}\label{sec:results}
\subsection{Morphology}\label{sec:morphology}
We show the all-sky map of the GMIMS-LBS polarized intensity at 408\,MHz in Fig.~\ref{fig:pimapallsky}. Here we have averaged our data to match the 3.5\,MHz \citet{Haslam1982} bandwidth. From this map we can see that G150$-$50 is the brightest region in polarized intensity on the sky at these frequencies. Part of the North Polar Spur (NPS) is present in both GMIMS-LBS and the \citet{Mathewson1965} map, but G150$-$50 is at least 50\,per\,cent brighter in polarized intensity than the NPS in both surveys.

In all panels of Fig.~\ref{fig:allpimapallsky}, we overlay the positions of the large-scale radio loops, as summarized by \citet{Vidal2015}. We use the nomenclature from that paper to refer to the loops. We pay particular attention to the location of Loop II, which was first described by \citet{Large1962}, noting that G150$-$50 lies along the position of this large-scale feature. We do not consider areas on the Galactic plane, nor the area near Centaurus A, as these regions suffer from Stokes $I$ leakage into $Q$ and $U$. The Stokes $I$ emission towards G150$-$50 is relatively low ($\sim25$\,K), and therefore the region will be negligibly affected by instrumental polarization. 

In Figs.~\ref{fig:pimap_610}--\ref{fig:pimapallsky_planck} we show the same region of sky in polarized intensity at 610, 820, 1411\,MHz, and 30\,GHz, respectively. We find that the G150$-$50 region becomes less distinct as a function of frequency, with no morphological trace at 1.4 or 30\,GHz. Notably, at 610\,MHz we see that some polarized emission extends away from both G150$-$50 and Loop II, along a great circle line connecting ${(l,b)}{\approx}{(150\degr,-50\degr)}$ and ${(l,b)}{\approx}{(180\degr,0\degr)}$. Due to the low resolution of the interpolated Dwingeloo images, it is difficult to make any definitive morphological conclusions about these more extended features. We do note, however, that the G150$-$50 region itself also appears extended along this line at 408\,MHz. Focusing on the 1.4\,GHz survey, we see that the high-longitude edge of Loop II is clearly present. This same feature is also present at 610 and 820\,MHz. There is also some polarized emission coincident with the Southern edge of the loop. The low-longitude edge is also highly polarized at 1.4\,GHz, but this emission has been associated with Loop VIIb~\citep[see][]{Wolleben2007,Vidal2015}, with curvature in the opposite direction to Loop II. At 30\,GHz, the high-longitude component of Loop II is not present, other than some emission possibly associated with the Fan Region, leaving only a thin strip of emission on the low-longitude border with Loop VIIb. We note that the \textit{Planck} data are noise limited, so it is possible that the diffuse polarized emission associated with Loop II at 30\,GHz is below the noise floor.

\begin{figure*}
	\centering
	\begin{subfigure}[b]{\textwidth}
		\includegraphics[width=\textwidth]{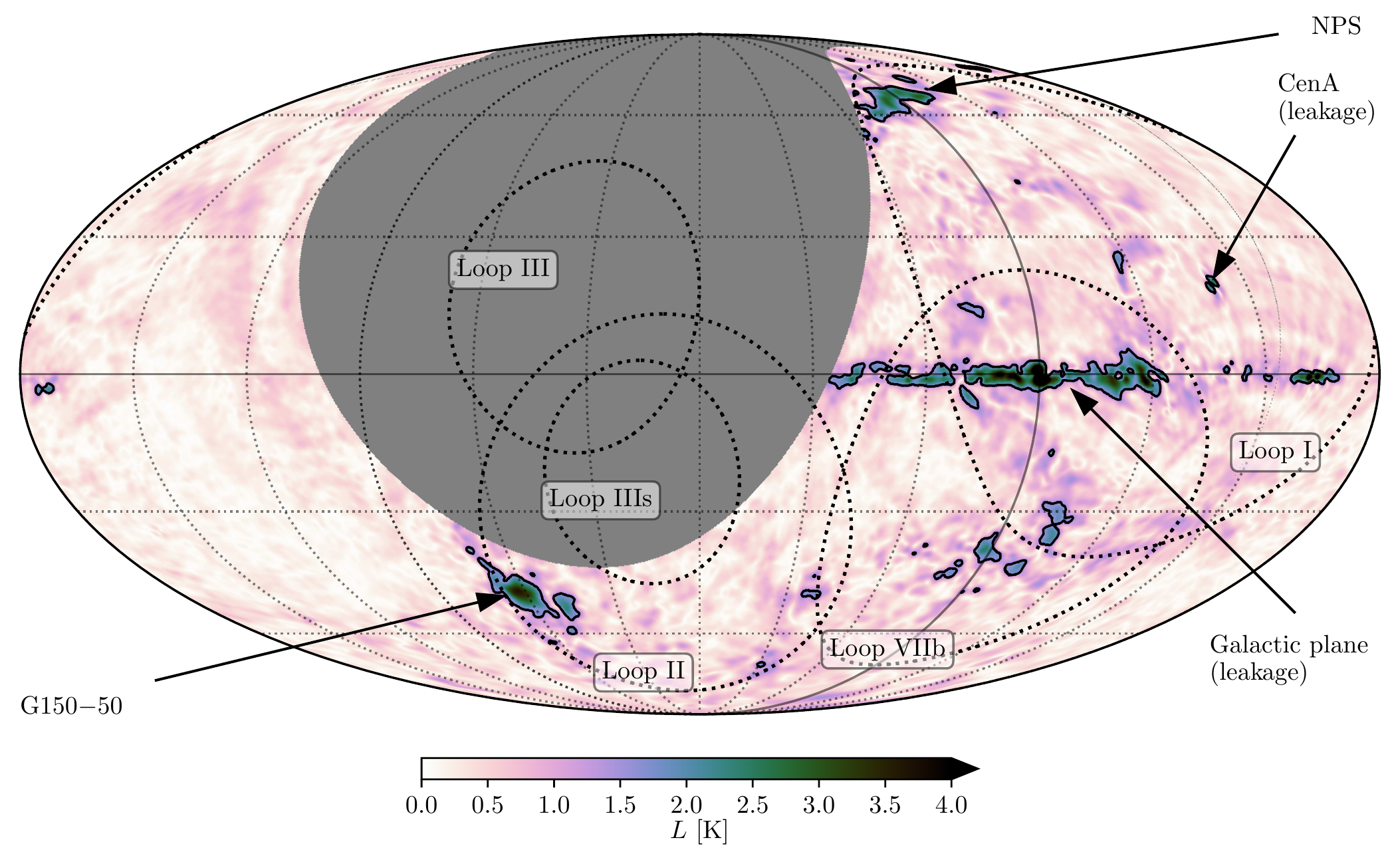}
		\caption{~408\,MHz}
		\label{fig:pimapallsky}
	\end{subfigure}
	\centering
	\begin{subfigure}[b]{\textwidth}
		\includegraphics[width=\textwidth]{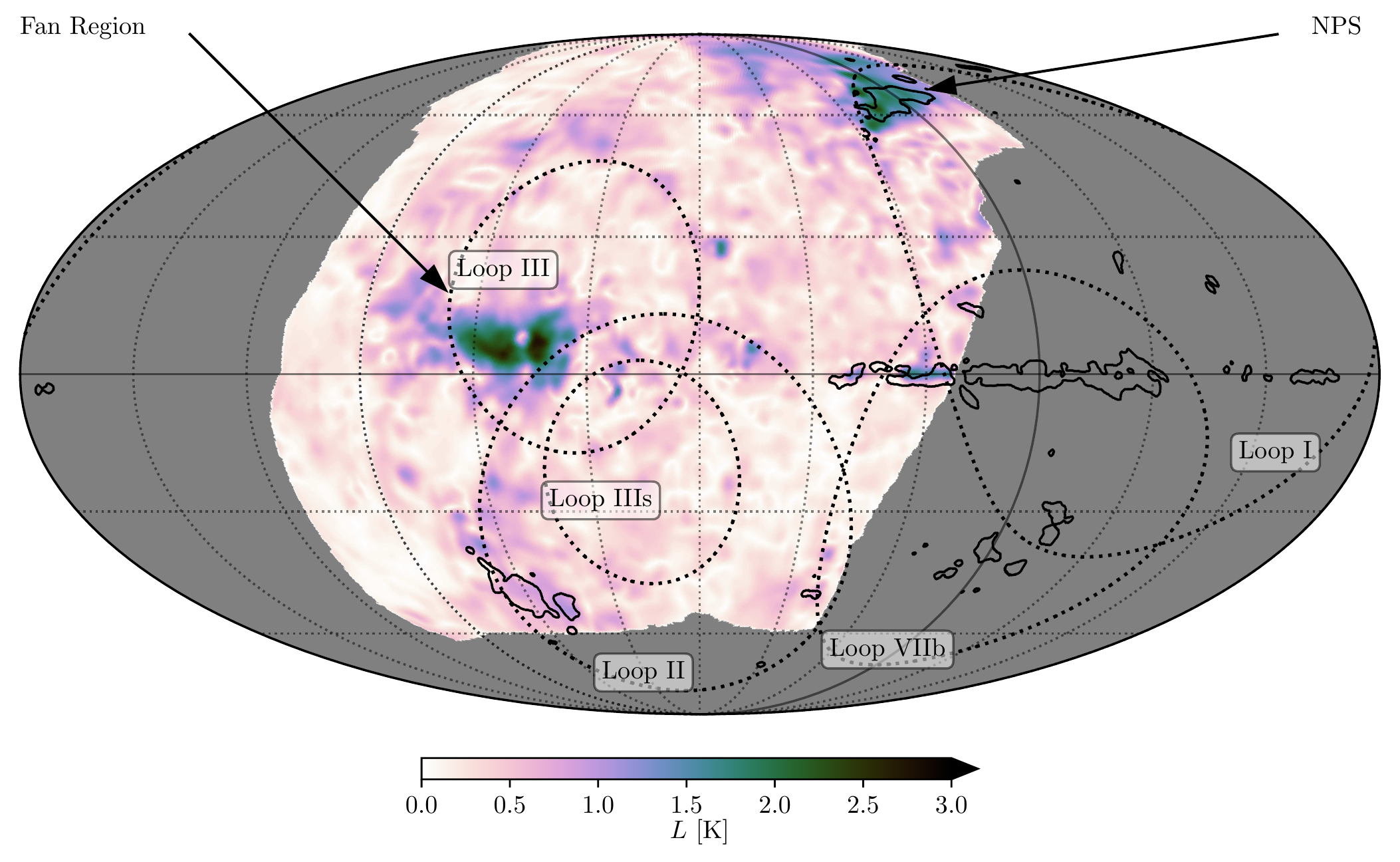}
		\caption{~610\,MHz}
		\label{fig:pimap_610}
	\end{subfigure}
\end{figure*}

\begin{figure*}\ContinuedFloat
	\begin{subfigure}[b]{\textwidth}
		\includegraphics[width=\textwidth]{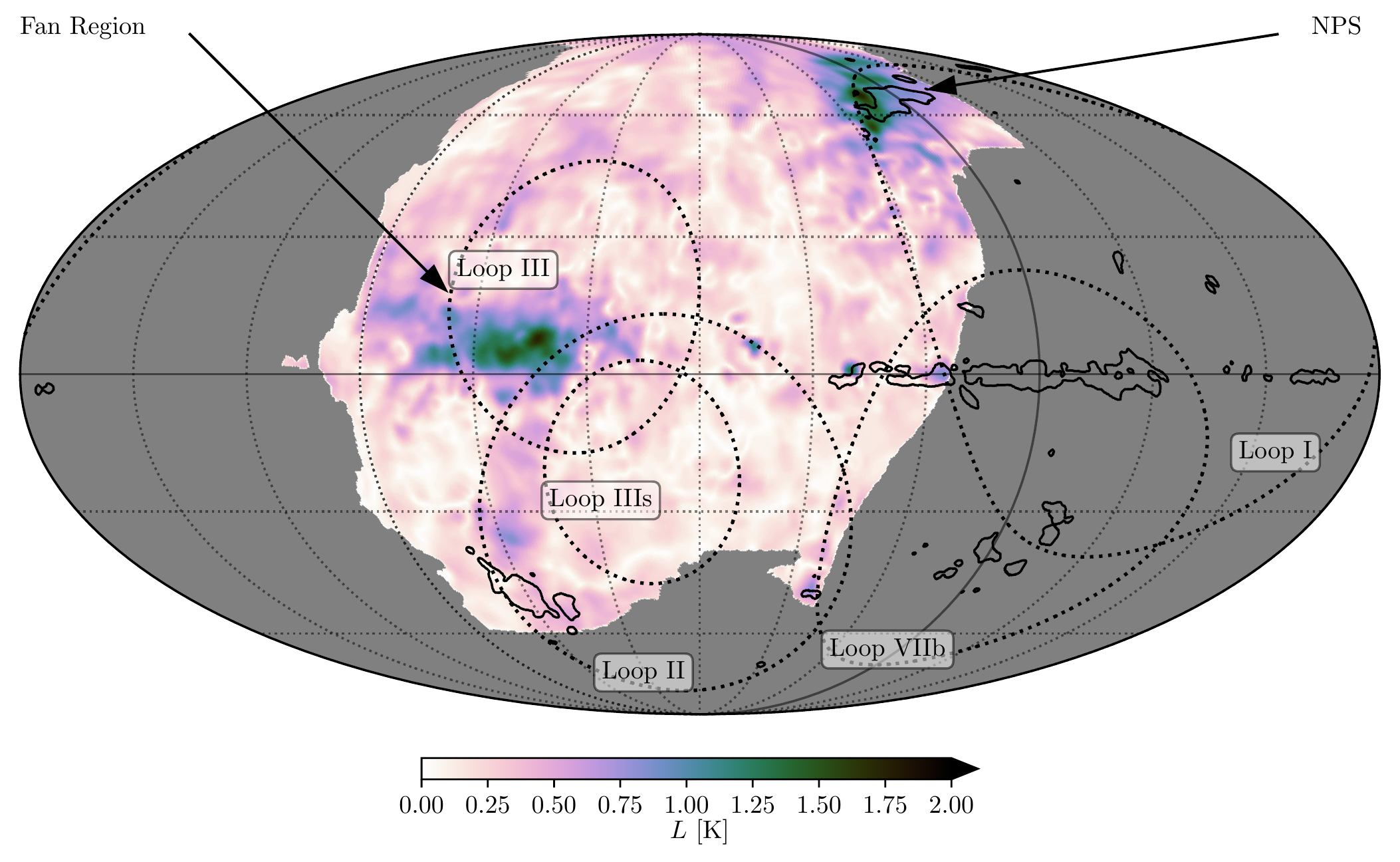}
		\caption{~820\,MHz}
		\label{fig:pimap_820}
	\end{subfigure}

	\begin{subfigure}[b]{\textwidth}
		\includegraphics[width=\textwidth]{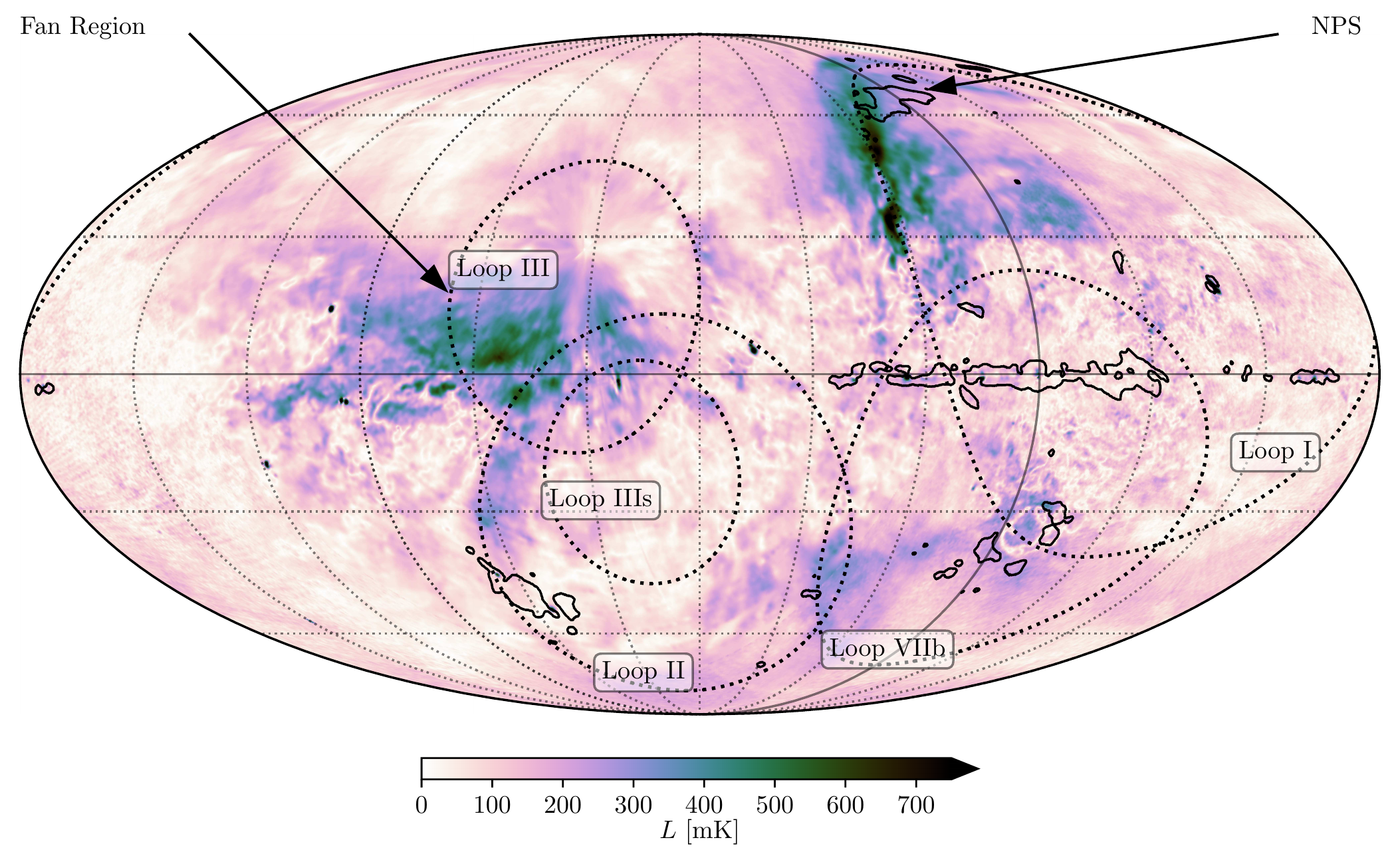}
		\caption{~1410\,MHz}
		\label{fig:pimapallsky_drao}
	\end{subfigure}
\end{figure*}

\begin{figure*}\ContinuedFloat
	\begin{subfigure}[b]{\textwidth}
		\includegraphics[width=\textwidth]{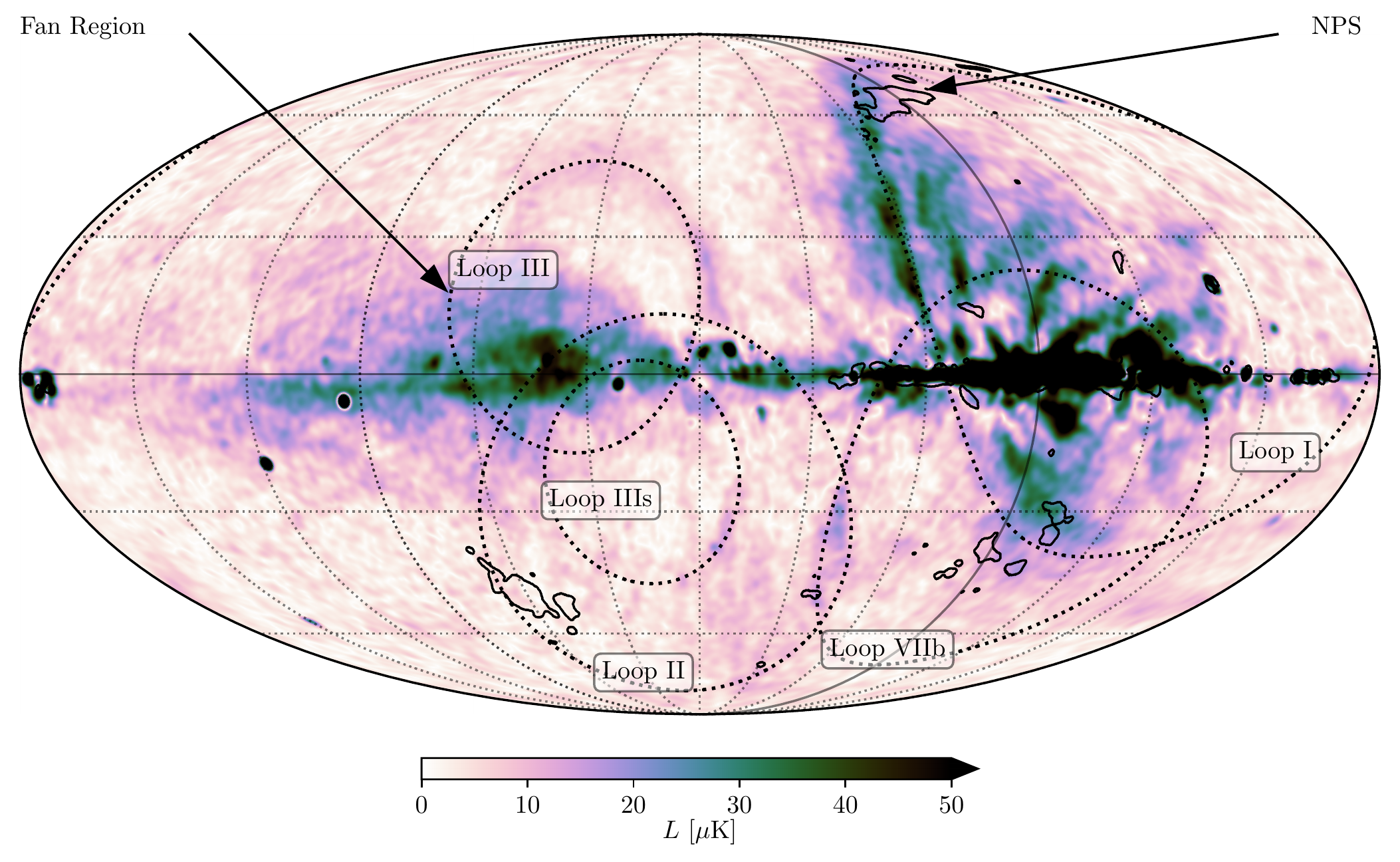}
		\caption{~30\,GHz}
		\label{fig:pimapallsky_planck}
	\end{subfigure}
	\caption{All-sky maps of polarized intensity ($L$) from \subref{fig:pimapallsky} GMIMS-LBS at 408\,MHz \citep[averaged to match the bandwidth of][]{Haslam1982}, \subref{fig:pimap_610} and \subref{fig:pimap_820} \citet{Brouw1976} at 610 and 820\,MHz, \subref{fig:pimapallsky_drao} \citet{Wolleben2006} and \citet{Testori2008} at 1410\,MHz, and \subref{fig:pimapallsky_planck} 30\,GHz from \textit{Planck}. For all maps we use Mollweide projection, centred on $l,b=(90\degr,0\degr)$. We also highlight the locations of notable polarized emission, such as G150$-$50 and the North Polar Spur (NPS), and label some of the large-scale radio loops as described by \citet{Vidal2015}. In black solid contours, we show the $25\sigma$ (1.5\,K) polarized intensity from GMIMS-LBS. We overlay graticules every $30\degr$ in longitude and latitude. Note that the Galactic plane and Centaurus A suffer from Stokes $I$ leakage in \subref{fig:pimapallsky}.}
	\label{fig:allpimapallsky}
\end{figure*}

\begin{figure}
	\centering
	\includegraphics[width=\columnwidth]{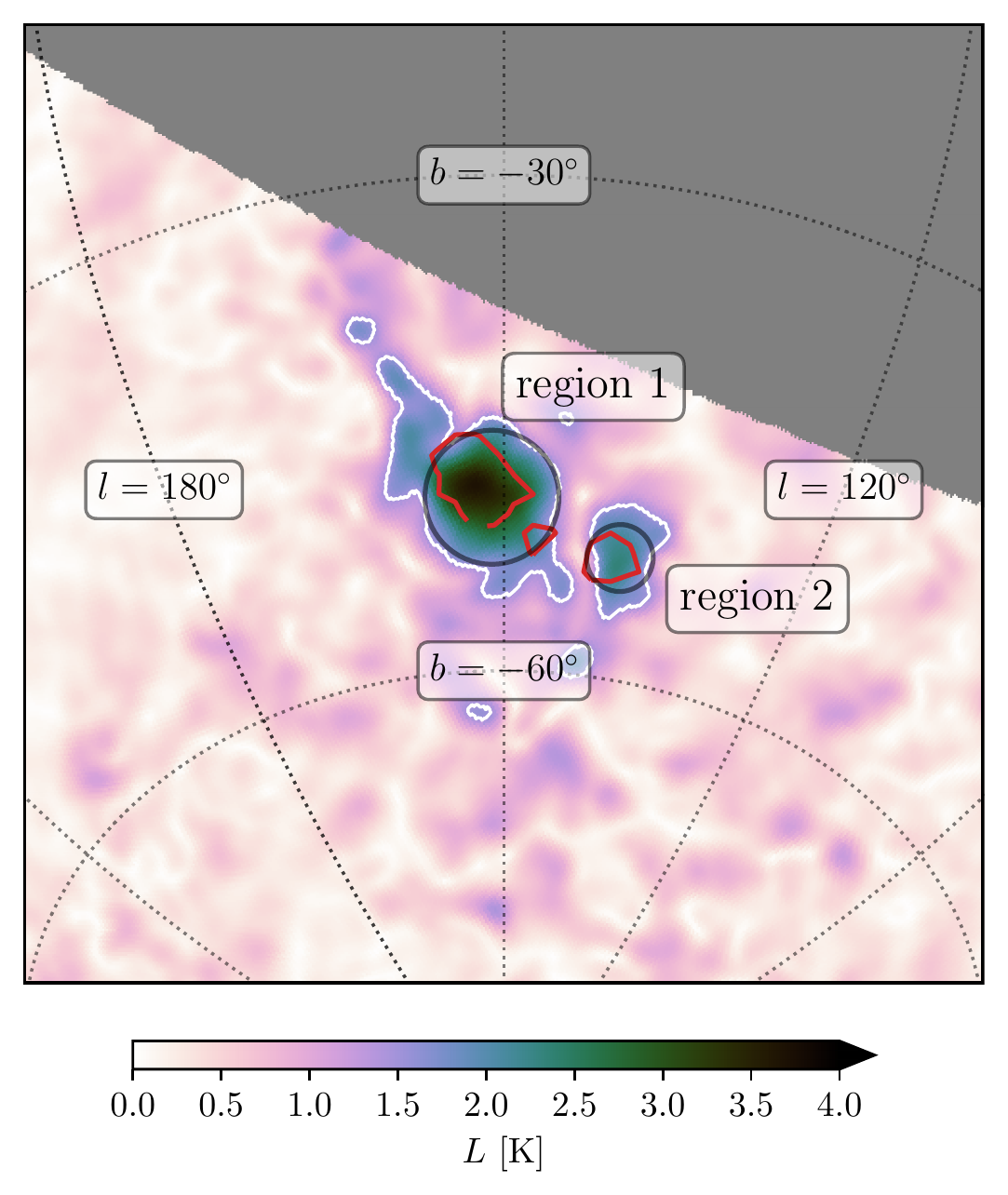}
	\caption{Map of polarized intensity ($L$) from GMIMS-LBS at 408\,MHz focusing on G150$-$50. We show this map using an orthographic projection, centred on $l,b=(150\degr,-50\degr)$. As per Fig.~\ref{fig:pimapallsky}, we average the GMIMS-LBS channels to match the bandwidth of \citet{Haslam1982}. The red contour shows the polarized intensity from \citet{Mathewson1965} at 3\,K. In black circles we denote the two regions we refer to as `region 1' (left) and `region 2' (right). We overlay graticules every $30\degr$ in Galactic longitude and latitude.} 
	\label{fig:pimap}
\end{figure}

In Fig.~\ref{fig:pimap} we zoom in on the polarized intensity image of G150$-$50 at 408\,MHz. We see that this area is divided into two bright regions. We highlight these two regions in Fig.~\ref{fig:pimap} with black circles, which are centred on $l,b\sim(151\degr,-50\degr)$ (region 1) and $l,b\sim(139\degr,-53\degr)$ (region 2), respectively. These same regions appear in the \citet{Mathewson1965} map, which we show in red contours. These correspondences reassure us that G150$-$50 is not an artefact in the GMIMS-LBS data. We can therefore be confident that this is a true feature of the polarized sky that needs to be investigated. 

To estimate the extragalactic contribution, we inspect the catalogue of polarized sources in regions 1 and 2 from \citet{Taylor2009}. The integrated flux density ($S$) of all sources in these regions is 0.5 and 0.2\,Jy, respectively. Scaling these values to 408\,MHz with spectral index $\alpha=-0.8$~\citep[e.g.][]{Condon1992, Mauch2003, Smolcic2017}, where $S\propto\nu^{\alpha}$, and then converting to brightness temperature with the GMIMS-LBS beam gives 0.4 and 0.2\,K, respectively. Integrating over the same area from GMIMS-LBS yields a value of 2400\,K in region 1, and 460\,K in region 2. The extragalactic sources are therefore contributing 0.02 and 0.03\% to the integrated polarized brightness temperature. We conclude that extragalactic contributions to the polarized emission in these regions is negligible.

\begin{figure}
	\centering
	\includegraphics[width=\columnwidth]{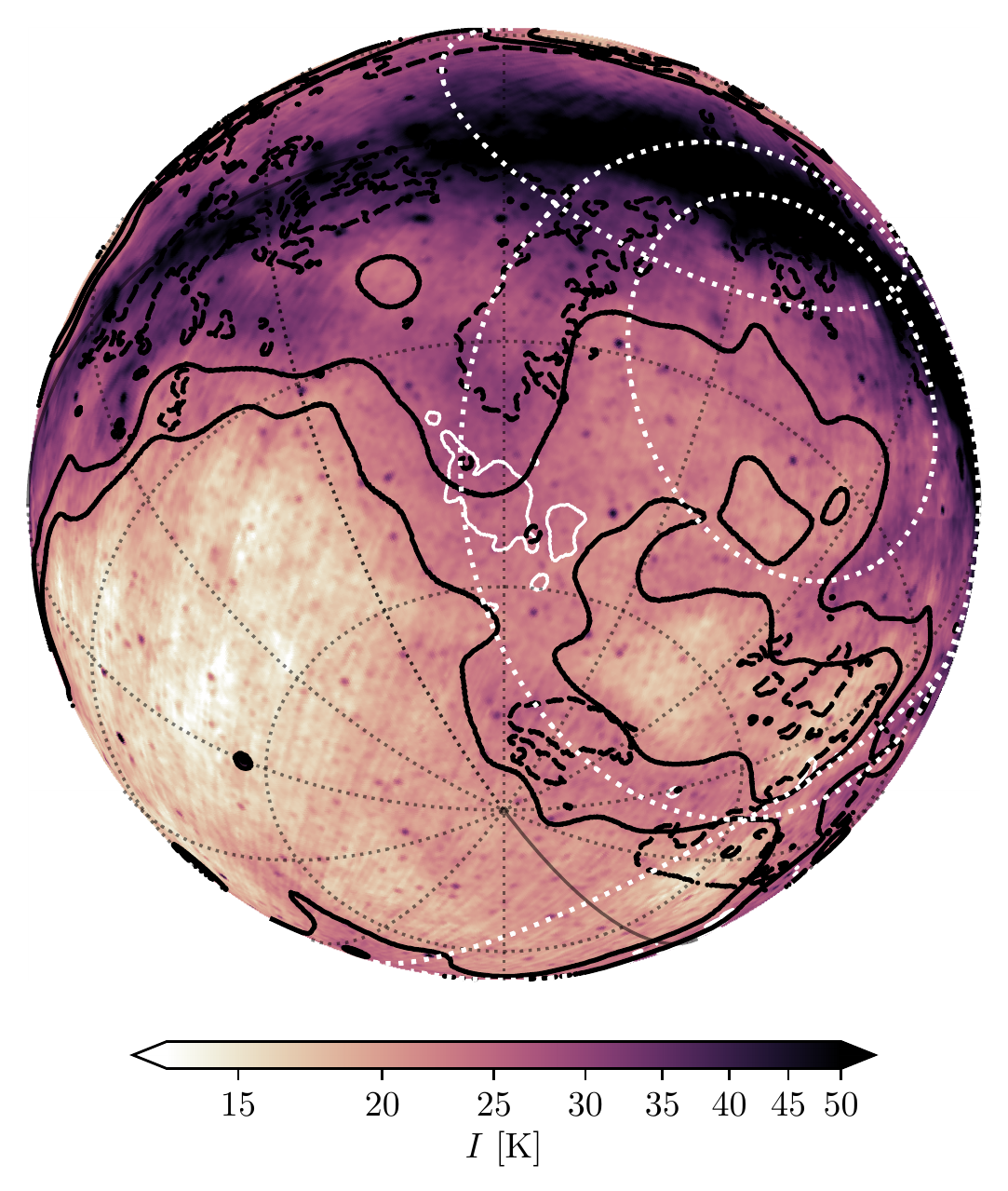}
	\caption{Map of total intensity at 408\,MHz from \citet{Haslam1982}, centred on $l,b=(150\degr,-50\degr)$ using an orthographic projection, and logarithmic colour-scale. In white, solid contours we show the GMIMS-LBS polarized intensity at 408\,MHz (as per Fig.~\ref{fig:pimap}) at the $25\sigma$ (1.5\,K) level. In addition we show the 0.2\,K ($\sim17\sigma$) polarized intensity at 1.4\,GHz in black, dashed contours. In white dotted lines we draw the same radio loops as in Fig.~\ref{fig:allpimapallsky}. We have selected a colour-scale such that the diffuse emission from Loop II is highlighted, and overlay contours of the smoothed Stokes I map at 22 and 26\,K in black to further highlight the loop.}
	\label{fig:imap}
\end{figure}

In Fig.~\ref{fig:imap} we show the 408\,MHz map using the \citet{Haslam1982} Stokes $I$ data. On this image we also overlay contours of linearly polarized intensity ($L$) at 408\,MHz and 1.4\,GHz, as well as the \citet{Haslam1982} smoothed map with a $5\degr$ Gaussian kernel. We use the smoothed Stokes $I$ contours to highlight the diffuse emission associated with Loop II. As discussed above, the G150$-$50 region appears to coincide with this emission region. This loop is described as a large circular feature on the sky, following a circle centred on  $l,b\sim(100.0\degr,-32.5\degr)$ with a radius of $45.5\degr$, which we show in white dashed lines. In the \citet{Haslam1982} map the loop is very diffuse, and appears to be dropping in total intensity, from North to South across the region of G150$-$50.

To better understand the nature of the polarized emission we calculate the polarization fraction at 408\,MHz and 1.4\,GHz. Performing such analysis comes with some difficulties, however. First, the total intensity images have a significant uncertainty in their zero level: $\pm3\,$K at 408\,MHz~\citep{Haslam1982} and $\pm0.5\,$K at 1420\,MHz~\citep{Reich1982}. Second, to accurately describe the emission, the polarization fraction should be calculated from the same source of emission. The total intensity measurements contain non-synchrotron emission from the cosmic microwave background (CMB), and additional unresolved emission from extragalactic background sources. At these frequencies and latitudes, the radio spectrum is dominated by synchrotron emission~\citep{Peterson2002}. As such, we do not consider the negligible contribution of free-free absorption and emission. The CMB has a very smooth brightness temperature of 2.7\,K~\citep{Mather1999}. The problem of correcting both the zero level offset and non-Galactic emission was investigated by \citet{Reich1988} on these exact data. They bootstrap a zero level correction by assuming a correct scale at 1420\,MHz and extrapolate to 408\,MHz. Combining the zero level correction with the extragalactic background estimate from \citet{Bridle1967} and \citet{Lawson1987}, \citet{Reich1988} give total offset values of $3.7\pm0.85\,$K and $2.8\pm0.03\,$K at 408\,MHz and 1.4\,GHz, respectively. We subtract these values from both the 408\,MHz and 1.4\,GHz Stokes $I$ images.

An additional factor to consider is the `polarization horizon'~\citep{Uyaniker2003}. The size of the telescope beam, combined with Faraday rotation in the MIM of the Galaxy, results in polarized emission beyond a particular distance becoming depolarized. \citet{Dickey2018} found the distance for this `horizon' to be $\lesssim500$\,pc for GMIMS-LBS at Galactic latitudes $||b||>25\degr$. We now consider the distribution of Galactic synchrotron emission, which was modelled by \citet{Beuermann1985} as a thin plus a thick disk, with the thick disk emitting 90\% of the total power. At the position of the Sun, the scale heights ($h$) of the thick and thin disk are 1.5\,kpc and 150\,pc, respectively \citep{Ferriere2001, Haverkorn2012a}. For a given latitude, the distance ($d$) to the edge of the emitting disk is given by:
\begin{equation}
    d  = \frac{h}{\sin{||b||}}.
\end{equation}
At a latitude of $-50\degr$, the distance to the edge of the thick synchrotron disk is $\sim2\,$kpc. If G150$-$50 is part of Loop II, however, a discrete emitting object, the emission may be coming from much closer than the edge of the disk. Taking the size and distance estimates of Loop II as 180\,pc and 100\,pc, respectively, we would expect the entirety of Loop II to be within the GMIMS-LBS polarization horizon. In any case, it is important to note that the polarization horizon affects the observed polarized intensity in a complicated manner. As discussed by \citet{Hill2018}, the polarization horizon is not necessarily a `polarization wall'. That is, if the polarization horizon causes a change in the measured polarized emissivity, as a function of $\lambda^2$, the resulting effect on the polarized intensity could be destructive or constructive. Therefore, if the GMIMS-LBS observations are not sensitive to some G150$-$50 polarized emission that may be beyond the polarization horizon, this would modulate the polarization fraction in a way that is very hard to predict. Here we are unable to make a correction for such an effect, as we lack a complete model of the synchrotron emissivity in this direction.

\begin{figure}
	\centering
	\begin{subfigure}[b]{\columnwidth}
	    \includegraphics[width=\columnwidth]{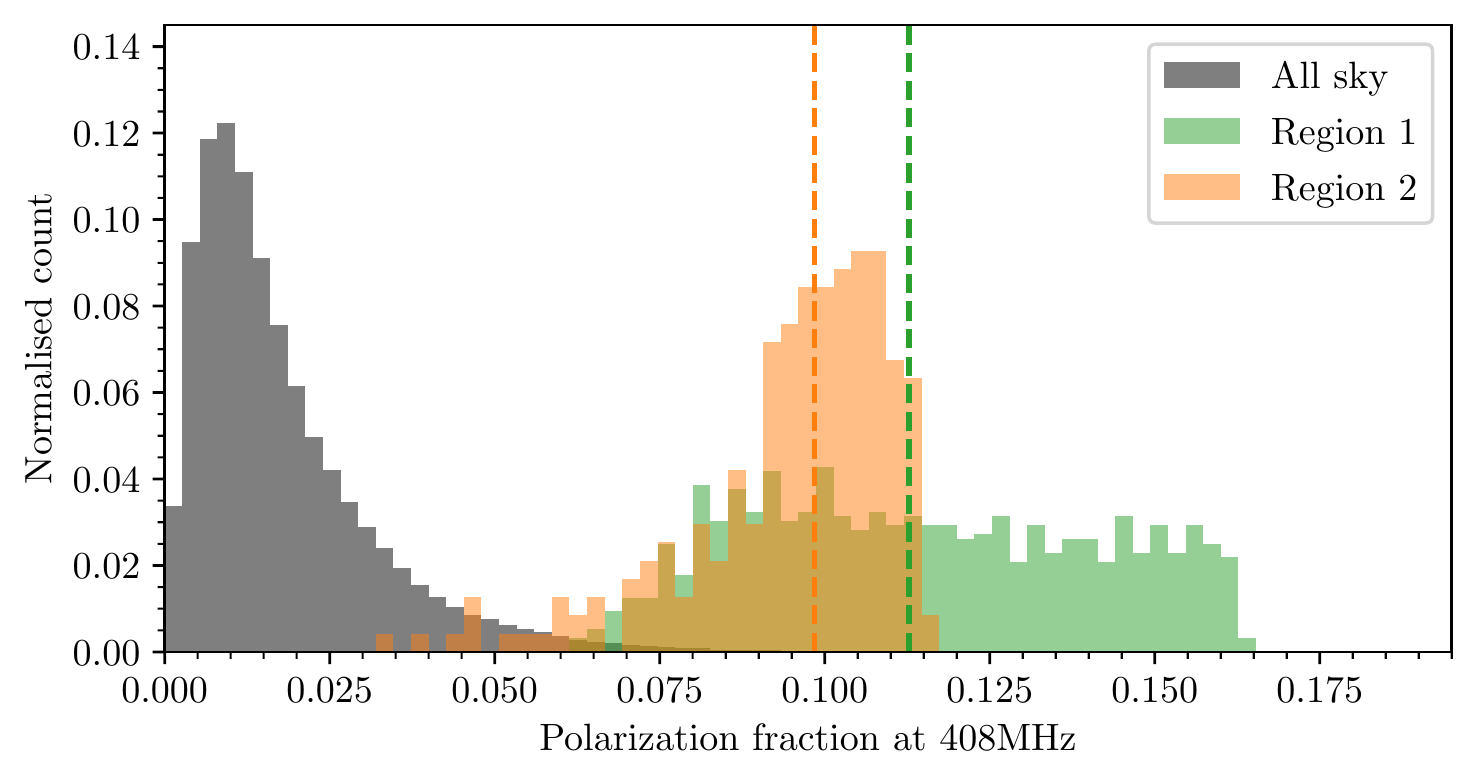}
	    \caption{}
	    \label{fig:408hist}
	\end{subfigure}
	\begin{subfigure}[b]{\columnwidth}
	    \includegraphics[width=\columnwidth]{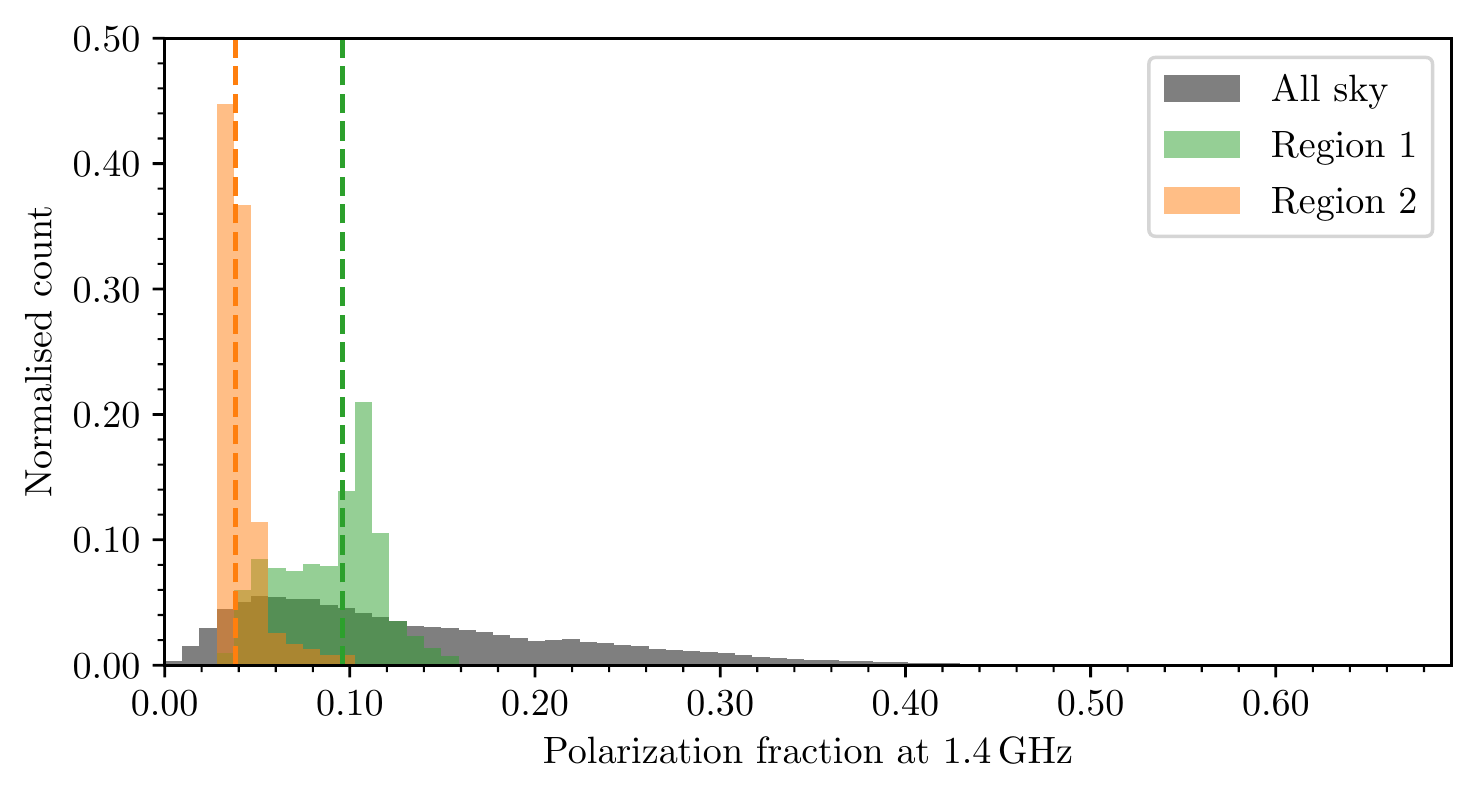}
	    \caption{}
	    \label{fig:1411hist}
	\end{subfigure}
	\caption{Histograms of polarization fraction ($p_0$) at \subref{fig:408hist} 408\,MHz and \subref{fig:1411hist} 1.4\,GHz, normalised such that the total area is unity. Grey: The entire sky as observed by each survey. Green: Region 1, a circular region centred on $l,b\sim(151\degr,-50\degr)$, with a radius of $4\degr$. Orange: Region 2, a circular region centred on $l,b\sim(139\degr,-53\degr)$, with a radius of $2\degr$. Dashed lines indicate the median value in each region.}
	\label{fig:hist}
\end{figure}

We compute the polarization fraction in both regions 1 and 2 of G150$-$50, and the entire observed sky, at both 408\,MHz and 1.4\,GHz. In Fig.~\ref{fig:hist} we show the distribution of polarization fraction across each of these areas. We extract data towards the two G150$-$50 regions from circular cut-outs, centred on the coordinates above, with radii of $4\degr$ and $2\degr$, respectively. The median polarization fractions in regions 1 and 2 are respectively $11.3\pm0.1$\% and $9.7\pm0.1$\% at 408\,MHz, and $9.0\pm0.3$\% and $4.1\pm0.5$\% at 1.4\,GHz. We note that our reported errors here are inclusive of zero level uncertainties. At 408\,MHz, this fraction is anomalously high; regions 1 and 2 have a higher fractional polarization than 99.9\% and 99.8\% of the observed sky, respectively. We also calculate the signal-to-noise ratio (SNR) in the polarized intensity in both regions for the GMIMS-LBS observations, with a noise per channel of 60\,mK~\citep{Wolleben2019}. We find that the SNR does not go below 12 across the GMIMS-LBS band. As such, we do not apply a correction for the negligible polarization bias \citep{Wardle1974}.

An outstanding question remains: why is G150$-$50 so prominent at 408\,MHz, but not at higher frequencies? From the morphological considerations alone, we can begin to draw some conclusions. G150$-$50 is unlikely to arise from a region of enhanced emissivity; in such a case we would expect to see a corresponding, localised, region of bright emission in total intensity. Whilst we do see Stokes $I$ emission associated with Loop II, this extends much further than G150$-$50. A caveat to this notion is that observed emission volumes in total and polarized intensity may be different. That is, there may be some polarized emission within the polarization horizon that only stands out from its surroundings because more distant emission is depolarized. That same emission would also be present in total intensity, but would be lost to confusion from background emission. We can dismiss this, however, as there is no corresponding bright polarized emission at higher frequencies, where the polarization horizon is much further away. This is compounded by the fact that depolarization models would predict higher polarization fraction at higher frequencies, whereas we observe the opposite. 

\begin{figure*}
    	\begin{subfigure}[b]{\columnwidth}
    	    \includegraphics[width=\columnwidth]{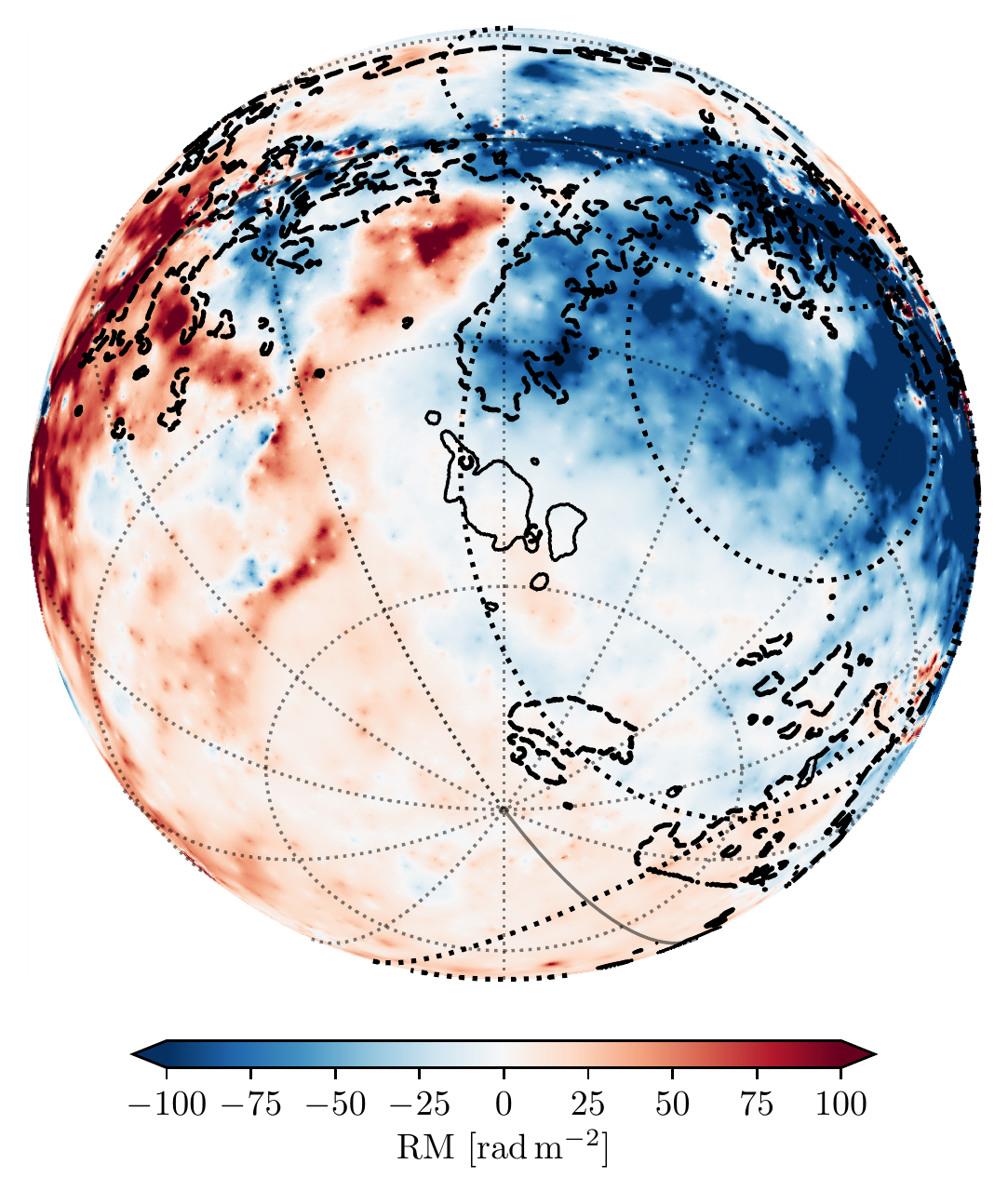}
    	    \caption{}
    	    \label{fig:exgalrm}
    	\end{subfigure}
    	\begin{subfigure}[b]{\columnwidth}
    	    \adjustbox{raise=5.65em}{\includegraphics[width=\columnwidth]{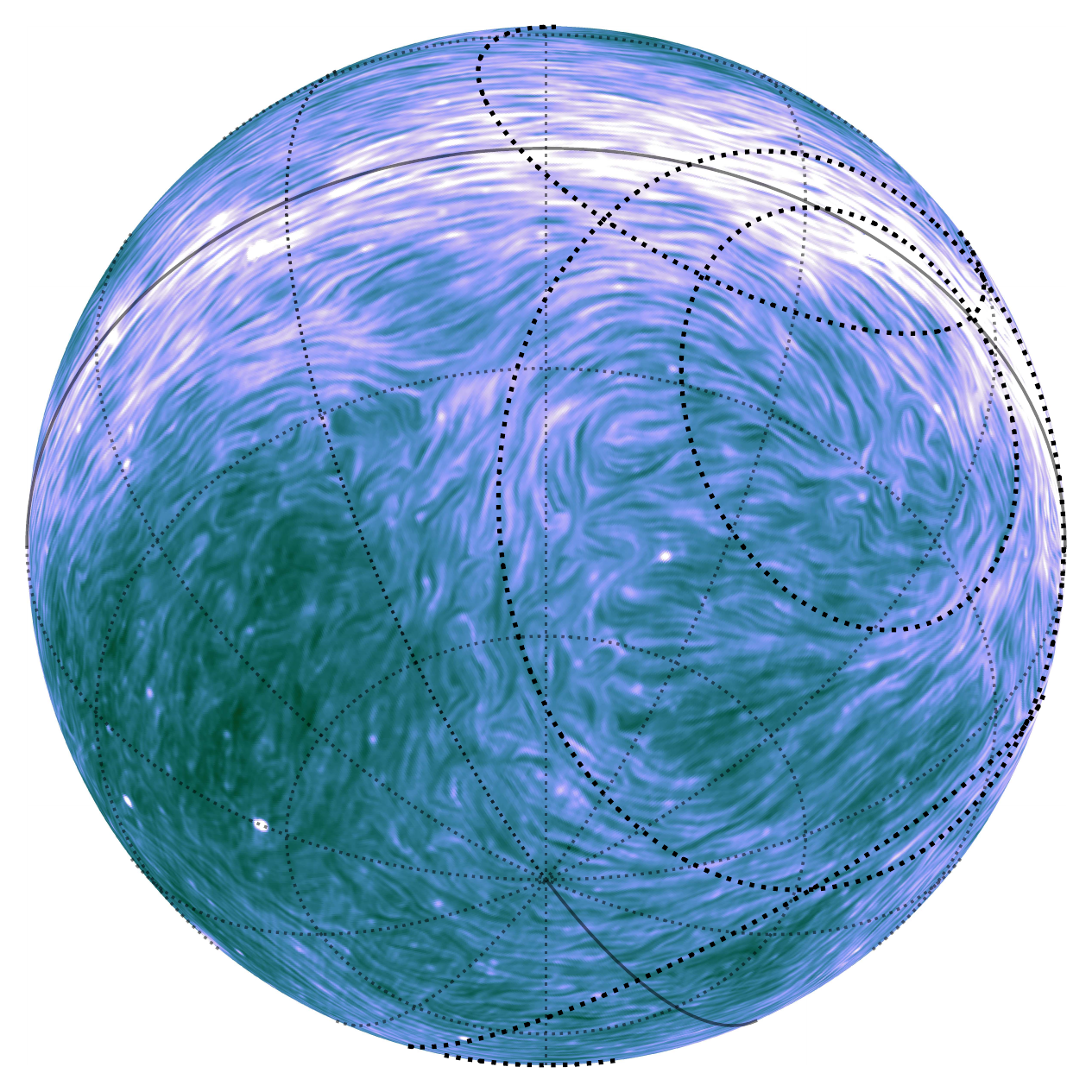}}
    	    \caption{}
    	    \label{fig:LIC}
    	\end{subfigure}
	\caption{\subref{fig:exgalrm} Galactic RM values from \citet{Hutschenreuter2021}. The line of Loop II follows a line of low RM. \subref{fig:LIC} Plane-of-sky magnetic field orientation from \citet{Planck2018-diffuse}. We note that the magnetic field in the plane of the sky primarily follows the path of Loop II. The vector field is visualised using line integral convolution \citep{Cabral1993}, weighted by the normalised Stokes $I$ emission from \citet{Haslam1982}. The projection and overlays are the same as in Fig.~\ref{fig:imap}, with the exception of total intensity contours which we do not show here.}
	\label{fig:bfields}
\end{figure*}

Bearing in mind the consistently high polarized fraction we find at both 408\,MHz and 1.4\,GHz towards G150$-$50, we now turn our attention to both the Galactic RM, as mapped by \citet{Hutschenreuter2021}, and the synchrotron polarization vectors from \textit{Planck}. These observations trace the LOS and plane-of-sky (POS) magnetic fields, respectively. In Fig.~\ref{fig:exgalrm} we show the map of RM, which is derived from extragalactic sources and therefore probes the entire LOS through the Milky Way. We find that the region where Loop II appears in Stokes $I$ emission corresponds to a ridge of low Galactic $||\text{RM}||$. Additionally, the region where we find G150$-$50 is coincident with a large region of almost 0\,\radms. Looking to the \textit{Planck} synchrotron polarization, we visualise the POS magnetic vector-field using line integral convolution \citep{Cabral1993}. Here we have taken the magnetic lines to be perpendicular to the electric field as traced by the polarization angles. In Fig.~\ref{fig:LIC} we show the magnetic vector-field, weighted by the Stokes $I$ emission from \citet{Haslam1982}. As reported by \citet{Planck2016-diffuseforeground}, the magnetic field vectors in the vicinity of Loop II follow its circular path. Similar configurations have been observed in other Galactic loops and filaments \citep[see e.g.][]{Vidal2015}. The combination of low Galactic RM, Stokes $I$ emission, and coherent POS magnetic field vectors together indicate magnetic fields that are primarily in the POS and coherent.

We need to seek a physical model that can efficiently explain these morphological observations. We now look to further inform this model using the broad-band spectro-polarimetric information through both Faraday tomography and $QU$-fitting.

\subsection{Faraday Tomography}\label{sec:tomography}
The G150$-$50 region exhibits remarkable spectral structure. Here we investigate this through both Faraday tomography and single-pixel RM-synthesis. First, we apply moment analysis of the GMIMS-LBS Faraday spectrum cubes. Moment analysis is a powerful way to inspect the results of Faraday tomography \citep{Dickey2018}. The zeroth, first, second, and third moments of Faraday spectra are defined respectively as the integrated polarized intensity, 
\begin{equation}
	M_0 \equiv \Delta\phi\sum_{i=1}^{n}L_i,
	\label{eqn:mom0}
\end{equation}
the polarized intensity-weighted peak Faraday depth,
\begin{equation}
	M_1 \equiv \frac{ \Delta\phi\sum_{i=1}^{n}L_i\phi_i}{M_0},
	\label{eqn:mom1}
\end{equation}
the polarized intensity-weighted width of the peak,
\begin{equation}
	M_2 \equiv \frac{\Delta\phi\sum_{i=1}^{n}L_i\left( \phi_i - M_1 \right)^2 }{M_0}
	\label{eqn:mom2}
\end{equation}
and the polarized intensity-weighted skewness of the peak,
\begin{equation}
	M_3 \equiv \frac{\Delta\phi\sum_{i=1}^{n}L_i\left( \phi_i - M_1 \right)^3 }{M_0}.
	\label{eqn:mom3}
\end{equation}
It is useful to use the root-normalized second and third moments, $m_2$ and $m_3$, which are simply defined as:
\begin{align*}
    m_2 = M_2^{1/2}, \\
    m_3 = M_3^{1/3}.
	\label{eqn:rtmom}
\end{align*}
Both of these moments have units of \radms, similar to $M_1$. From here we will refer to $m_2$ and $m_3$ as the second and third moments, respectively. For the higher moments, we mask the data using signal-to-noise ratio of 3 from the zeroth moment.

\begin{figure*}
	\centering
	\begin{subfigure}[b]{0.245\textwidth}
	    \includegraphics[height=1.3\columnwidth]{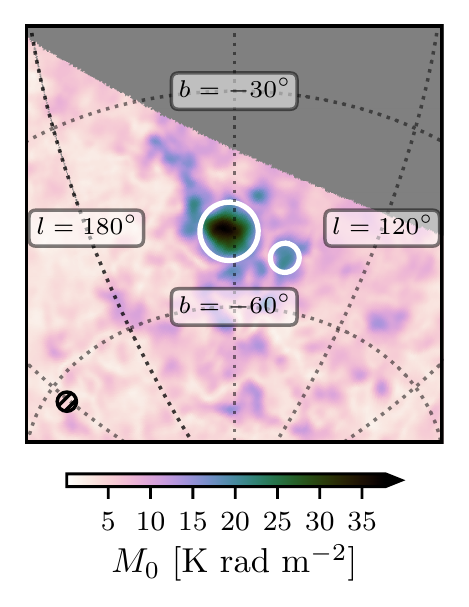}
	    \caption{}
	    \label{fig:m0map}
	\end{subfigure}
	\begin{subfigure}[b]{0.245\textwidth}
	    \includegraphics[height=1.3\columnwidth]{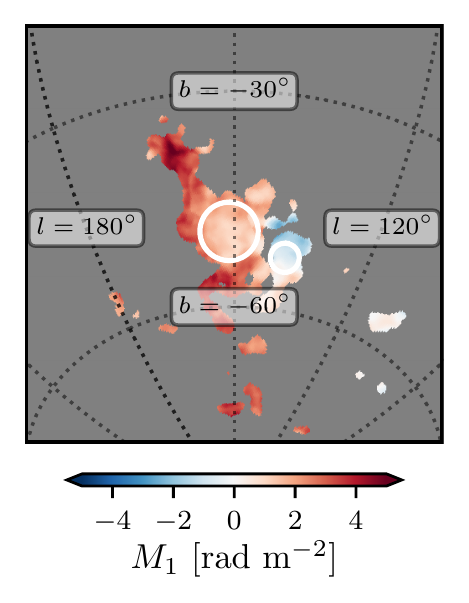}
	    \caption{}
	    \label{fig:m1map}
	\end{subfigure}
	\begin{subfigure}[b]{0.245\textwidth}
	    \includegraphics[height=1.3\columnwidth]{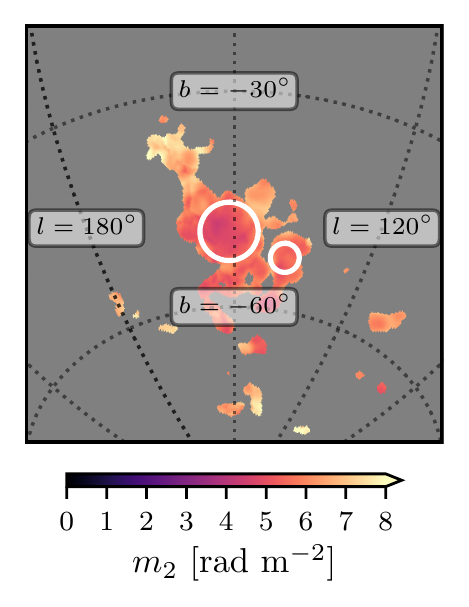}
	    \caption{}
	    \label{fig:m2map}
	\end{subfigure}
	\begin{subfigure}[b]{0.245\textwidth}
	    \includegraphics[height=1.3\columnwidth]{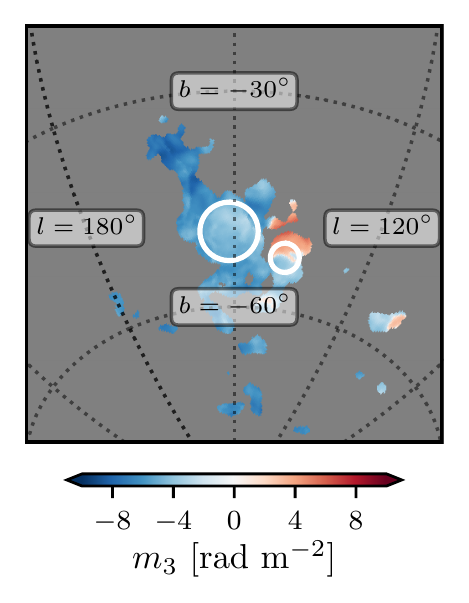}
	    \caption{}
	    \label{fig:m3map}
	\end{subfigure}
	\caption{Faraday moments of the Faraday spectra from GMIMS-LBS for $|\phi|\leq20$\,\radms. \subref{fig:m0map} The zeroth moment ($M_0$); the sum of the Faraday spectrum multiplied by the Faraday depth channel width ($0.5\,$\radms). \subref{fig:m1map} The first moment ($M_1$); the intensity-weighted mean of the Faraday spectrum. \subref{fig:m2map} The square root of the second moment, $m_2 = {M_2}^{1/2}$; the intensity-weighted width of the Faraday spectrum. \subref{fig:m3map} The cube root of the third moment, $m_3 = {M_3}^{1/3}$; the intensity-weighted skewness of the Faraday spectrum, where positive values indicate a skew to positive Faraday depth and \textit{vice versa}. We use the same projection as in Fig.~\ref{fig:pimap} and show regions 1 and 2 by white circles. In \subref{fig:m0map} we also overlay the size of the beam at FWHM as a hatched circle in the lower left of the frame.}
	\label{fig:moments}
\end{figure*}

The G150$-$50 region stands out in the moment analysis by \citet{Dickey2018} (see their fig. 5). However, inspecting the Faraday depth spectra of G150$-$50, we find corresponding structure only within $||\phi||\leq20$\,\radms. Here we compute the moments using this range in $\phi$, capturing the primary Faraday depth structure we observe towards G150$-$50, which we show in Fig.~\ref{fig:moments}.

The zeroth moment, as shown in Fig.~\ref{fig:m0map}, closely resembles the 408\,MHz $L$ map (Fig.~\ref{fig:pimap}). Again we see a depolarized feature separating regions 1 and 2. In Fig.~\ref{fig:m1map} we show the first moment, masking out regions with a SNR$<3$ in the zeroth moment. Regions 1 and 2 are at Faraday depths of opposite sign with a strong gradient between them. Gradients in Faraday depth can generate a depolarization canal~\citep[e.g.][]{Haverkorn2000,Fletcher2007}. For a resolved, linear gradient across a Gaussian beam, the degree of depolarization (DP) is given by \citet{Sokoloff1998} as:
\begin{equation}
	\text{DP} \equiv \frac{L}{L_0} = \left|\left| e^{2i\phi_0\lambda^2  - 2(\Delta\phi\lambda^2)^2} \right|\right|,
	\label{eqn:dpbeam}
\end{equation}
where $L$ and $L_0$ are respectively the observed and intrinsic ($\lambda^2=0$) polarized intensity, $\phi_0$ is the Faraday depth at the centre of the beam with a variation of $\Delta\phi$. Across G150$-$50 we find an average gradient of $0.2$\,\radms\ $\deg^{-1}$, which is not sufficient to cause any significant depolarization. At the location of the depolarized feature, however, we find a gradient of $\Delta\phi\approx3\,$\radms\ in the GMIMS beam. Taking the mean frequency of 390\,MHz and $\phi_0=0$ at the location of the canal, we find a DP of 0.4\,per\,cent. That is, the gradient at this location reduces the polarized intensity to 0.4\,per\,cent of its original value. We therefore still consider G150$-$50 as a single object with changing Faraday depth structure across the sky. The appearance of the depolarization canal is, in part, due to the resolution of the observations.

Inspecting the second moment map, we find a relatively consistent value of around 5\,\radms; with the average values on regions 1 and 2 being 5 and 5.5\,\radms, respectively. For a single, noiseless, Gaussian feature, the value of $m_2$ corresponds to the standard deviation of that Gaussian~\citep{Dickey2018}. Converting the average values for regions 1 and 2 to the equivalent FWHM gives 11.9 and 12.9\,\radms, respectively. These widths are approximately twice that of the Faraday resolution for GMIMS-LBS. We also note that these are larger than the $\phi_\text{max-scale}$ of 8\,\radms\ for GMIMS-LBS. Therefore, if this feature corresponds to a single, broad feature, GMIMS-LBS is sensitive to less than half of its emission. The corresponding `true' peak of such a feature would therefore be much greater than measured by the GMIMS-LBS Faraday spectra.

We show the third moment map in Fig.~\ref{fig:m3map}. We find that regions 1 and 2 mostly share a common, negative value of $m_3$, with a sharp sign-change within region 2.  Care should be taken in the interpretation of this value, however, as the \texttt{RM-CLEAN} algorithm, as well as noise, can introduce false or non-physical features in the Faraday spectrum, especially in the higher moments.

Inspecting the original Faraday spectra of these regions, we find that Faraday spectra agree with the indications from their moments. These spectra resemble the ones we produce later in Section~\ref{sec:singlepix} and show in Figs.~\ref{fig:spec} and \ref{fig:FDF}. The spectra show broad Faraday structure relative to the RM spread function (RMSF), indicating that the Faraday spectra are either Faraday thick or multi-component. Additionally, the broadening appears to be skewed towards negative $\phi$ across the entirety of G150$-$50. In the part of region 2 exhibiting positive values of $m_3$, we find a minor peak at positive Faraday depths (separate from the broad structure) which is contributing to the third moment. The primary feature, however, still shows a skew towards negative Faraday depths.

\subsubsection{Single-pixel RM synthesis}\label{sec:singlepix}
\begin{figure}
	\centering
	\includegraphics[width=\columnwidth]{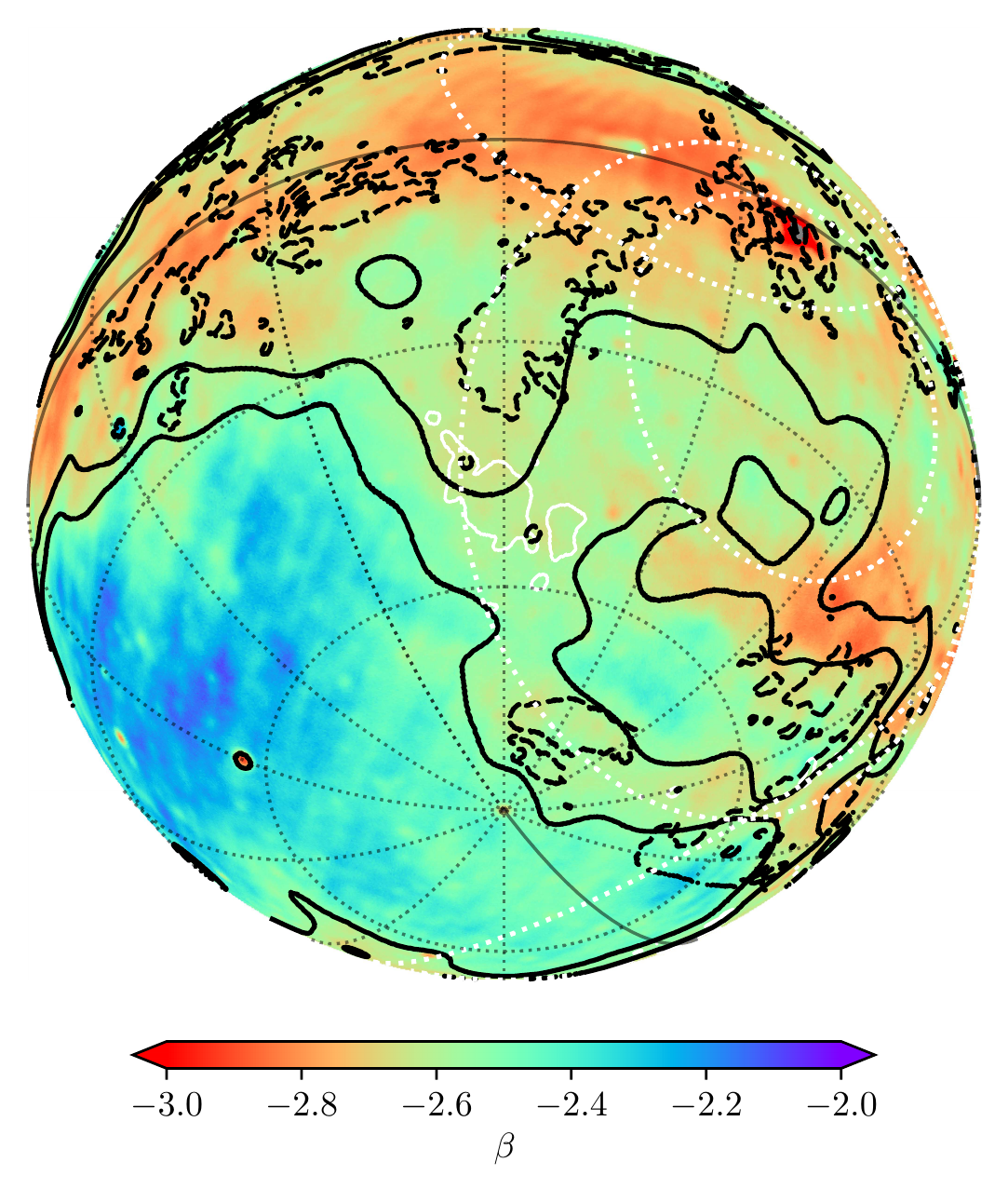}
	\caption{The brightness temperature spectral index ($\beta$) between 408\,MHz and 1.4\,GHz in total intensity. These indices were computed using the background-corrected \citet{Haslam1982} and \citet{Reich1982,Reich1986,Reich2001} surveys. }
	\label{fig:beta}
\end{figure}

We now turn our attention to the spectra as a function of $\lambda^2$, including both GMIMS-LBS and the additional, higher-frequency data. We wish to obtain the fractional polarized spectra across the entire $\lambda^2$ range. To this end, we assume a power-law Stokes $I$ model with spectral index $\beta$ ($T_b\propto\nu^\beta$). For maps such as these, errors in the spectral index are usually dominated by systematic uncertainty in the zero level of each survey. As discussed in Section~\ref{sec:morphology}, we have adopted the corrections of \citet{Reich1988}, and we are therefore assuming a correct total intensity scale at 1.4\,GHz. Using the background-corrected total intensity maps at 408\,MHz and 1.4\,GHz we compute the spectral index across the sky, which we show in Fig.~\ref{fig:beta}. Along the path of Loop II we find a consistent value of $\beta=-2.6$ ($\beta=-2.62\pm0.01$ and $\beta=-2.61\pm0.01$ in regions 1 and 2, respectively -- see Fig.~\ref{fig:beta}).

\begin{figure}
	\centering
	\includegraphics[width=\columnwidth]{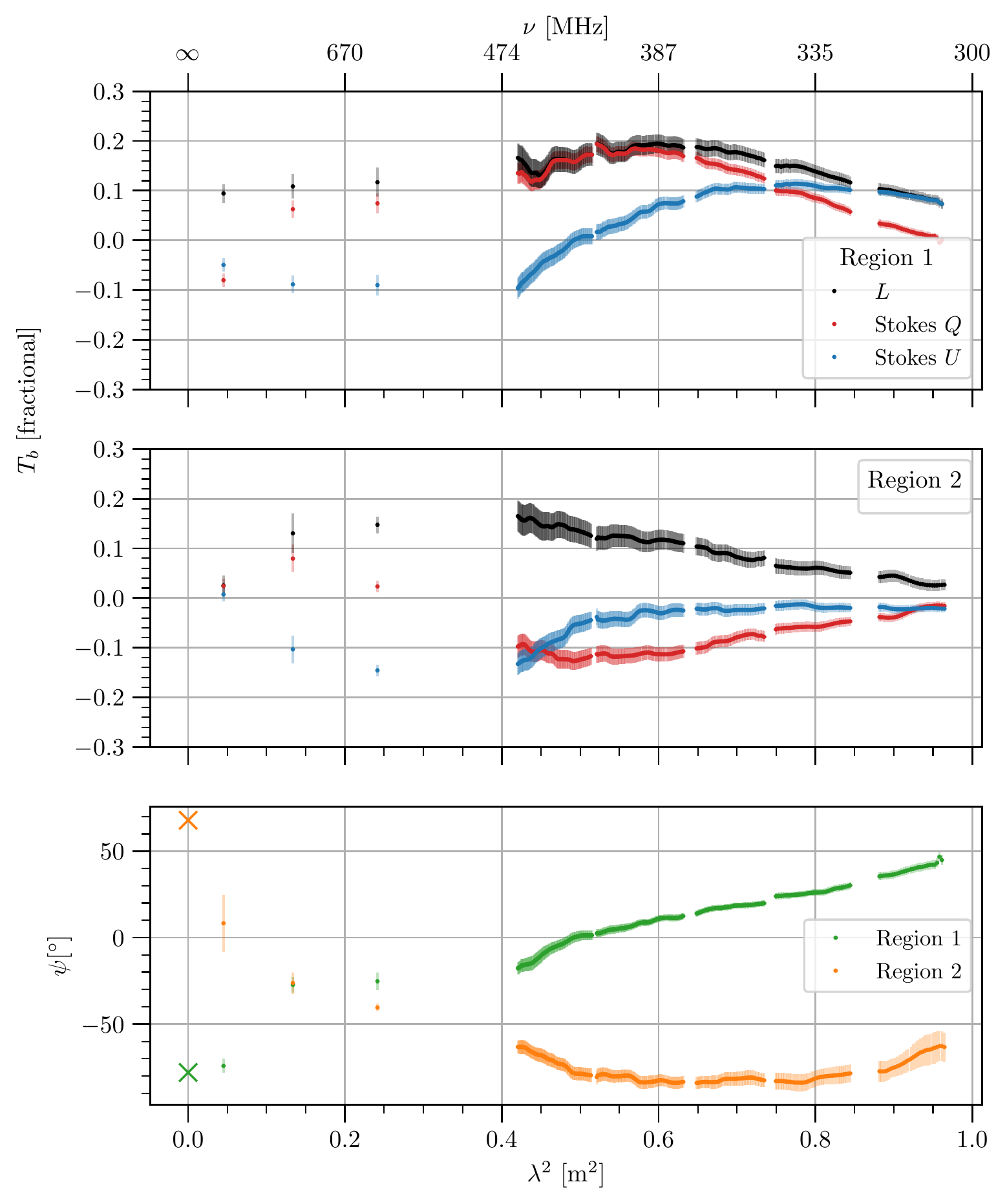}
	\caption{The linear polarization spectra from the centres of regions 1 and 2 as a function of $\lambda^2$. Here we present the linear polarization as a fraction of a power-law fit to the median Stokes $I$ spectrum. Upper panel: Stokes $Q$, $U$, and polarized intensity ($L$) in region 1. Middle panel: Same as upper panel for region 2. Lower panel: Polarization angle ($\psi$) in both regions 1 (green) and 2 (orange). We show the polarization angles from the $Planck$ synchrotron data as crosses.}
	\label{fig:spec}
\end{figure}

In Fig.~\ref{fig:spec}, we show the fractional polarization spectra from the centres of regions 1 and 2. Even though we do not compute fractional $Q$ and $U$ at 30\,GHz, we do show the polarization angles for these data. In both regions, we note the smooth continuity across the entire band, encompassing four unique surveys, again reassuring us of the accuracy of the polarization observations. Additionally, the spectra in both regions have a `hump-like' appearance in polarized intensity, peaking around 0.6\,m$^2$ and 0.4\,m$^2$ in regions 1 and 2, respectively. Taking the angles from $Planck$ to correspond to $\lambda\sim0$\,m$^2$, we find intrinsic polarization angles of $-75$\degr($=+105$\degr) and $+59$\degr for regions 1 and 2. Most notably, the polarization angles vary non-linearly with $\lambda^2$, which clearly demonstrates that a single RM cannot describe these spectra. Therefore we must apply more sophisticated analysis methods to these spectra.

Firstly, we apply RM-synthesis to the GMIMS-LBS component of these fractional data using \textsc{rm-tools}~\footnote{\url{https://github.com/CIRADA-Tools/RM-Tools}}~\citep{Purcell2020}.  Our aim here is to minimise potentially spurious effects of \texttt{RM-CLEAN}. We do not use the additional high-frequency data, as the large $\lambda^2$ gaps create large side-lobes in the resulting RMSF. In our RM-synthesis application, we take a number of steps beyond the original data release: we use the fractional spectra for RM-synthesis (rather than absolute $Q$ and $U$), we have removed potential RFI artefacts, we stop \texttt{RM-CLEAN} at a higher cut-off of 180\,mK ($\sim3\sigma_{Q,U}$), and we use much finer Faraday depth channels ($\Delta\phi=0.01\,$\radms). A higher cutoff avoids any risk of over-cleaning, and finer Faraday depth channels allows for \texttt{RM-CLEAN} to fit model components more precisely, which we find improves the performance of the algorithm.

\begin{figure*}
	\centering
	\includegraphics[width=\textwidth]{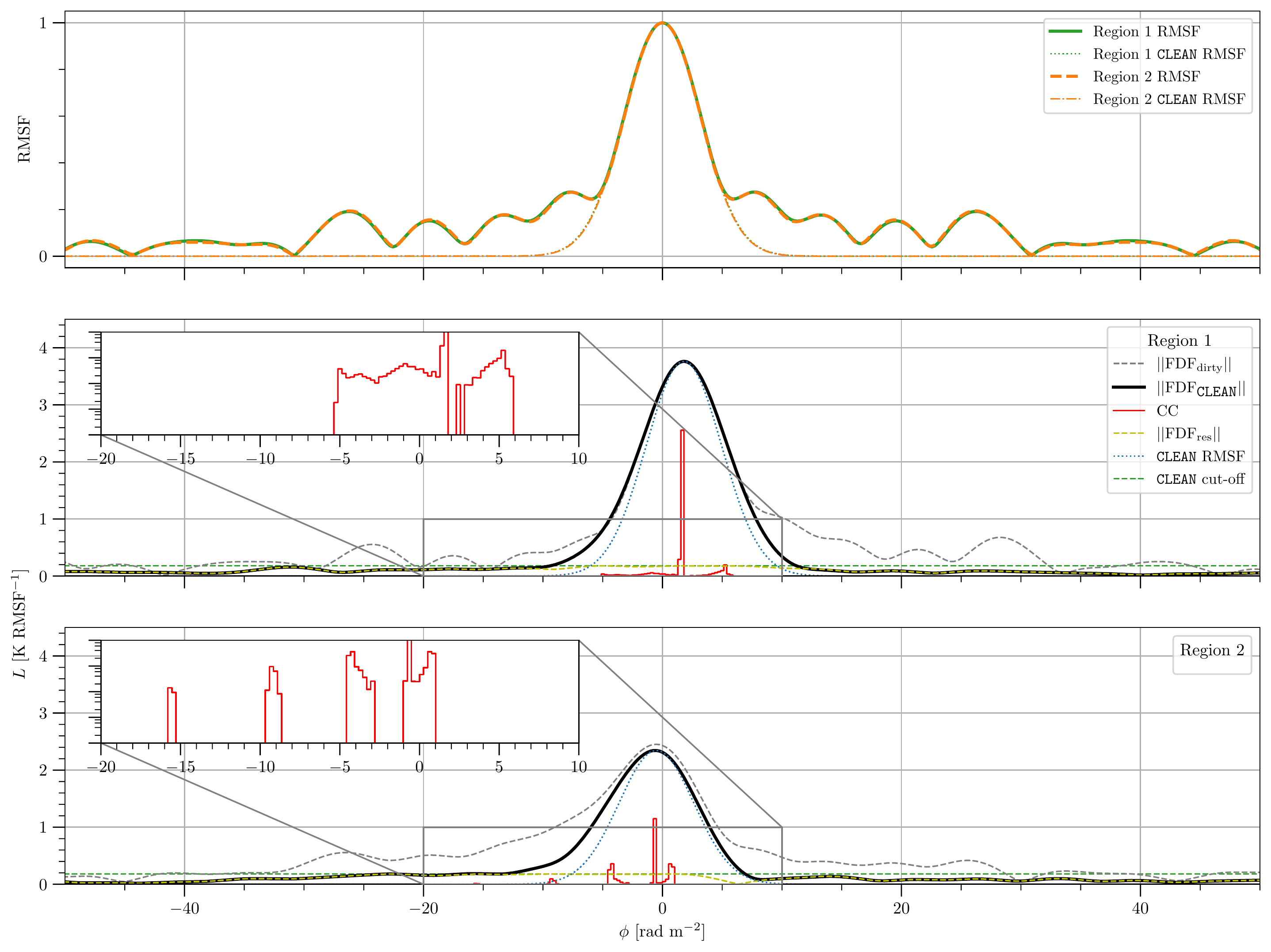}
	\caption{The Faraday dispersion function ($||\text{FDF}||$) and RMSF (upper panel) in regions 1 (middle panel) and 2 (lower panel). For each region we show the Faraday spectra before (`dirty') and after (`\texttt{CLEAN}') the application of \texttt{RM-CLEAN} in grey dashed and black solid lines, respectively. We also show the \texttt{RM-CLEAN} model components (CC), the \texttt{RM-CLEAN} intensity cut-off (\texttt{CLEAN} cut-off), and the residual (`res') Faraday spectrum after the application of \texttt{RM-CLEAN} in red solid, green dashed, and olive dashed lines, respectively. We show the \texttt{CLEAN} Gaussian RMSF in blue-dotted lines, which we place to match the peak of each Faraday spectrum. In the inset panels we show the CC with a logarithmic scale. These inset panels span $\phi$ from $-20$ to $+10$\,\radms\ and $L$ from $10^{-4}$ to $10^0$\,K\,RMSF$^{-1}$.}
	\label{fig:FDF}
\end{figure*}

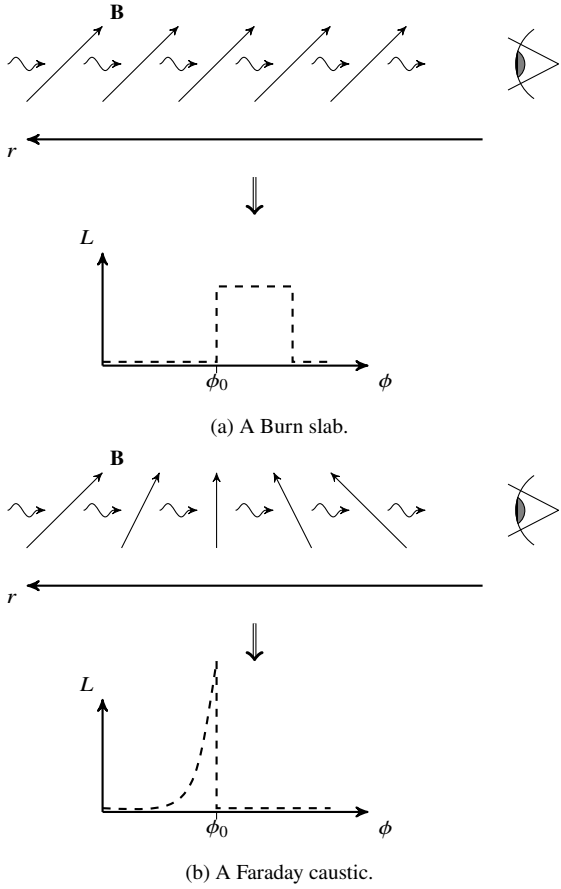
\begin{figure}
    \centering
    \begin{subfigure}[b]{\columnwidth}
        \centering
        \begin{tikzpicture}[
          >=stealth',
          pos=.8,
          photon/.style={decorate,decoration={snake,post length=1mm}}
        ]
                \draw (1,3) node[anchor=south west] {$\mathbf{B}$};
                \draw [->,photon] (-0.25,2.5) -- (0.25,2.5);
            	\draw [->] (0,2) -- (1,3);
            	\draw [->,photon] (0.75,2.5) -- (1.25,2.5);
            	\draw [->] (1,2) -- (2,3);
            	\draw [->,photon] (1.75,2.5) -- (2.25,2.5);
            	\draw [->] (2,2) -- (3,3);
            	\draw [->,photon] (2.75,2.5) -- (3.25,2.5);
                \draw [->] (3,2) -- (4,3);
                \draw [->,photon] (3.75,2.5) -- (4.25,2.5);
                \draw [->] (4,2) -- (5,3);
                \draw [->,photon] (4.75,2.5) -- (5.25,2.5);
                \eye{0.75}{7}{2.5}{180};
                \draw[thick,->] (6,1.5) -- (0,1.5) node[anchor=north east] {$r$};
        
                \draw[-to, double] (3,1) -- (3,0.5);
        
                \draw[thick,->] (9-8,2-3.5) -- (12.5-8,2-3.5) node[anchor=north west] {$\phi$};
                \draw[thick,->] (9-8,2-3.5) -- (9-8,3.5-3.5) node[anchor=south east] {$L$};
                \draw[black,dashed, thick] (9-8,2.05-3.5) -- (10.5-8,2.05-3.5) -- (10.5-8,3.05-3.5) -- (11.5-8,3.05-3.5)-- (11.5-8,2.05-3.5) -- (12-8,2.05-3.5);
                \draw (10.5-8,1.9-3.5) -- (10.5-8,2-3.5) node[anchor=north] {$\phi_0$};
                
        \end{tikzpicture}
        \caption{~A Burn slab.}
        \label{fig:subburn}
    \end{subfigure}
    \begin{subfigure}[b]{\columnwidth}
        \centering
        \begin{tikzpicture}[
          >=stealth',
          pos=.8,
          photon/.style={decorate,decoration={snake,post length=1mm}}
        ]
                \draw (1,3) node[anchor=south west] {$\mathbf{B}$};
                \draw [->,photon] (-0.25,2.5) -- (0.25,2.5);
            	\draw [->] (0,2) -- (1,3);
            	\draw [->,photon] (0.75,2.5) -- (1.25,2.5);
            	\draw [->] (1+0.25,2) -- (2-0.25,3);
            	\draw [->,photon] (1.75,2.5) -- (2.25,2.5);
            	\draw [->] (2.5,2) -- (2.5,3);
            	\draw [->,photon] (2.75,2.5) -- (3.25,2.5);
                \draw [->] (4-0.25,2) -- (3+0.25,3);
                \draw [->,photon] (3.75,2.5) -- (4.25,2.5);
                \draw [->] (5,2) -- (4,3);
                \draw [->,photon] (4.75,2.5) -- (5.25,2.5);
                \eye{0.75}{7}{2.5}{180}
                \draw[thick,->] (6,1.5) -- (0,1.5) node[anchor=north east] {$r$};

				\draw[-to, double] (3,1) -- (3,0.5);
                
                \draw[thick,->] (9-8,2-3.5) -- (12.5-8,2-3.5) node[anchor=north west] {$\phi$};
                \draw[thick,->] (9-8,2-3.5) -- (9-8,3.5-3.5) node[anchor=south east] {$L$};
                \draw[black,dashed, thick] (10.5-8,4-3.5) -- (10.5-8,2.05-3.5) -- (12-8,2.05-3.5);
                \draw[black,dashed, thick] (9-8,2.05-3.5) .. controls (10.2-8,2-3.5) .. (10.5-8,4-3.5);
                \draw (10.5-8,1.9-3.5) -- (10.5-8,2-3.5) node[anchor=north] {$\phi_0$};
                
        \end{tikzpicture}
        \caption{~A Faraday caustic.}
        \label{fig:subcaustic}
    \end{subfigure}
    \caption{Schematic configurations for both \subref{fig:subburn} a Burn slab / differential Faraday rotation, and \subref{fig:subcaustic} a Faraday caustic. On the top we show the magnetic field ($\mathbf{B}$) configuration. In each case there is polarized emission coming to the observer from along the entire line-of-sight. On the bottom we show the linearly polarized intensity ($L$) Faraday dispersion function produced by each configuration. The Burn slab produces a top-hat function in Faraday depth ($\phi$), whereas a caustic follows a $1/\sqrt{\phi}$ curve~\citep{Bell2011}. In both cases $\phi_0$ is the Faraday depth of the medium between the observer and each feature.}
    \label{fig:schematic}
\end{figure}

We show the resulting Faraday spectra, and RMSF, in Fig.~\ref{fig:FDF}. Broadly, the appearance of these spectra is very similar to the original Faraday cubes from the survey paper, as described above. Surviving the additional measures we have applied, the asymmetric peak in Faraday depth remains, with \texttt{CLEAN} components cascading to more negative $\phi$ in both regions. This structure in Faraday depth is what is predicted by \citet{Bell2011} for a Faraday caustic: a strong peak in Faraday depth with an asymmetric tail. The direction of the tail corresponds to the direction of the gradient and reversal in $B_\parallel$, with a tail towards negative $\phi$ indicating $dB_\parallel/dr<0$, where $r$ is the distance along the LOS. A caustic will occur when the LOS component of the magnetic field has a gradient along the LOS and crosses $0\,\mu$G. A Faraday caustic is essentially a Burn slab \citep{Burn1966} that is bent in such a way that somewhere along the line-of-sight through the emitting/rotating volume the field lines become perpendicular to the line-of-sight. We show a schematic model for both a Faraday caustic and a Burn slab in Fig.~\ref{fig:schematic}. A Burn slab is itself a simple model of a Faraday dispersive medium, with a uniform magnetic field, free electron density, and synchrotron emissivity. In Faraday-depth-space, a Burn slab is described by a top-hat function. 

It remains possible, however, that this structure is caused by multiple Faraday depth components which are blending into a broader feature. On the scale of the large beam-width, it would not be improbable to see different Faraday depth features blended together. For example, if there were a spatial gradient in $\phi$ on the plane of the sky, or simply two adjacent features, a large beam would blend those two features together. The feature we observe, however, extends much further than the beam-width.

The spatially extended nature of G150$-$50 is indicative of a dispersive feature, rather than many blended components. This narrows the possible physical description of G150$-$50 to one of several options. It could be a Faraday caustic, as already discussed, or some other naturally Faraday thick feature such as a Burn slab, or it could arise from multiple Faraday components (see Sec. \ref{sec:qufit} for more detail). We consider Faraday thick models to be more realistic than multiple thin models in the context of the Galactic ISM, given that emitting and rotating regions are probably mixed in the diffuse Galactic medium. However, we are unable to discriminate between these possibilities through RM synthesis alone.

\subsection{A proposed physical model}\label{sec:physical-model}
Using the GMIMS-LBS data we wish to obtain some physical insights. Here we make the assumption that these data arise from within a consistent polarization horizon. The radio loops have been described previously using expanding shell models of varying complexity \citep[e.g.][]{Berkhuijsen1971,Berkhuijsen1973,Spoelstra1972,Wolleben2007,Vidal2015}. Here we adopt a model similar to the one applied by \citet{Vidal2015} to Loop I. This is a simple spherical-shell configuration of the B-field, which approximates more complex models such as those of \citet{VanDerLaan} or \citet{Whiteoak1968}. This model simply takes the B-field to be tangent to a spherical surface, following lines of constant longitude. The underlying idea is that this spherical surface is expanding into an ambient, coherent magnetic field. We consider this type of model to be useful for qualitative comparisons with our observations. This model diverges from a more realistic construction at the `poles' of the spherical shell, where the field lines themselves either converge or diverge from a single point.

\begin{figure}
	\centering
	\includegraphics[width=\columnwidth]{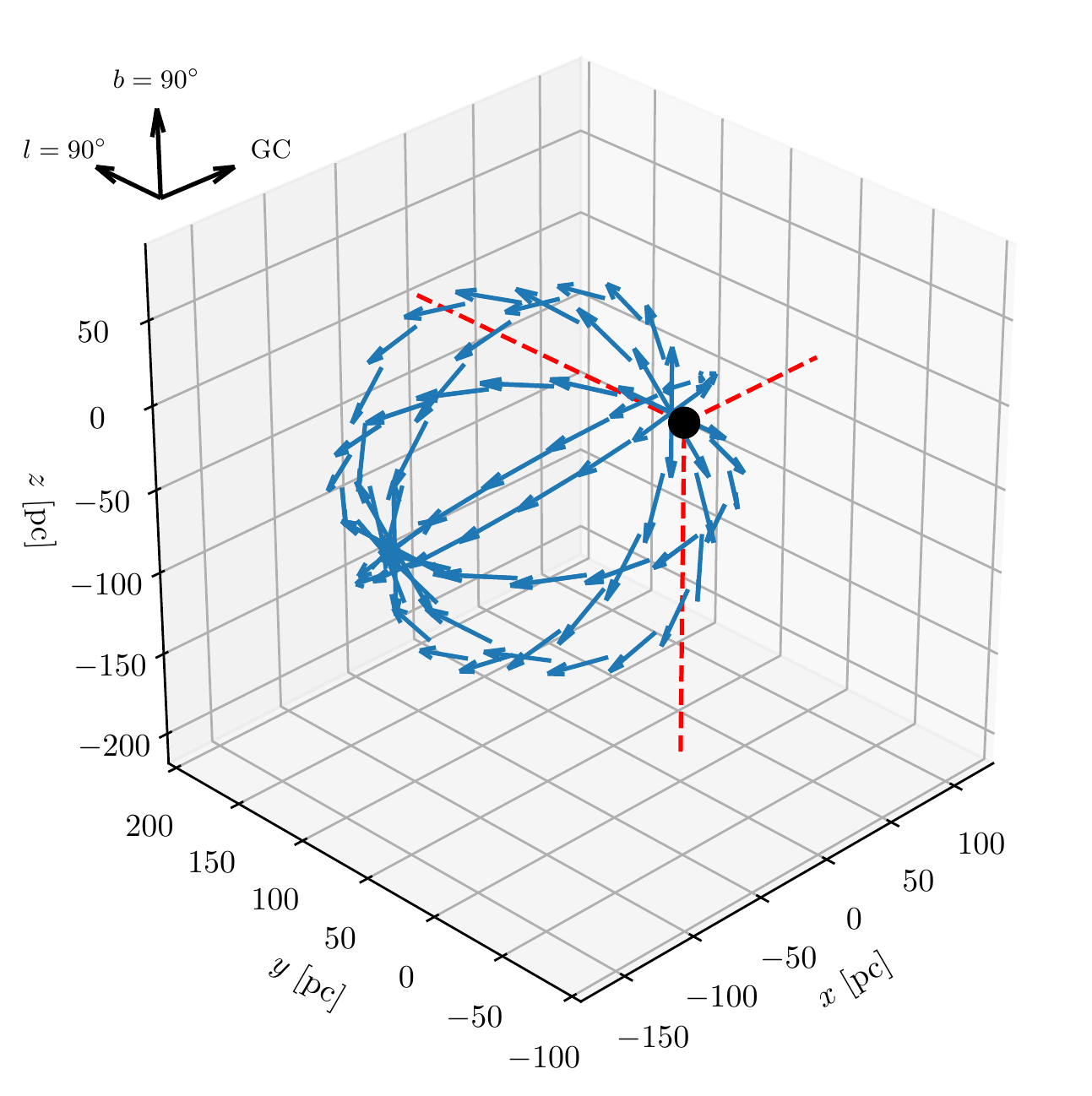}
	\caption{Magnetic field vectors on a simple spherical shell model similar to \citet{Vidal2015}. This model is constructed to fit Loop II, with a distance from the Sun to the centre of 100\,pc, a central coordinate of $(l,b)=(100\degr,-32.5\degr)$, and an angular size of $45.5\degr$. Following \citet{Vidal2015}, we show the model in a Cartesian coordinate system, centred on the Sun at $(0,0,0)$ (shown as a black point, with red-dashed guides). Here the $x$-axis points towards the Galactic centre (GC), the $y$-axis towards $l=90$\degr, and the $z$-axis towards $b=90$\degr.}
	\label{fig:bmodel}
\end{figure}

\citet{Spoelstra1972} fitted both a spherical shell and an expanding shell model to Loop II. Here we take a similar approach, using the more modern measurements from \citet{Vidal2015}: A distance to Loop II of 100\,pc, a central coordinate of $(l,b)=(100\degr,-32.5\degr)$, and an angular size of $45.5\degr$. Since we compute the physical radius of the shell from its projected angular size, the actual distance to the centre of the shell does not affect how the field appears as projected on the sky, and we base our conclusions on that appearance. We now have a choice as to the orientation of the shell. This orientation corresponds to the same orientation the ambient mean field would have had before the shell expanded, diverting the field lines. We find that simply choosing an initial mean field direction away from the Galactic centre and towards the Sun, as shown in Fig.~\ref{fig:bmodel}, provides a series of striking correspondences which we discuss below. We do note, however, that this orientation is rotated by $90\degr$, about the line towards the Galactic pole, in comparison to the \citeauthor{Vidal2015} model for Loop II. We note that the mean, large-scale field near the Sun is known to follow the spiral arm, and runs counter-clock-wise as viewed from the North Galactic pole~\citep{Manchester1972,Manchester1974,Heiles1996,Brown2007,Hutschenreuter2021}. This discrepancy can be explained if the ambient, `seed' field into which shell was blown had an orientation counter to the large-scale field, which has now become frozen-in to the shell structure. In any case, this orientation is required to match the shell model to the observed magnetic field structure of Loop II.

\begin{figure*}
	\centering
		\begin{subfigure}[b]{\textwidth}
    		\includegraphics[width=\textwidth]{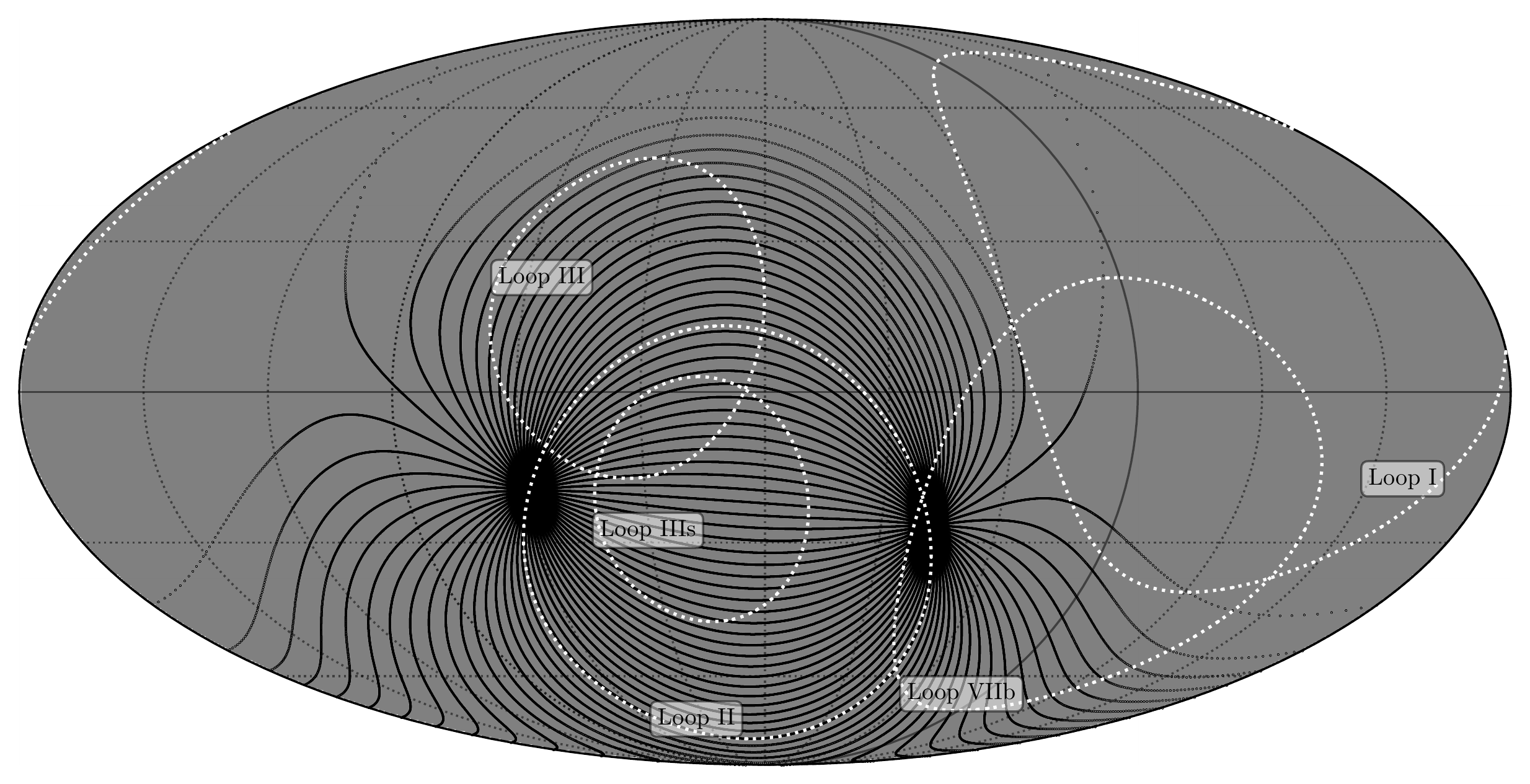}
    		\caption{}
    		\label{fig:bmodel_proj}%
	    \end{subfigure}
    	\begin{subfigure}[b]{0.49\textwidth}
    		\includegraphics[width=\textwidth]{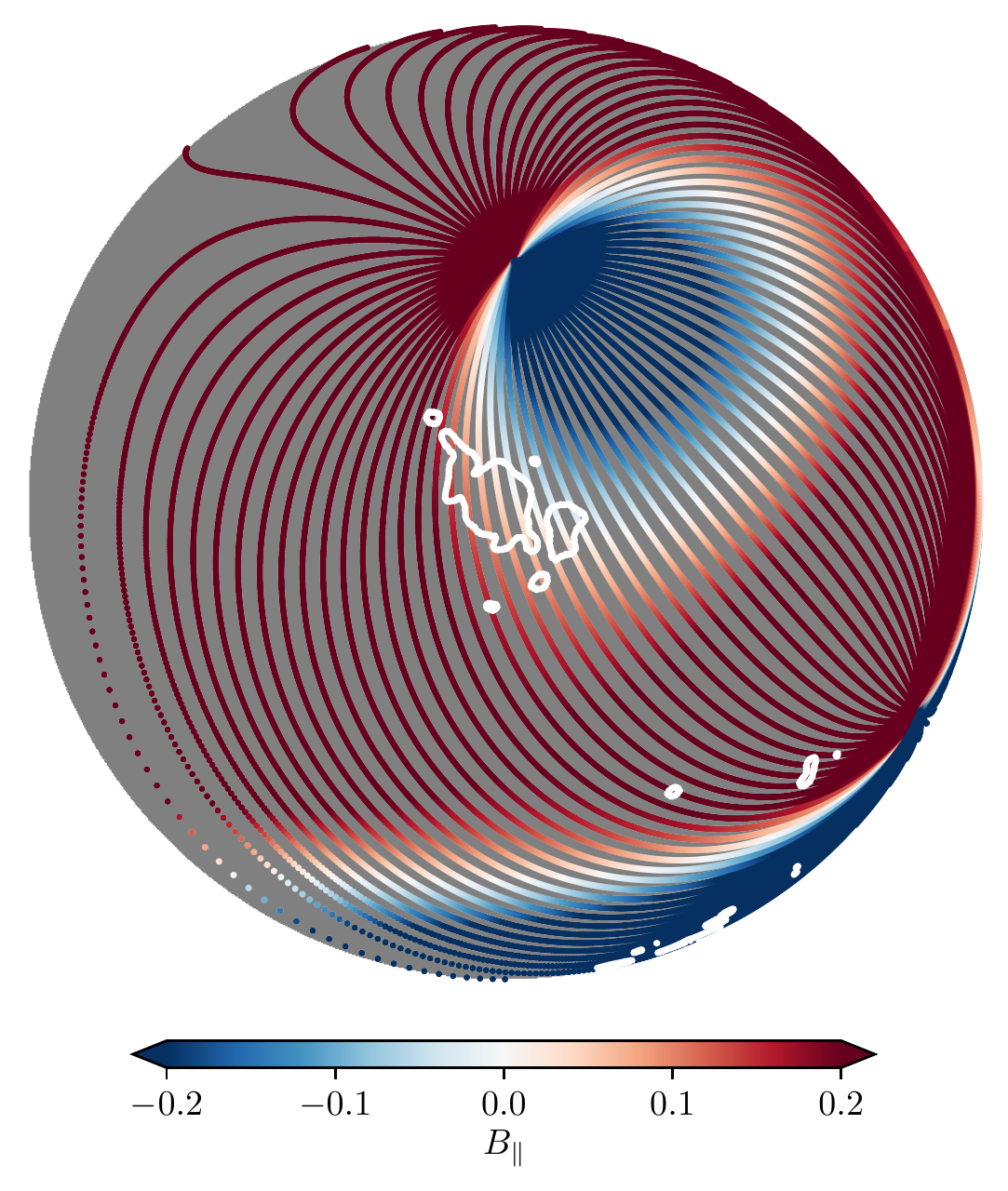}
    		\caption{}
    		\label{fig:bmodel_proj_para}%
        \end{subfigure}
    	\begin{subfigure}[b]{0.49\textwidth}
    		\includegraphics[width=\textwidth]{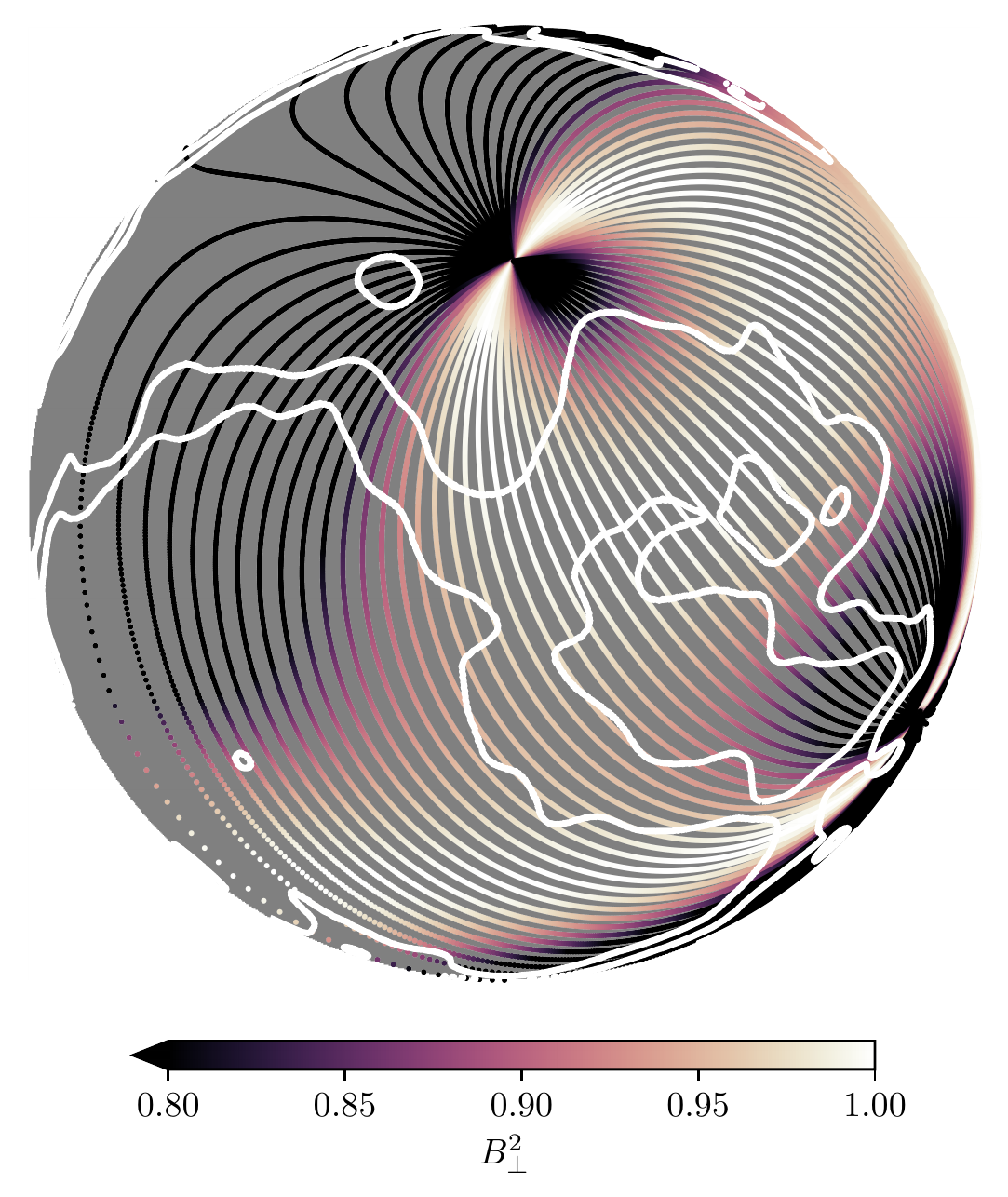}
    		\caption{}
    		\label{fig:bmodel_proj_perp}%
        \end{subfigure}
	\caption{The spherical shell B-field model projected on the sky. \subref{fig:bmodel_proj} The projected field lines and Radio Loops. Here we use the same projection as Fig.~\ref{fig:allpimapallsky}. \subref{fig:bmodel_proj_para} and \subref{fig:bmodel_proj_perp} show the same projected field lines, coloured by $B_\parallel$ and $B_\perp$ for a unit field, respectively. Here we use the same projection as Fig.~\ref{fig:imap}. We also overlay the same $L$ contour as Fig.~\ref{fig:pimap} in \subref{fig:bmodel_proj_para}, and the same Stokes $I$ contour from Fig.~\ref{fig:imap} in \subref{fig:bmodel_proj_perp}.}
	\label{fig:bmodel_sky}
\end{figure*}

In Fig.~\ref{fig:bmodel_sky} we show the field configuration from Fig.~\ref{fig:bmodel} projected on the sky. In Fig.~\ref{fig:bmodel_proj} we show the projected B-field lines in comparison to the position of Loop II. As in Fig.~\ref{fig:LIC}, we find that the field lines follow the path of Loop II along the plane of the sky. 

We now compute the fraction of a unit magnetic field, projected along the LOS ($B_\parallel$) and in the POS ($B_\perp$). The $B_\parallel$ is simply found using:
\begin{equation}
    B_\parallel = \hat{\mathbf{r}}\cdot\mathbf{B},
\end{equation}
where $\hat{\mathbf{r}}$ is the radial unit vector, originating at the Sun, and $\mathbf{B}$ is the total magnetic field vector, the magnitude of which we take to be 1. The perpendicular component is then:
\begin{equation}
    B_\perp = \sqrt{1 - B_\parallel^2}.
\end{equation}
In Fig.~\ref{fig:bmodel_proj_para}, we show the model, projected $B_\parallel$, which produces Faraday rotation. At the location of G150$-$50 we find that the B-field is mostly in the POS, with a gradient in $B_\parallel$ from positive, pointing towards us; to negative, pointing away from us. This is very close to what we see in the moment 1 data (Fig.~\ref{fig:m1map}).

Turning our attention to $B_\perp$, we now consider the total intensity data. The total synchrotron intensity ($I$) of a source with depth $l$ is:
\begin{equation}
I \propto N_0 B_\perp^{(1-\gamma)/2}l,
\label{eqn:synchi}
\end{equation}
where $N_0$ is the density of cosmic-ray electrons per energy interval~\citep{Beck2013}. Since the cosmic-rays themselves follow a power-law distribution with respect to their energy $E$, $N(E) = N_0E^{\gamma}$, this formulation provides the familiar power-law spectrum in frequency ($\nu$), $I\propto\nu^\alpha$, where $\alpha=(\gamma+1)/2 $. In the case where $\alpha\sim-1$ ($\beta=-3$) we get $I\propto B_\perp^2$. In Fig.~\ref{fig:bmodel_proj_perp} we show $B_\perp^2$ from our model. It is important to note that the $B_\perp$ term in Equation~\ref{eqn:synchi} is a function of both the computed $B_\perp$, and the density of field lines on the sky. The latter should be read from Fig.~\ref{fig:bmodel_proj}, which uses a Mollweide projection, rather than Figs.~\ref{fig:bmodel_proj_para} or \ref{fig:bmodel_proj_perp}, which use an Orthographic projection. Whilst both projections have distortion, the Mollweide projection does not stretch parallel lines of longitude in the latitudinal direction. With this consideration, we again find a remarkable correspondence. We find that the field lines are densest along the path of Loop II, with a significant fraction being in the POS, as seen in \ref{fig:bmodel_proj_perp}.

This simple shell model is capable of reproducing, at least qualitatively, observations of both Stokes $I$ emission and Faraday depth, as observed by GMIMS-LBS. This result further strengthens the conclusion that G150$-$50 is, in fact, part of Loop II. We suggest that we do not see more extended polarized emission from Loop II because it is too depolarized. G150$-$50 is at a point along Loop II where the magnetic field is primarily in the plane of the sky, and the Faraday rotation and dispersion are low. If we assume a Burn slab depolarization mode, that is a uniform magnetic field and electron density distribution, we can estimate the Faraday depth at which the intrinsic emission is depolarized below the noise threshold of GMIMS-LBS. The polarized intensity of a Burn slab, as function of $\lambda^2$, is given by \citep{Burn1966,Sokoloff1998}
\begin{equation}
    L(\lambda^2) = L_0 \left|\left|\frac{\sin{\left(\phi\lambda^2\right)}}{\phi\lambda^2} \right|\right|,
    \label{eqn:burnslab}
\end{equation}
where $L_0$ is the intrinsic polarized intensity, and $\phi$ is the Faraday depth through the slab. At 408\,MHz the average polarized intensity in region 1, the brighter of the two region, is 2.5\,K. We numerically solve Equation~\ref{eqn:burnslab} to find the Faraday depth at which 2.5\,K will depolarize to 60\,mK, the RMS noise in GMIMS-LBS. We find that absolute Faraday depths of $||\phi||\gtrsim5.7\,$\radms\ will depolarize emission from G150$-$50 below this threshold. This value of Faraday depth is in line with what we see from the Faraday moments, with no polarized emission with first moment $||M_1||\gtrsim5\,$\radms.

The simplicity of the shell model, however, does not allow us to explore much detail of the B-field along the line-of-sight, nor the nature of the source of polarized emission. To reveal this detail, we can look to $QU$-fitting. 

\subsection{\texorpdfstring{\boldmath{$QU$}}{QU}-fitting}\label{sec:qufit}
Here we utilise the $QU$-fitting routines of \textsc{rm-tools} to fit a number of models to the average spectra of regions 1 and 2, and compare their performance in representing the data. The routines of \textsc{rm-tools} allow for a Bayesian model fitting comparison. Whilst we fit the spectra of regions 1 and 2 independently, we require that the same model should apply to both regions as we consider G150$-$50 to arise from a single physical region on the sky. Here it is important to note that we are assuming both that our model of Stokes $I$ is correct, and that the measured total emission and polarized emission arise from the same volume. We also increase the input errors in the GMIMS-LBS band to cover some small-scale fluctuations in those data. Such ripples would correspond to a high value of Faraday depth, which we do not wish to include in our modelling. These errors are shown in Fig.~\ref{fig:spec}. 

The full broad-band spectra on both regions 1 and 2, as shown in Fig.~\ref{fig:spec}, show a high degree of complexity. We note, however, that the spectra from 0.5 to 1\,m$^2$ appear to be simpler in both regions. This can be seen clearly in the polarization angle spectra, which exhibit relatively linear relationships with $\lambda^2$ in this sub-band. For the sake of comparison, we fit to both the full band, through to 0.046\,m$^2$ (1.4\,GHz), and this simpler sub-band.

Here we adopt and expand upon the model framework of \citet{OSullivan2017}, which is derived from \citet{Burn1966} and \citet{Sokoloff1998}. This functional form allows for a description of a Faraday simple or dispersive medium, as well as multiple components. We summarize all the models we use in Table~\ref{tab:modelsum}. Our models 1 through 12 are identical to those from \citet{OSullivan2017}, whose full functional form gives the complex polarization fraction of a source as:
\begin{equation}
    P_{j} =  p_{0,j}
    \underbrace{e^{2 i (\psi_{0,j}+\phi_{0,j} \lambda^2)}}_\text{A}  
    \overbrace{\frac{\sin \Delta \phi_j \lambda^2}{\Delta \phi_j \lambda^2}}^\text{B}
    \underbrace{e^{-2\sigma^2_{\phi_j} \lambda^4}}_\text{C}.
    \label{eqn:osul}
\end{equation}
Here, $j$ refers to the $j^{\text{th}}$ polarized component within the beam with intrinsic polarized fraction $p_{0,j}$. Multi-component models with $N_\text{comp}$ components can therefore be constructed using:
\begin{equation}
    P = \sum_{j=1}^{N_\text{comp}}P_{j},
\end{equation}
where we fit only up to $N_\text{comp}=3$. We have labelled the three factors of Equation~\ref{eqn:osul} relating to the nature of the emitting medium as A, B, C. A describes a purely rotating, or Faraday thin, medium where polarized emission and Faraday rotation originate in separate volumes. The foreground rotating volume contains a uniform magnetic field and thermal electron density. B describes differential Faraday rotation, or a `Burn slab', where the emitting volume additionally contains a uniform LOS B-field component which causes Faraday rotation through the volume. This term can also describe RM gradients across the source~\citep{Schnitzeler2015}. C refers to external Faraday dispersion. In such a case, emission and Faraday rotation originate in separate volumes, and the foreground rotating volume contains a turbulent medium, comprised of multiple cells which fall within the telescope beam. We refer the reader to \citet{OSullivan2017} and \citet{Anderson2019} for additional discussions of this formulation.

Given the `humped' structure we see in the $L$ spectra on G150$-$50, we are motivated to seek a modification to Equation~\ref{eqn:osul}. By itself, this form describes types of depolarization, which usually implies that $L$ should peak globally at $\lambda^2=0$. This trend can be modulated given the sum of many terms. Such a formulation, however, requires many free parameters. Instead, \citet{Horellou2014} provide a simple formulation that can result in peak polarization at $\lambda^2>0$ through the effect of helical magnetic fields. Helical magnetic fields are predicted to occur in the magnetic fields of Galaxies through dynamo action \citep{Beck1996,Subramanian2002}. Evidence for large-scale helicity has recently been reported by \citet{West2020}. Given a large-scale helical field, turbulent dynamo theory also predicts helicity on small scales \citep{Subramanian2002}. \citet{Horellou2014} provide a detailed derivation and description, which we will not reproduce here. For our purposes, we use their quantification of a linear variation in $\psi$, as a function of $\phi$, due to the presence of a helical magnetic field in the plane of the sky, with a constant LOS B-field. They define a parameter which we shall refer to as $\eta$ \citep[note this is $\beta$ in][]{Horellou2014}:
\begin{equation}
    \eta = \frac{k_h}{0.812n_eB_\parallel},
\end{equation}
where $k_h$ is the winding parameter of the helical field in rad\,kpc$^{-1}$, and $\eta$ therefore has units of m$^2$. We can consider $\eta$ as a quantification of the Faraday rotation due to helicity. \citet{Horellou2014} go on to show that, for a given model of the complex polarization, $P(\lambda^2)$, the effect of a helical field can be included using:
\begin{equation}
    P(\lambda^2, \eta) = P(\lambda^2 + \eta) e^{2i\phi_0(\lambda^2+\eta)}.
    \label{eqn:helicity}
\end{equation}
This is simply a translation of the complex polarization as a function of $\lambda^2$. The effect of helical magnetic fields, in the case of a Faraday-dispersive medium, is to shift the peak polarized intensity away from $\lambda^2=0$. We therefore adapt the \citet{OSullivan2017} terms that are not constructed from Faraday screens alone to include a term for helicity. We refer to these as models 13 to 21.

Lastly, we include a model of a Faraday caustic. In their derivation \citet{Bell2011} assumed a flat spectrum where $\alpha=-1$, and find that for a caustic: 
\begin{equation}
	P = D e^{\left(2i\phi_0\lambda^2\right) } \frac{\left( -1 -i \right) \sqrt{\pi}}{2\sqrt{\lambda^2}},
	\label{eqn:Bellm11}
\end{equation}
where $D$ is a complex number which encapsulates information of both the line-of-sight and parallel magnetic fields. For GMIMS-LBS \citep{Wolleben2019} the brightness temperature spectral index, $\beta$, was consistent with $-2.5$. Since $\alpha=\beta+2$ this corresponds to $\alpha\sim-0.5$. \citet{Bell2011} also note that this model is divergent as $\lambda$ becomes small. In their full formulation (see their Eqn. 13), the $D$ term acts against this divergence. In our case, however, we are simply fitting $D$ as a parameter. Therefore we consider this formulation suitable only for longer wavelengths. Since $D$ is just a complex number, we can write:
\begin{equation}
    D = p_0e^{2i\psi_0}.
\end{equation}
Therefore, complex polarization is:
\begin{equation}
	P = p_0e^{2i\psi_0} e^{\left(2i\phi_0\lambda^2\right) } \frac{\left( -1 -i \right) \sqrt{\pi}}{2\sqrt{\lambda^2}},
	\label{eqn:m22}
\end{equation}
which is our model 22.

\begin{table*}
	\centering
	\caption{A summary of the models we use in $QU$-fitting of the Faraday spectra.  The definitions for these models are given by Equations~\ref{eqn:osul}, \ref{eqn:helicity}, and \ref{eqn:m22}. Column (1): The number we assign to each model. Column (2): A broad description of physical nature of each model. Column (3): The number of free parameters ($N_\text{free}$) within each model. Column (4): The number of emitting components ($N_\text{comp}$) along the line-of-sight. Column (5): Which terms from Equation~\ref{eqn:osul} are included.}
	\label{tab:modelsum}
	\begin{tabular}{clcccc}
		\hline
		Model \# & Description                                             & $N_\text{free}$ & $N_\text{comp}$ & Eqn.~\ref{eqn:osul} term(s) &  \\ \hline
		1     & Faraday thin screen                                     &        3        &        1        &              A               &  \\
		2     & Differential Faraday rotation w/ a foreground screen              &        4        &        1        &             A,B              &  \\
		3     & External dispersion w/ a foreground screen              &        4        &        1        &             A,C              &  \\
		4     & Internal and external dispersion w/ a foreground screen &        5        &        1        &            A,B,C             &  \\
		5     & Faraday thin screen                                     &        6        &        2        &              A               &  \\
		6     & Differential Faraday rotation w/ a foreground screen              &        8        &        2        &             A,B              &  \\
		7     & External dispersion w/ a foreground screen              &        8        &        2        &             A,C              &  \\
		8     & Internal and external dispersion w/ a foreground screen &       10        &        2        &            A,B,C             &  \\
		9     & Faraday thin screen                                     &        9        &        3        &              A               &  \\
		10    & Differential Faraday rotation w/ a foreground screen              &       12       &        3        &             A,B              &  \\
		11    & External dispersion w/ a foreground screen              &       12       &        3        &             A,C              &  \\
		12    & Internal and external dispersion w/ a foreground screen &       15        &        3        &            A,B,C             &  \\
		13    & Model 2 w/ helicity                                     &        5        &        1        &             A,B              &  \\
		14    & Model 3 w/ helicity                                     &        5        &        1        &             A,C              &  \\
		15    & Model 4 w/ helicity                                     &        6        &        1        &            A,B,C             &  \\
		16    & Model 6 w/ helicity                                     &       10        &        2        &             A,B              &  \\
		17    & Model 7 w/ helicity                                     &       10        &        2        &             A,C              &  \\
		18    & Model 8 w/ helicity                                     &       12        &        2        &            A,B,C             &  \\
		19    & Model 10 w/ helicity                                    &       15       &        3        &             A,B              &  \\
		20    & Model 11 w/ helicity                                    &       15       &        3        &             A,C              &  \\
		21    & Model 12 w/ helicity                                    &       18       &        3        &            A,B,C             &  \\
		22    & Faraday caustic                                         &        3        &        1        &              --              &  \\ \hline
	\end{tabular}
\end{table*}

The $QU$-fitting of \textsc{rm-tools} utilizes the MultiNest algorithm~\citep{Feroz2008,Feroz2009,Feroz2014}, implemented as \textsc{pymultinest}~\citep{Buchner2014}. The details of this algorithm and its implementation are in the aforementioned references and Purcell et al. (in prep.). We provide the data, error estimates, and priors for each parameter to \textsc{pymultinest}, which returns the prior distribution for each model parameter, and the Bayesian evidence ($\mathcal{Z}$) of each model, which allows for a number of goodness-of-fit quantities to be evaluated Typically, one could consider the reduced-$\chi^2$ ($ \chi^2_\text{red} $) to determine goodness-of-fit:
\begin{equation}
\chi^2_\text{red} = \frac{\chi^2}{\text{DoF}},
\label{eqn:chisqred}
\end{equation}
where DoF is the number of degrees-of-freedom:
\begin{equation}
\text{DoF} = N_\text{data} - N_\text{free} - 1,
\label{eqn:dof}
\end{equation}
where $ N_\text{free} $ is the number of free parameters and $\chi^2$ is:
\begin{equation}
\chi^2 = \sum_{i=1}^{N_\text{data}}\left[  \left( \frac{Q_{\text{data},i} - Q_{\text{model},i}}{\sigma_{QU,i}}\right)^2 + \left( \frac{U_{\text{data},i} - U_{\text{model},i}}{\sigma_{QU,i}}\right)^2\right].
\label{eqn:chisq}
\end{equation}
Instead. one could consider the Bayesian evidence. In model selection, say between models `$a$' and `$b$', we are interested in the Bayes odds ratio ($\Delta\ln{\mathcal{Z}}$):
\begin{equation}
	\Delta \ln{\mathcal{Z}} =  \ln{\mathcal{Z}_a} - \ln{\mathcal{Z}_b} = \ln{\left(\frac{\mathcal{Z}_a}{\mathcal{Z}_b}\right)},
	\label{eqn:bayes}
\end{equation}
where $\mathcal{Z}_a$ and $\mathcal{Z}_b$ are the Bayesian evidence of models `$a$' and `$b$', respectively. It is important to note that the value of the evidence itself, rather we require relative values between different models. Here we adopt the standard set out by \citet{Kass1995}, who describe ranges of $2\times\Delta \ln{\mathcal{Z}}$. For values of 0 to 2, the evidence of model `$a$' over `$b$' is ``not worth more than a bare mention'', for 2 to 6 the evidence is positive, for 6 to 10 the evidence is strong, and for $>10$ the evidence of `$a$' over `$b$' is very strong.

We can evaluate the performance of $QU$-fitting from three perspectives: the goodness-of-fit metrics, visual comparison of the best-fitting model to the data, and best-fitting parameters such as the predicted polarization angle at $\lambda^2=0$. In Table~\ref{tab:goodnes} we give both the reduced $\chi^2$ ($\chi^2_\text{red}$), and Bayesian odds ratios. Here we compute the odds ratio between the model with the highest evidence and all other models.

\begin{table*}
	\caption{A summary of the best-fitting metrics for each model we use in $QU$-fitting of the polarized spectra across \subref{tab:goodnes_full} the full band and \subref{tab:goodnes_sub} the $\lambda^2=[0.5,1]$\,m$^2$ sub-band. We sort the rows of this Table by the $2\Delta\ln(\mathcal{Z})$ value. Col. (1): The model number. Col. (2): The number of free parameters (N$_\mathrm{free}$) in the mode. Col. (3): The number of degrees of freedom (see Equation~\ref{eqn:dof}). Col. (4): The reduced $\chi^2$ of the model (see Equation~\ref{eqn:chisqred}). Col. (5): The natural log of the Bayesian evidence ($\ln(\mathcal{Z})$). Col. (6): Twice the natural logarithm of the Bayesian odds ratio ($2\Delta\ln(\mathcal{Z})$). Note that differences greater than 10 in $2\Delta\ln(\mathcal{Z})$ are considered the most significant.}
	\label{tab:goodnes}
		\begin{subtable}[t]{0.49\textwidth}
			\scriptsize
			\caption{~Full band}
			\label{tab:goodnes_full}
			\begin{tabular}[t]{ccccccccc}
				\hline
				Model \# & N$_\mathrm{free}$ & DoF & $\chi^2_\text{red}$ & $\ln(\mathcal{Z})$ & $2\Delta\ln(\mathcal{Z})$ \\
				\hline
				Region 1 &                 &  \\ \hline
				19 & 15 & $554$ & $1.1 \times 10^{-1}$ & $1684.9\pm0.3$ & $0.0\pm0.0$ \\
				20 & 15 & $554$ & $2.6 \times 10^{-1}$ & $1665.4\pm0.3$ & $38.9\pm0.8$ \\
				16 & 10 & $559$ & $2.7 \times 10^{-1}$ & $1664.1\pm0.2$ & $41.6\pm0.7$ \\
				17 & 10 & $559$ & $2.8 \times 10^{-1}$ & $1652.0\pm0.3$ & $65.7\pm0.8$ \\
				12 & 15 & $554$ & $2.6 \times 10^{-1}$ & $1641.6\pm0.3$ & $86.5\pm0.8$ \\
				21 & 18 & $551$ & $7.9 \times 10^{-1}$ & $1629.7\pm0.3$ & $110.3\pm0.8$ \\
				18 & 12 & $557$ & $4.0 \times 10^{-1}$ & $1612.1\pm0.3$ & $145.4\pm0.8$ \\
				14 & 5 & $564$ & $6.3 \times 10^{-1}$ & $1586.4\pm0.2$ & $196.9\pm0.7$ \\
				15 & 6 & $563$ & $6.4 \times 10^{-1}$ & $1581.0\pm0.2$ & $207.7\pm0.7$ \\
				13 & 5 & $564$ & $7.3 \times 10^{-1}$ & $1559.5\pm0.2$ & $250.7\pm0.7$ \\
				8 & 10 & $559$ & $7.0 \times 10^{-1}$ & $1541.0\pm0.2$ & $287.8\pm0.8$ \\
				6 & 8 & $561$ & $7.3 \times 10^{-1}$ & $1539.4\pm0.2$ & $290.9\pm0.7$ \\
				9 & 9 & $560$ & $7.0 \times 10^{-1}$ & $1532.2\pm0.3$ & $305.3\pm0.8$ \\
				7 & 8 & $561$ & $8.9 \times 10^{-1}$ & $1489.0\pm0.2$ & $391.7\pm0.8$ \\
				5 & 6 & $563$ & $2.2 \times 10^{0}$ & $1142.1\pm0.2$ & $1085.4\pm0.7$ \\
				2 & 4 & $565$ & $2.3 \times 10^{0}$ & $1109.0\pm0.2$ & $1151.7\pm0.7$ \\
				4 & 5 & $564$ & $2.4 \times 10^{0}$ & $1101.2\pm0.2$ & $1167.3\pm0.7$ \\
				3 & 4 & $565$ & $2.5 \times 10^{0}$ & $1056.3\pm0.2$ & $1257.0\pm0.7$ \\
				23 & 3 & $566$ & $2.7 \times 10^{0}$ & $1021.7\pm0.1$ & $1326.3\pm0.6$ \\
				1 & 3 & $566$ & $9.9 \times 10^{0}$ & $-1029.3\pm0.2$ & $5428.3\pm0.7$ \\
				22 & 3 & $566$ & $2.2 \times 10^{1}$ & $-4391.6\pm0.2$ & $12152.9\pm0.7$ \\
				\hline
				Region 2&                 &  \\ \hline
				19 & 15 & $554$ & $3.9 \times 10^{1}$ & $1671.5\pm0.3$ & $0.0\pm0.0$ \\
				16 & 10 & $559$ & $1.6 \times 10^{-1}$ & $1663.6\pm0.2$ & $15.7\pm0.7$ \\
				17 & 10 & $559$ & $1.7 \times 10^{-1}$ & $1651.0\pm0.3$ & $40.9\pm0.7$ \\
				20 & 15 & $554$ & $1.9 \times 10^{-1}$ & $1643.4\pm0.3$ & $56.1\pm0.8$ \\
				21 & 18 & $551$ & $1.9 \times 10^{1}$ & $1637.2\pm0.3$ & $68.5\pm0.8$ \\
				12 & 15 & $554$ & $2.5 \times 10^{-1}$ & $1634.7\pm0.3$ & $73.6\pm0.8$ \\
				6 & 8 & $561$ & $3.0 \times 10^{-1}$ & $1628.2\pm0.2$ & $86.5\pm0.7$ \\
				8 & 10 & $559$ & $4.1 \times 10^{-1}$ & $1581.9\pm0.3$ & $179.2\pm0.8$ \\
				7 & 8 & $561$ & $9.0 \times 10^{-1}$ & $1462.0\pm0.2$ & $418.9\pm0.7$ \\
				18 & 12 & $557$ & $1.3 \times 10^{0}$ & $1437.2\pm0.2$ & $468.6\pm0.7$ \\
				14 & 5 & $564$ & $1.2 \times 10^{0}$ & $1402.0\pm0.2$ & $538.9\pm0.6$ \\
				15 & 6 & $563$ & $1.2 \times 10^{0}$ & $1402.0\pm0.2$ & $539.0\pm0.7$ \\
				13 & 5 & $564$ & $1.2 \times 10^{0}$ & $1385.9\pm0.2$ & $571.2\pm0.6$ \\
				2 & 4 & $565$ & $1.5 \times 10^{0}$ & $1304.3\pm0.2$ & $734.3\pm0.6$ \\
				4 & 5 & $564$ & $1.6 \times 10^{0}$ & $1296.8\pm0.2$ & $749.4\pm0.6$ \\
				3 & 4 & $565$ & $1.8 \times 10^{0}$ & $1233.1\pm0.2$ & $876.7\pm0.6$ \\
				22 & 3 & $566$ & $4.9 \times 10^{0}$ & $369.6\pm0.1$ & $2603.7\pm0.6$ \\
				23 & 3 & $566$ & $6.0 \times 10^{0}$ & $61.2\pm0.1$ & $3220.6\pm0.6$ \\
				1 & 3 & $566$ & $6.1 \times 10^{0}$ & $24.4\pm0.1$ & $3294.1\pm0.6$ \\
				5 & 6 & $563$ & $6.3 \times 10^{0}$ & $-55.7\pm0.2$ & $3454.3\pm0.7$ \\
				9 & 9 & $560$ & $9.6 \times 10^{0}$ & $-1001.0\pm0.3$ & $5344.9\pm0.8$ \\
				\hline
			\end{tabular}
			\hfill
		\end{subtable}
		\begin{subtable}[t]{0.49\textwidth}
			\scriptsize
			\caption{~$\lambda^2=[0.5,1]$\,m$^2$ sub-band}
			\label{tab:goodnes_sub}
			\begin{tabular}[t]{ccccccccc}
				\hline
				Model \# & N$_\mathrm{free}$ & DoF & $\chi^2_\text{red}$ & $\ln(\mathcal{Z})$ & $2\Delta\ln(\mathcal{Z})$ \\
				\hline
				Region 1 &                 &  \\ \hline
				20 & 15 & $394$ & $9.0 \times 10^{1}$ & $1284.7\pm0.2$ & $0.0\pm0.0$ \\
				19 & 15 & $394$ & $7.4 \times 10^{1}$ & $1284.7\pm0.2$ & $0.0\pm0.6$ \\
				17 & 10 & $399$ & $1.8 \times 10^{-1}$ & $1284.5\pm0.2$ & $0.5\pm0.5$ \\
				14 & 5 & $404$ & $1.7 \times 10^{-1}$ & $1284.0\pm0.2$ & $1.5\pm0.5$ \\
				16 & 10 & $399$ & $4.8 \times 10^{-1}$ & $1282.2\pm0.2$ & $5.0\pm0.6$ \\
				21 & 18 & $391$ & $7.3 \times 10^{1}$ & $1279.4\pm0.2$ & $10.6\pm0.5$ \\
				15 & 6 & $403$ & $1.7 \times 10^{-1}$ & $1278.9\pm0.2$ & $11.6\pm0.5$ \\
				18 & 12 & $397$ & $1.8 \times 10^{-1}$ & $1278.7\pm0.2$ & $11.9\pm0.5$ \\
				6 & 8 & $401$ & $1.1 \times 10^{-1}$ & $1277.7\pm0.2$ & $14.0\pm0.6$ \\
				13 & 5 & $404$ & $2.2 \times 10^{-1}$ & $1277.3\pm0.2$ & $14.9\pm0.5$ \\
				12 & 15 & $394$ & $1.1 \times 10^{-1}$ & $1265.8\pm0.3$ & $37.7\pm0.6$ \\
				7 & 8 & $401$ & $1.9 \times 10^{-1}$ & $1260.3\pm0.2$ & $48.8\pm0.6$ \\
				8 & 10 & $399$ & $2.4 \times 10^{-1}$ & $1257.2\pm0.3$ & $55.1\pm0.6$ \\
				2 & 4 & $405$ & $7.1 \times 10^{-1}$ & $1179.2\pm0.2$ & $210.9\pm0.5$ \\
				4 & 5 & $404$ & $7.2 \times 10^{-1}$ & $1176.0\pm0.2$ & $217.5\pm0.5$ \\
				3 & 4 & $405$ & $9.2 \times 10^{-1}$ & $1140.9\pm0.2$ & $287.6\pm0.5$ \\
				23 & 3 & $406$ & $2.4 \times 10^{0}$ & $849.7\pm0.1$ & $869.9\pm0.5$ \\
				22 & 3 & $406$ & $3.6 \times 10^{0}$ & $600.3\pm0.2$ & $1368.7\pm0.5$ \\
				9 & 9 & $400$ & $3.6 \times 10^{0}$ & $564.6\pm0.3$ & $1440.3\pm0.6$ \\
				1 & 3 & $406$ & $7.1 \times 10^{0}$ & $-107.1\pm0.1$ & $2783.6\pm0.5$ \\
				5 & 6 & $403$ & $1.8 \times 10^{1}$ & $-2259.5\pm0.2$ & $7088.5\pm0.6$ \\
				\hline
				Region 2&                 &  \\ \hline
				19 & 15 & $394$ & $8.6 \times 10^{-1}$ & $1266.7\pm0.2$ & $0.0\pm0.0$ \\
				16 & 10 & $399$ & $2.4 \times 10^{0}$ & $1264.7\pm0.2$ & $3.9\pm0.6$ \\
				6 & 8 & $401$ & $1.3 \times 10^{-1}$ & $1259.1\pm0.2$ & $15.1\pm0.6$ \\
				7 & 8 & $401$ & $1.8 \times 10^{-1}$ & $1247.8\pm0.2$ & $37.7\pm0.6$ \\
				12 & 15 & $394$ & $1.5 \times 10^{1}$ & $1240.8\pm0.2$ & $51.8\pm0.5$ \\
				2 & 4 & $405$ & $3.1 \times 10^{-1}$ & $1240.6\pm0.2$ & $52.1\pm0.5$ \\
				8 & 10 & $399$ & $2.1 \times 10^{0}$ & $1240.4\pm0.2$ & $52.4\pm0.5$ \\
				17 & 10 & $399$ & $3.6 \times 10^{0}$ & $1239.9\pm0.2$ & $53.6\pm0.5$ \\
				20 & 15 & $394$ & $4.7 \times 10^{0}$ & $1239.8\pm0.2$ & $53.7\pm0.5$ \\
				13 & 5 & $404$ & $3.1 \times 10^{-1}$ & $1239.3\pm0.2$ & $54.7\pm0.5$ \\
				4 & 5 & $404$ & $3.1 \times 10^{-1}$ & $1239.2\pm0.2$ & $54.9\pm0.5$ \\
				14 & 5 & $404$ & $3.0 \times 10^{-1}$ & $1238.5\pm0.2$ & $56.2\pm0.5$ \\
				3 & 4 & $405$ & $3.5 \times 10^{-1}$ & $1236.0\pm0.2$ & $61.2\pm0.5$ \\
				21 & 18 & $391$ & $3.7 \times 10^{0}$ & $1235.0\pm0.2$ & $63.3\pm0.5$ \\
				18 & 12 & $397$ & $4.2 \times 10^{-1}$ & $1234.7\pm0.2$ & $63.9\pm0.5$ \\
				15 & 6 & $403$ & $3.3 \times 10^{-1}$ & $1233.4\pm0.2$ & $66.4\pm0.5$ \\
				9 & 9 & $400$ & $1.3 \times 10^{0}$ & $1002.5\pm0.3$ & $528.3\pm0.7$ \\
				5 & 6 & $403$ & $1.5 \times 10^{0}$ & $969.4\pm0.2$ & $594.5\pm0.6$ \\
				22 & 3 & $406$ & $3.0 \times 10^{0}$ & $695.3\pm0.1$ & $1142.7\pm0.5$ \\
				23 & 3 & $406$ & $3.8 \times 10^{0}$ & $517.6\pm0.2$ & $1498.1\pm0.6$ \\
				1 & 3 & $406$ & $4.6 \times 10^{0}$ & $380.0\pm0.1$ & $1773.3\pm0.5$ \\
				\hline
			\end{tabular}
		\end{subtable}
\end{table*}

\begin{figure*}
	\centering
	\begin{subfigure}[b]{0.49\textwidth}
		\includegraphics[width=\textwidth]{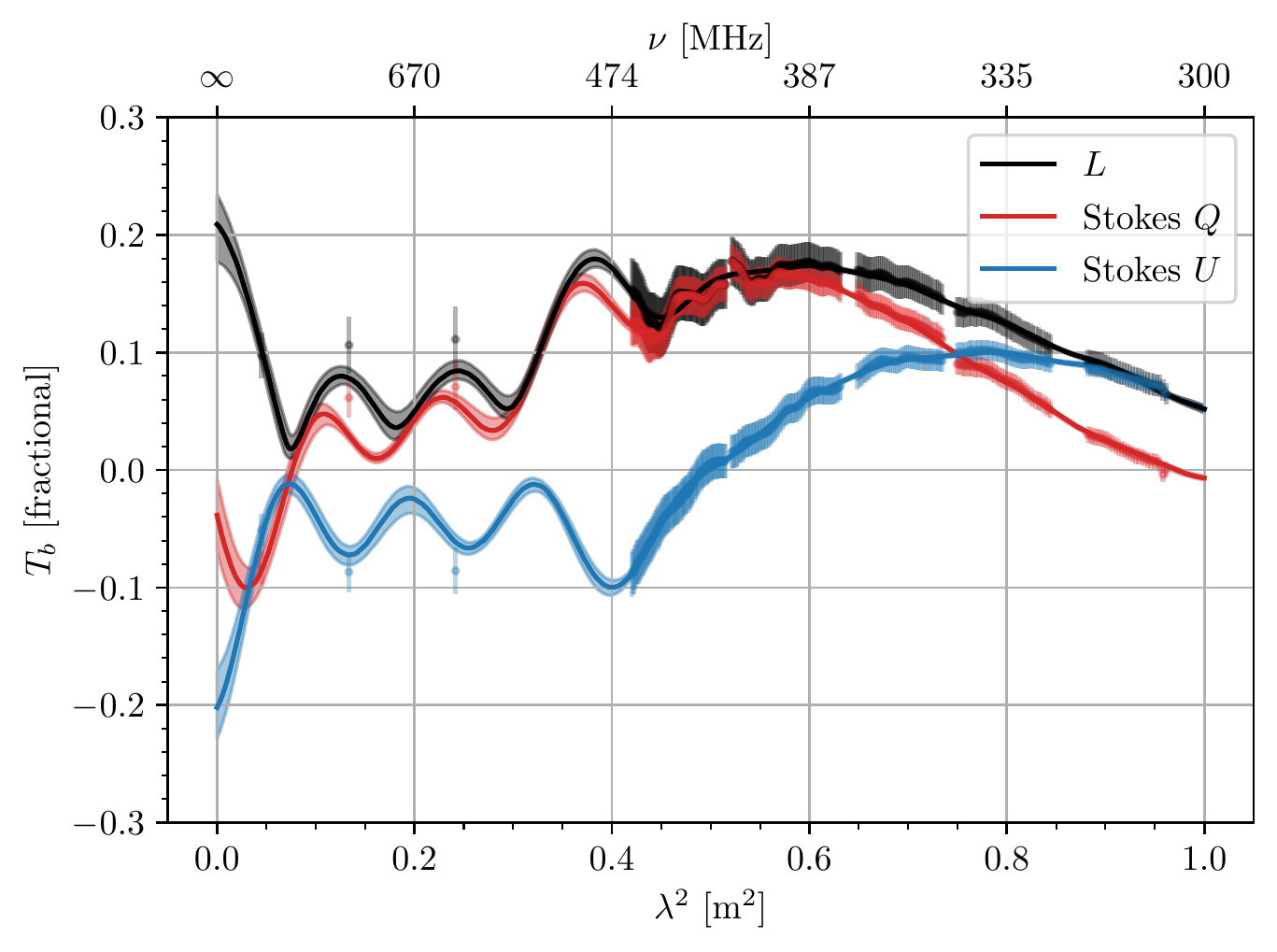}
		\label{fig:pi_model_red_big_19}%
	\end{subfigure}
	~ 
	\begin{subfigure}[b]{0.49\textwidth}
		\includegraphics[width=\textwidth]{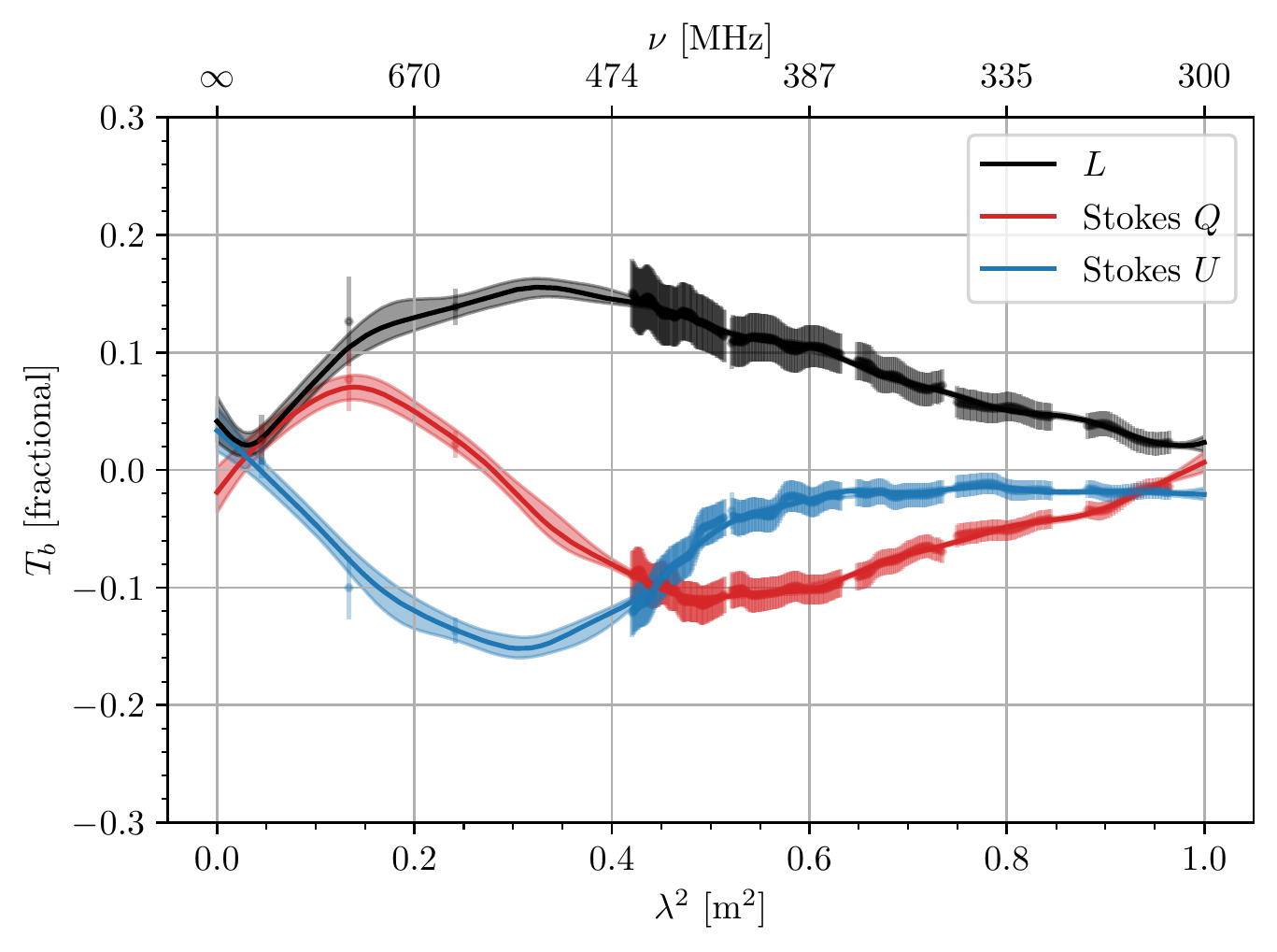}
		\label{fig:pi_model_blue_big_19}%
	\end{subfigure}
	\begin{subfigure}[b]{0.49\textwidth}
		\includegraphics[width=\textwidth]{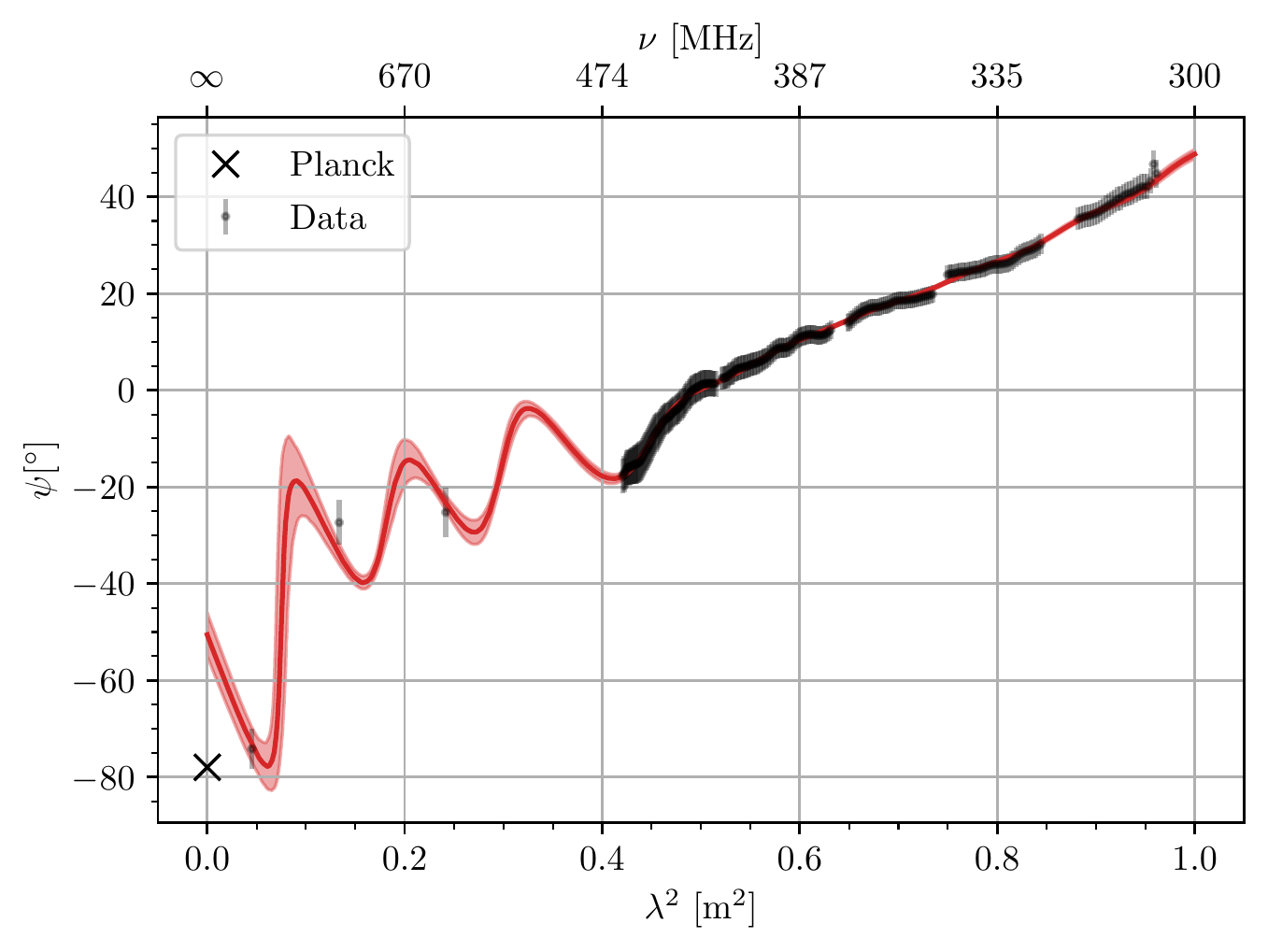}
		\label{fig:ang_model_red_big_19}%
	\end{subfigure}
	~ 
	\begin{subfigure}[b]{0.49\textwidth}
		\includegraphics[width=\textwidth]{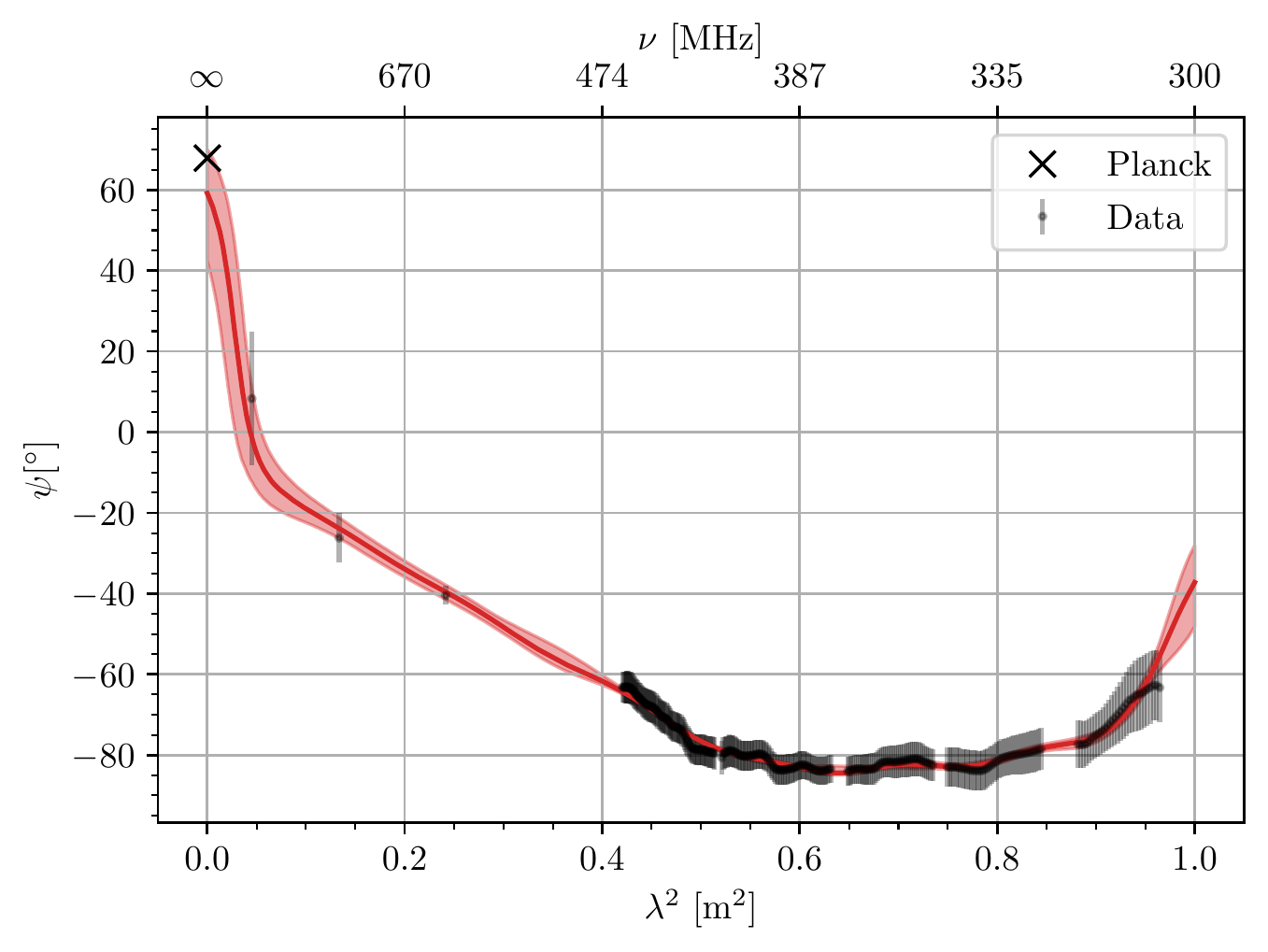}
		\label{fig:ang_model_blue_big_19}%
	\end{subfigure}
	\begin{subfigure}[b]{0.49\textwidth}
		\includegraphics[width=\textwidth]{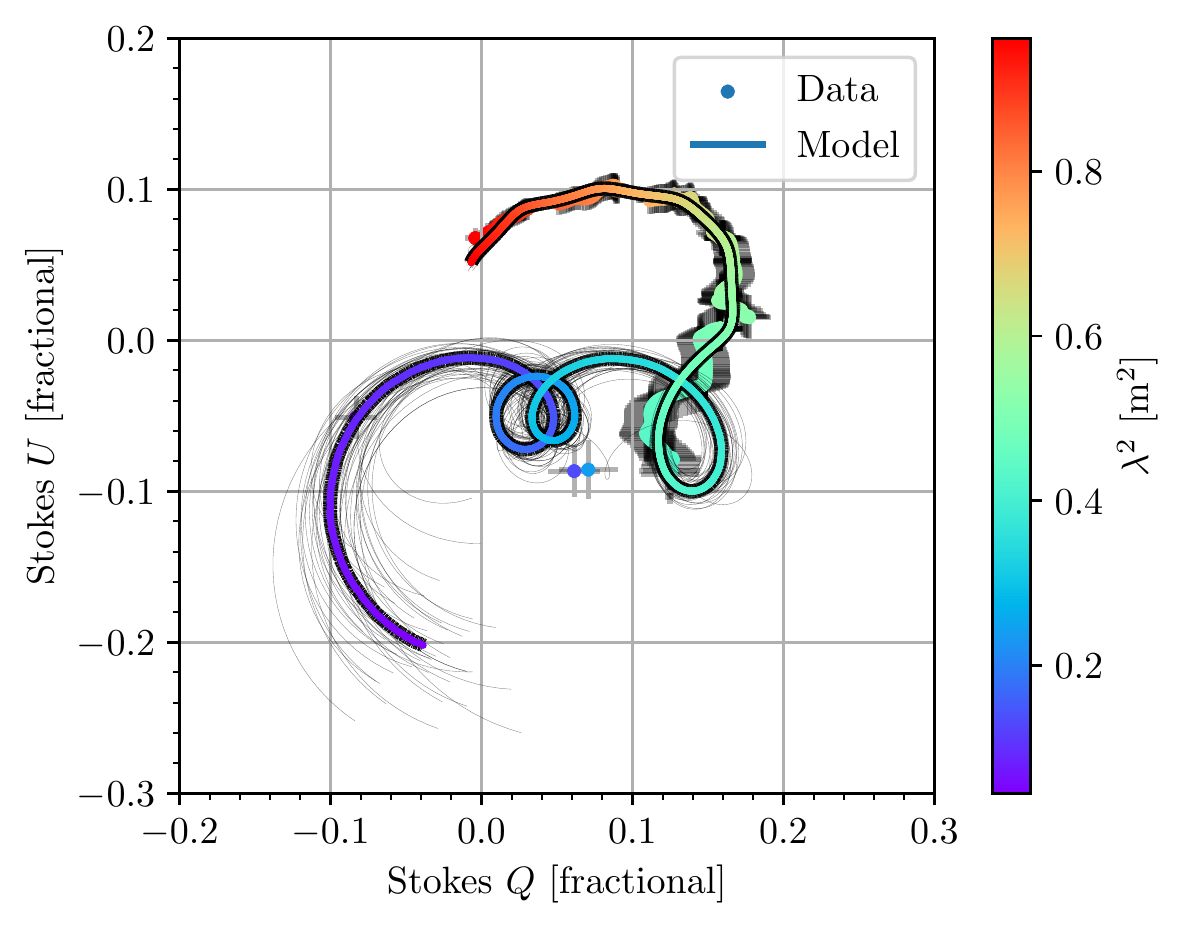}
		\label{fig:qu_model_red_big_19}%
	\end{subfigure}
	~ 
	\begin{subfigure}[b]{0.49\textwidth}
		\includegraphics[width=\textwidth]{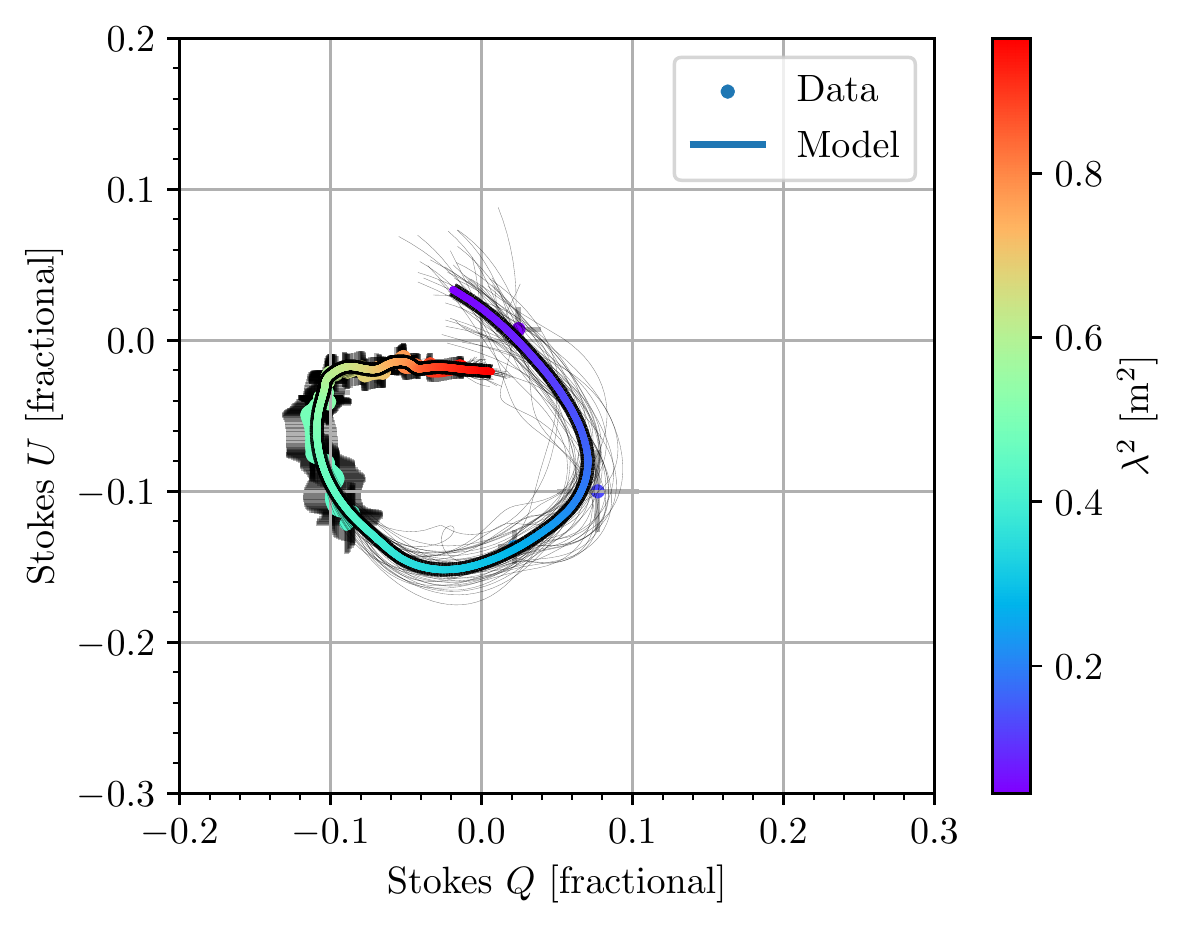}
		\label{fig:qu_model_blue_big_19}%
	\end{subfigure}
	
	\caption{Differential Faraday rotation with a foreground screen, helicity, and three emitting components (model 19). Left column -- Region 1, right column -- Region 2. Top -- Stokes $Q$, $U$, and polarized intensity ($L$) against $\lambda^2$. Middle -- Polarization angle ($\psi$) against $\lambda^2$. Bottom -- Stokes $Q$ against Stokes $U$. The data and the model are coloured by $\lambda^2$. In all panels we show the data as scatter points with error bars. The solid lines are the median best-fitting line, with translucent or traced lines showing the model error.}
	\label{fig:qu_model_19}
\end{figure*}

We first consider the best-fitting models across the entire band available to us. We list the goodness of fit parameters for this band in Table~\ref{tab:goodnes_full}. We find that models 10 and 11 both fail to converge in both regions, despite sufficient computational resources, indicating that they do not provide adequate reproduction of the data. From the odds ratio values alone, there is one model which outperforms all others: model 19, which describes 3 sources within the beam, each with differential Faraday rotation and a helicity term. We show the fit of this model to the data in Fig.~\ref{fig:qu_model_19}, and list the best-fit parameters in Table~\ref{tab:bestfit_full}. Inspecting the spectral fits, we find that in the GMIMS-LBS band, where we have many samples in $\lambda^2$, the fit is relatively constrained. Towards lower values of $\lambda^2$, however, the fitting routine presents a model that fluctuates strongly in-between the points constrained by the data, particularly in region 1. This, in combination with the complexity of this model, leads us to believe that there is an additional underlying physical component that is not being properly described by our parameterisations. The only model capable of reproducing the data is one that is multi-component and highly complex, and is potentially over-fitting the data. This interpretation is further supported by the small ($<<1$) values of the reduced $\chi^2$, which also indicates over-fitting.

We can also focus on the values of the best-fit parameters, recalling that the polarization angles at 30\,GHz are $-75\degr$ and $+59\degr$ in regions 1 and 2, respectively. Model 19 predicts an overall angle at $\lambda^2=0$ of $(-50\substack{+5 \\ -5})\degr$ and $(+60\substack{+17 \\ -11})\degr$ in regions 1 and 2, respectively. These values, under this parameterisation, are a result of the superposition of three emitting sources along the LOS. None of these components have intrinsic position angles close to the 30\,GHz measurements. More worryingly, the model requires very high intrinsic polarization fractions. In particular, in region 2 one component has a polarization fraction of $0.7\substack{+ 0.08 \\ -0.2}$. This is consistent, within the errors, with the theoretical maximum fraction, but is still physically very unlikely in the MIM of the Milky Way.

\begin{figure*}
	\centering
	\begin{subfigure}[b]{0.49\textwidth}
		\includegraphics[width=\textwidth]{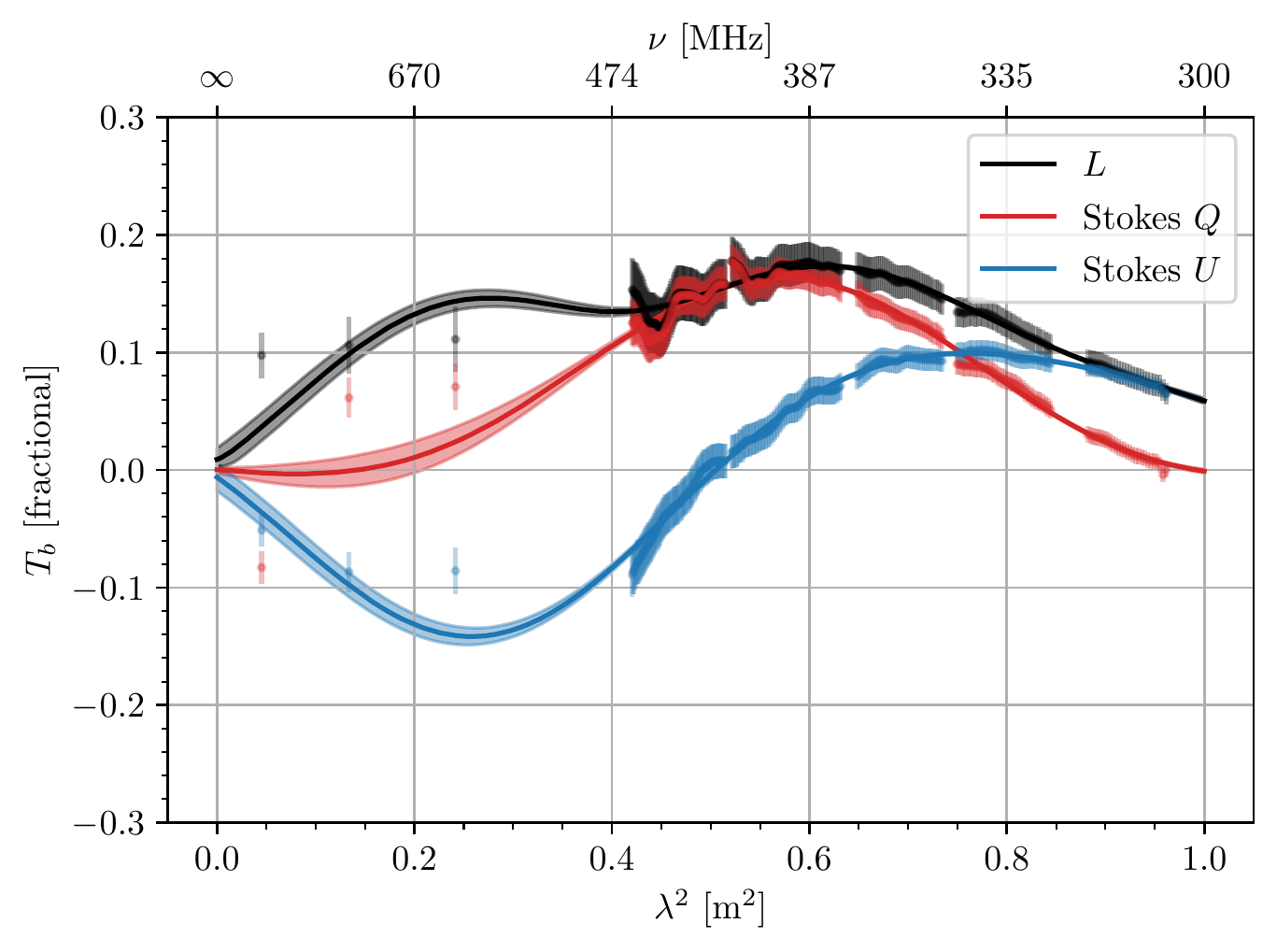}
		\label{fig:pi_model_red_big_16}%
	\end{subfigure}
	~ 
	\begin{subfigure}[b]{0.49\textwidth}
		\includegraphics[width=\textwidth]{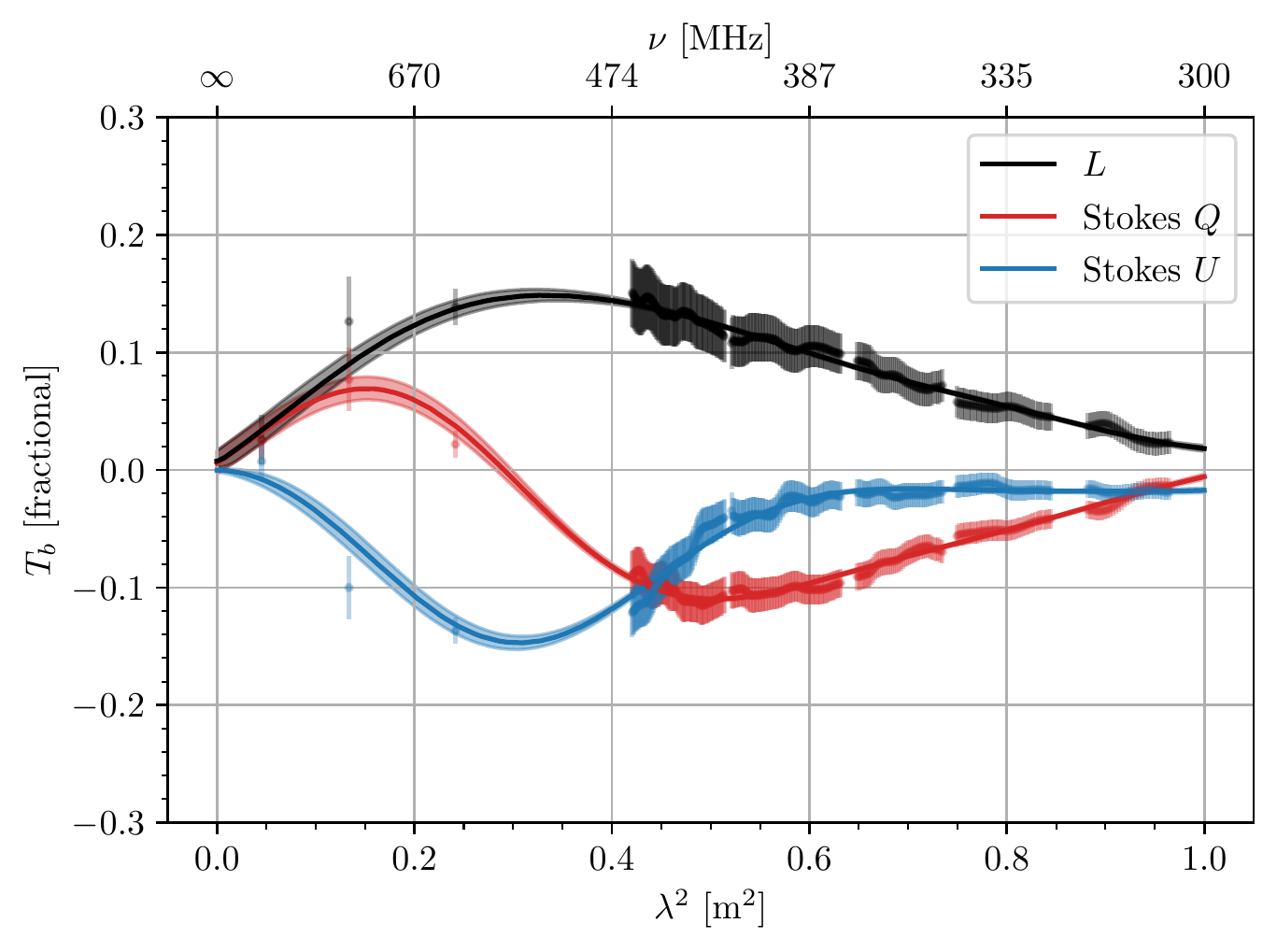}
		\label{fig:pi_model_blue_big_16}%
	\end{subfigure}
	\begin{subfigure}[b]{0.49\textwidth}
		\includegraphics[width=\textwidth]{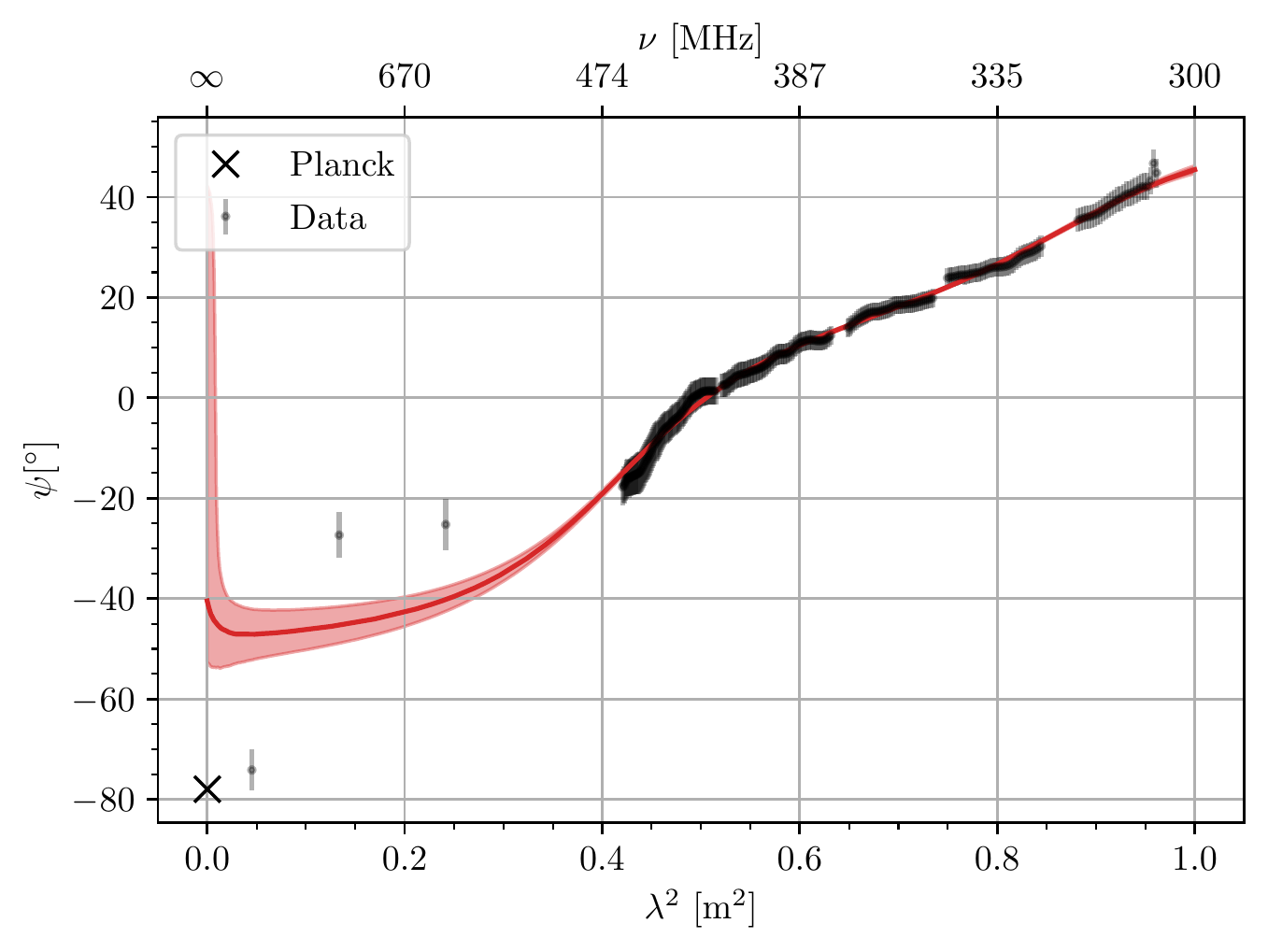}
		\label{fig:ang_model_red_big_16}%
	\end{subfigure}
	~ 
	\begin{subfigure}[b]{0.49\textwidth}
		\includegraphics[width=\textwidth]{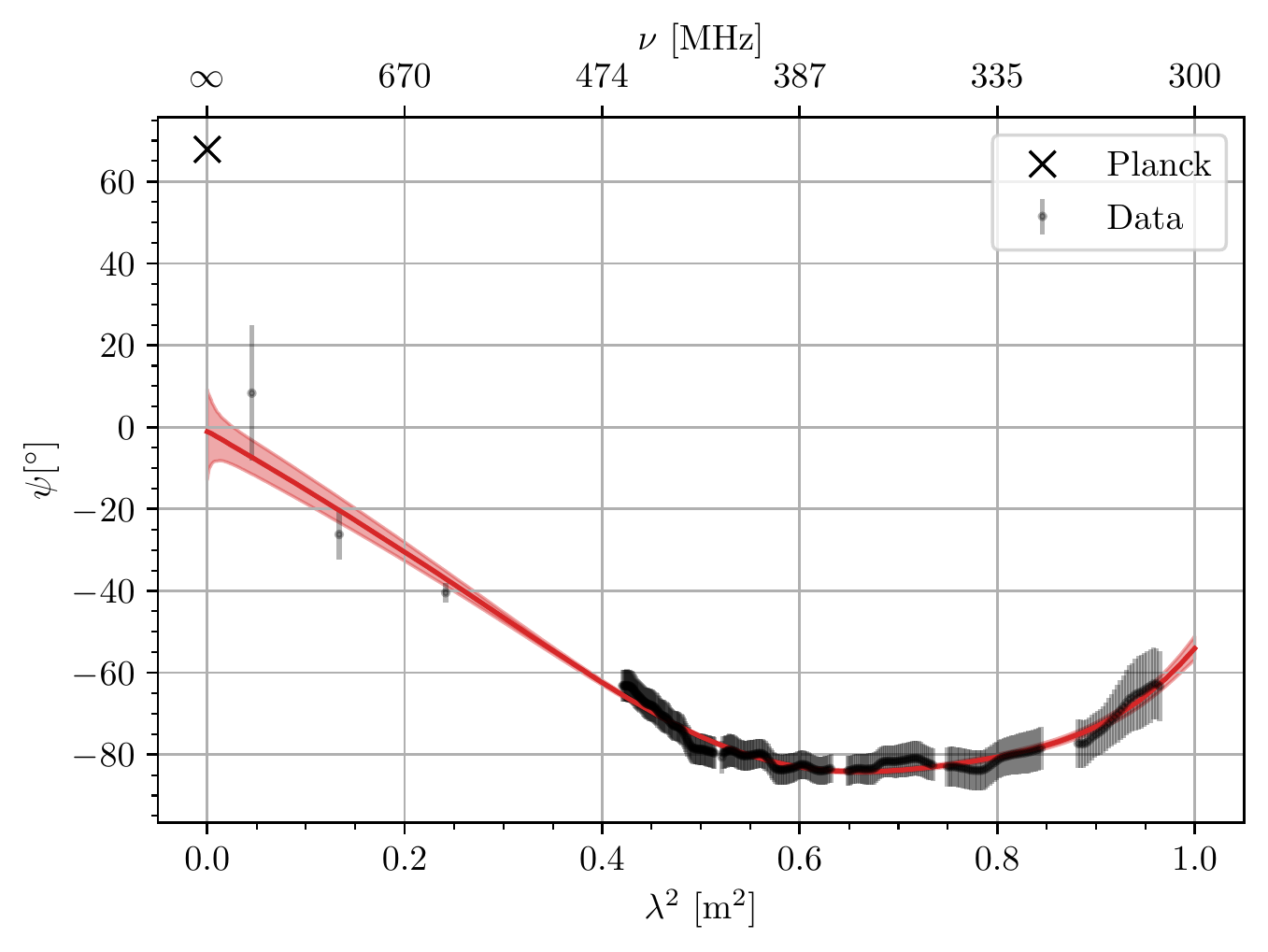}
		\label{fig:ang_model_blue_big_16}%
	\end{subfigure}
	\begin{subfigure}[b]{0.49\textwidth}
		\includegraphics[width=\textwidth]{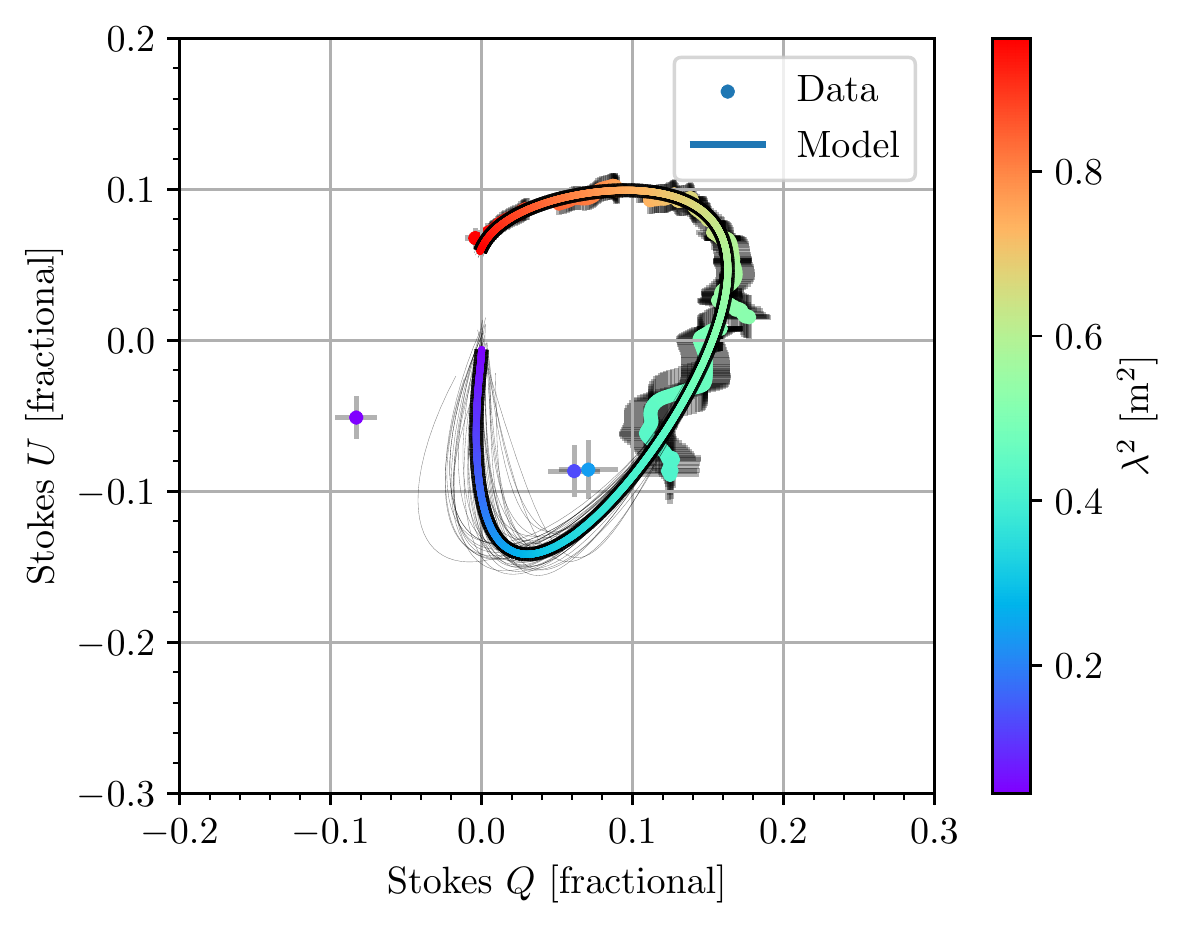}
		\label{fig:qu_model_red_big_16}%
	\end{subfigure}
	~ 
	\begin{subfigure}[b]{0.49\textwidth}
		\includegraphics[width=\textwidth]{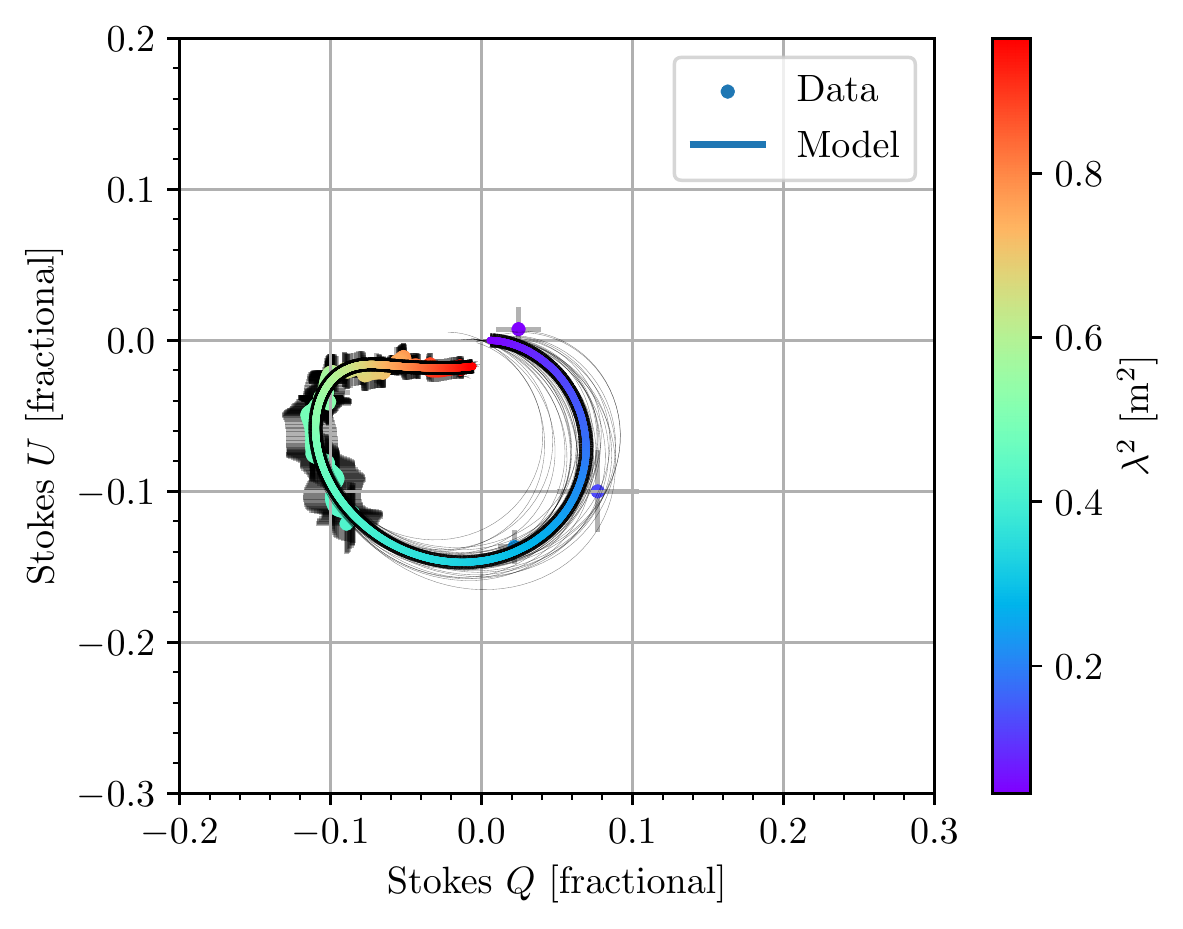}
		\label{fig:qu_model_blue_big_16}%
	\end{subfigure}
	
	\caption{Differential Faraday rotation with a foreground screen, helicity, and two emitting components (model 16). Arrangement is the same as Fig.~\ref{fig:qu_model_19}.}
	\label{fig:qu_model_16}
\end{figure*}

Looking to the next best-fitting model, we find that model 16 is the next best in both regions. This model describes only two sources of emission within the beam, each with internal Faraday dispersion and helicity.  Inspecting the fit of this model to the data in Fig.~\ref{fig:qu_model_16}, we find that this model does not produce the same high-frequency ripples as model 19. The fit to region 2 appears to be better than region 1 for this model, however overall the model appears to be much more reasonable. Similar to the previous model 19, model 16 is still highly complex, with 10 free parameters. Further, the largest deviations between the models occur where we do not have data to constrain them. Here the predicted polarization angles at $\lambda^2=0$ are $(-40\substack{+83 \\ -12})\degr$ and $(-1\substack{+10 \\ -12})\degr$, in regions 1 and 2, respectively; inconsistent with the measured values from \textit{Planck}. The low polarization fraction at $\lambda^2=0$ leads to large uncertainties in the recovered polarization angle. For this model, however, the required polarization fraction of the individual components is significantly lower than for model 19.

We are therefore left with two possible conclusions:
\begin{enumerate}
	\item The MIM in this direction is indeed multi-component, Faraday dispersive, and helical. In such a case we require more intermediate frequency ($\sim0.5$--$2$\,GHz) data to conclusively constrain the nature of this medium, or;
	\item there is an additional factor that is influencing the appearance of the broad-band spectra as a function of $\lambda^2$. We further discuss this option in Section~\ref{sec:physical}.
\end{enumerate}

\begin{table*}
		\centering
	\caption{Upper: Best fit parameters across the full band for models 19 and 16, as described in Table~\ref{tab:modelsum} and Equations~\ref{eqn:osul} and \ref{eqn:helicity}. Lower: Best fit parameters across the $\lambda^2=[0.5,1]$\,m$^2$ sub-band for models 6 and 2, as described in Table~\ref{tab:modelsum} and Equation~\ref{eqn:osul}.}
	\label{tab:bestfit_full}
\begin{tabular}{cccc|cc}
	\hline
	 & \multicolumn{5}{c}{~Full band} \\
	Region & \multicolumn{3}{c}{1} & \multicolumn{2}{c}{2} \\
		  \hline
	 Model \# &   &  19 &  16 &   19 &  16 \\
	 $p_{0,1}$ & fractional & $0.23\substack{+ 0.03 \\ -0.04}$ & $0.088\substack{+ 0.007 \\ -0.007}$ & $0.3\substack{+ 0.1 \\ -0.1}$ & $0.162\substack{+ 0.005 \\ -0.005}$ \\
	 $p_{0,2}$ & fractional & $0.052\substack{+ 0.006 \\ -0.006}$ & $0.159\substack{+ 0.002 \\ -0.001}$ & $0.7\substack{+ 0.08 \\ -0.2}$ & $0.074\substack{+ 0.003 \\ -0.003}$ \\
	 $p_{0,3}$ & fractional & $0.168\substack{+ 0.001 \\ -0.001}$ & -- & $0.167\substack{+ 0.008 \\ -0.008}$ & -- \\
	 $\phi_{0,1}$ & $\mathrm{rad\,m^{-2}}$ & $-0.5\substack{+ 0.1 \\ -0.1}$ & $-0.5\substack{+ 0.09 \\ -0.1}$ & $-2.0\substack{+ 2 \\ -0.4}$ & $-0.84\substack{+ 0.07 \\ -0.06}$ \\
	 $\phi_{0,2}$ & $\mathrm{rad\,m^{-2}}$ & $-8.8\substack{+ 0.4 \\ -0.4}$ & $0.79\substack{+ 0.04 \\ -0.03}$ & $-0.39\substack{+ 0.03 \\ -0.06}$ & $-1.0\substack{+ 0.1 \\ -0.09}$ \\
	 $\phi_{0,3}$ & $\mathrm{rad\,m^{-2}}$ & $0.83\substack{+ 0.01 \\ -0.01}$ & -- & $-0.61\substack{+ 0.09 \\ -0.06}$ & -- \\
	 $\psi_{0,1}$ & $\mathrm{{}^{\circ}}$ & $-30.0\substack{+ 10 \\ -10}$ & $-41.0\substack{+ 7 \\ -7}$ & $-90.0\substack{+ 60 \\ -50}$ & $-75.0\substack{+ 3 \\ -3}$ \\
	 $\psi_{0,2}$ & $\mathrm{{}^{\circ}}$ & $-4.0\substack{+ 3 \\ -6}$ & $8.0\substack{+ 1 \\ -1}$ & $78.0\substack{+ 6 \\ -5}$ & $-86.0\substack{+ 1 \\ -1}$ \\
	 $\psi_{0,3}$ & $\mathrm{{}^{\circ}}$ & $7.3\substack{+ 0.7 \\ -0.8}$ & -- & $-67.0\substack{+ 2 \\ -2}$ & -- \\
	 $\eta_1$ & $\mathrm{m^{2}}$ & $0.012\substack{+ 0.006 \\ -0.008}$ & $-0.25\substack{+ 0.02 \\ -0.02}$ & $0.8\substack{+ 0.7 \\ -0.6}$ & $-0.35\substack{+ 0.01 \\ -0.01}$ \\
	 $\eta_2$ & $\mathrm{m^{2}}$ & $-0.394\substack{+ 0.008 \\ -0.008}$ & $-0.34\substack{+ 0.04 \\ -0.04}$ & $0.0\substack{+ 1 \\ -0.7}$ & $-0.38\substack{+ 0.01 \\ -0.01}$ \\
	 $\eta_3$ & $\mathrm{m^{2}}$ & $-0.202\substack{+ 0.009 \\ -0.007}$ & -- & $-0.83\substack{+ 0.03 \\ -0.02}$ & -- \\
	 $\Delta\phi_{1}$ & $\mathrm{rad\,m^{-2}}$ & $35.3\substack{+ 0.8 \\ -0.7}$ & $12.3\substack{+ 0.4 \\ -0.4}$ & $30.0\substack{+ 2 \\ -10}$ & $8.8\substack{+ 0.2 \\ -0.2}$ \\
	 $\Delta\phi_{2}$ & $\mathrm{rad\,m^{-2}}$ & $21.0\substack{+ 1 \\ -2}$ & $5.3\substack{+ 0.3 \\ -0.3}$ & $8.6\substack{+ 0.2 \\ -0.3}$ & $12.6\substack{+ 0.3 \\ -0.4}$ \\
	 $\Delta\phi_{3}$ & $\mathrm{rad\,m^{-2}}$ & $5.8\substack{+ 0.2 \\ -0.2}$ & -- & $8.2\substack{+ 0.5 \\ -0.5}$ & -- \\
	\hline
	 & \multicolumn{5}{c}{~$\lambda^2=[0.5,1]$\,m$^2$ sub-band} \\
	Region & \multicolumn{3}{c}{1} & \multicolumn{2}{c}{2} \\
	\hline
	Model \# & &   6 & 2 &   6&  2\\
	$p_{0,1}$ & fractional & $0.17\substack{+ 0.05 \\ -0.01}$ & $0.249\substack{+ 0.003 \\ -0.003}$ & $0.171\substack{+ 0.004 \\ -0.006}$ & $0.179\substack{+ 0.003 \\ -0.003}$ \\
	$p_{0,2}$ & fractional & $0.221\substack{+ 0.006 \\ -0.006}$ & -- & $0.16\substack{+ 0.01 \\ -0.02}$ & -- \\
	$\phi_{0,1}$ & $\mathrm{rad\,m^{-2}}$ & $0.3\substack{+ 0.1 \\ -0.2}$ & $1.54\substack{+ 0.02 \\ -0.02}$ & $0.15\substack{+ 0.05 \\ -0.05}$ & $0.12\substack{+ 0.05 \\ -0.05}$ \\
	$\phi_{0,2}$ & $\mathrm{rad\,m^{-2}}$ & $1.47\substack{+ 0.02 \\ -0.02}$ & -- & $4.2\substack{+ 0.4 \\ -0.4}$ & -- \\
	$\psi_{0,1}$ & $\mathrm{{}^{\circ}}$ & $9.0\substack{+ 9 \\ -6}$ & $-43.2\substack{+ 0.9 \\ -0.8}$ & $-87.0\substack{+ 2 \\ -2}$ & $-86.0\substack{+ 2 \\ -2}$ \\
	$\psi_{0,2}$ & $\mathrm{{}^{\circ}}$ & $-41.0\substack{+ 1 \\ -1}$ & -- & $-80.0\substack{+ 20 \\ -6}$ & -- \\
	$\Delta\phi_{1}$ & $\mathrm{rad\,m^{-2}}$ & $11.4\substack{+ 0.1 \\ -0.1}$ & $2.43\substack{+ 0.02 \\ -0.02}$ & $2.89\substack{+ 0.03 \\ -0.03}$ & $2.95\substack{+ 0.02 \\ -0.02}$ \\
	$\Delta\phi_{2}$ & $\mathrm{rad\,m^{-2}}$ & $2.27\substack{+ 0.05 \\ -0.05}$ & -- & $22.2\substack{+ 0.2 \\ -0.2}$ & -- \\
\hline
	\end{tabular}
\end{table*}

We will next consider the best-fitting models on the $\lambda^2=[0.5,1]$\,m$^2$ sub-band, where the spectra appear most simple. We list the goodness of fit parameters for this band in Table~\ref{tab:goodnes_sub}. In this sub-band, we again find that models 10 and 11 fail to converge. The Bayesian odd ratios inform us that the best fit to both regions are again provided by models 19 and 16. In region 1, several models provide a comprably good fit, including the single-component model 14. Similar to the fit of model 19 to the full band, in region 2 this model appears to capture some extra complexity that we would consider unreasonable. This appears to be driven by the slight change in gradient of the position angle at the longest $\lambda^2$ values. It is important to note that this occurs at low polarized fraction, and is therefore subject to lower polarized signal-to-noise. The models fitted to just this sub-band do not extrapolate to the higher frequencies. We are not concerned by this result, particularly if we follow conclusion (ii) as described above.

The almost linear position angle against $\lambda^2$ relationship in the $\lambda^2=[0.5,1]$\,m$^2$ sub-band drives us to consider simpler models. Whilst the goodness of fit parameters would otherwise drive us to consider more complex models, we find that inspecting these fits shows a reasonably good description of the data in this sub-band. From inspection we find that simple, single-component models, such as 2 and 3, provide very reasonable fits to the low-band data. We believe that this is not reflected in the goodness-of-fit criteria due to depolarization between 0.5 and 0.6\,m$^2$ in region 1, and the polarization angle structure in region 2 and the longest $\lambda^2$. Outside of the sub-band however, the fits are again divergent from the high-frequency data. Lastly, the predicted polarization fractions are all $<30\,$per\,cent; which is not unreasonably high for the MIM of the Milky Way~\citep{Page2007,Kogut2007,Vidal2015}.

\section{Discussion}\label{sec:discussion}
\subsection{Physical feasibility of the models}\label{sec:physical}
A model that provides a reasonably good fit to the data comprises a complex, multi-component model, with various terms describing Faraday dispersion, and helicity. This is physically unlikely, however, because emission and Faraday rotation are mixed in the MIM, they do not occur in separate places. In other words, the MIM is more like a Burn slab, a region of mixed emission and rotation. This picture may start to break down at low radio frequencies ($\lesssim100\,$MHz) due to depolarization from highly ionised components of the ISM~\citep{VanEck2017}. In the GMIMS-LBS band, however, it is reasonable to assume that we can detect emission from most of the ISM within the polarization horizon \citep{Thomson2019}. We do find, moreover, that a simple Burn slab model does provide a reasonable description of the longest $\lambda^2$ data.

A Burn slab, while more representative of the true nature of the MIM, is also not very realistic because neither the field nor the distribution of thermal electrons is likely to be entirely uniform. In the case of an expanding shell, however, the MIM may approach a multi-component configuration. Additionally, a multi-component slab model could mimic the behaviour of a slab with non-uniform parameters. A Faraday caustic is one step more realistic than a Burn slab by including a linearly varying magnetic field. We note the recent work of \citet{Basu2019}, who investigated the Faraday spectra from simulated, turbulent, synchrotron emitting, magneto-ionic media. They found that broad structures, such as the features we observed towards G150$-$50, arise from gradients in $\phi$ at constant synchrotron emissivity. Such a scenario is what is described, at least in broad terms, by a Burn slab or a Faraday caustic. However, we note that changes in the emissivity produce substructure in the observed Faraday spectrum.

We have implemented a model of a Faraday caustic along the lines suggested by \citet{Bell2011}. Using their simplified equations produces a singularity at zero wavelength because of the $1/{\lambda}$ term in Equation~\ref{eqn:m22}, which automatically generates very strong polarized emission at short wavelengths. As \citeauthor{Bell2011} point out, their approximation ignores the actual spectrum of synchrotron emissivity, which decreases strongly towards short wavelengths, eliminating this singularity. At short wavelengths, however, Faraday rotation becomes less effective, and in all the models that we have considered the polarized intensity should remain high at short wavelengths, when in fact observations \citep[e.g.][]{Wolleben2006} show that there is no polarized feature at 1.4 GHz corresponding to G150$-$50.

Similarly, the Dwingeloo data \citep{Brouw1976} show that G150$-$50 is detected as a polarization feature only at low frequencies. Inspecting these surveys we can see evidence of G150$-$50 at 465 and 408 MHz (albeit with coverage of only part of the region at the lower frequency), but we do not see it in the Dwingeloo data at 610 or 820 MHz. The region is not covered by the Dwingeloo 1411 MHz data, but it is covered by the 1410\,MHz data of \citet{Wolleben2006}.

How can this be reconciled with our modelling? At the frequencies of GMIMS-LBS, the polarization horizon is at a distance of about 500\,pc \citep{Dickey2018}. At shorter wavelengths, the polarization horizon becomes progressively more distant. It must be the case that at the shorter wavelengths the high degree of polarization from the Faraday caustic is greatly diluted by the superposition of more distant emission.

This direction, $(150\degr,-50\degr)$, is towards the anticentre and well below the Galactic plane. As we previously discussed, the scale heights of cosmic ray electrons and of the regular magnetic field are both on the order of a few kpc. We observe a fractional polarization of 10\,per\,cent. This high percentage automatically implies that the emission we are seeing in G150$-$50 is generated along a fairly long path. The theoretical maximum polarized percentage of synchrotron emission is 70\,per\,cent. To first order, 10/70, about 15\,per\,cent, of the path must be contributing to the received signal. Taking a distance of 2\,kpc to the edge of the thick disk, the received signal is coming from $\sim300$\,pc of the total path. This is, of course, weighted by the varying synchrotron emissivity along the line-of-sight. In this direction it seems probable that the emissivity falls off with distance. It is therefore likely that the complexity we observe in the polarized spectra is caused by geometric depolarization, the degree of which is co-dependent on distance and $\lambda^2$.

Lastly, we can try to quantify the extent of the emitting path. First, we take the average foreground synchrotron emissivity ($\varepsilon$) at 76.2\,MHz \citep{Su2018},
\begin{align}
    \varepsilon(76.2) = 1\pm0.5\,\mathrm{K}\,\mathrm{pc}^{-1}.
\end{align}
Scaling this to 408\,MHz taking $\varepsilon\propto\nu^\beta$, with $\beta=-2.5$, gives
\begin{align}
    \varepsilon(408) = 0.015\pm0.008\,\mathrm{K}\,\mathrm{pc}^{-1}.
\end{align}
Now, the median polarized brightness temperature of region 1 of G150$-$50 is 2.5\,K. Again, if we assume the volume was emitting at the maximum fraction of 70\,per\,cent, this would correspond to 3.6\,K of total emission. Using the emissivity from above, this would correspond to an emitting path of $240\substack{+270 \\ -80}$\,pc, which is consistent with our estimate above. Naturally, if the intrinsic polarization fraction were lower, the emitting path would be longer. We can conclude from this that the path producing the polarized emission towards G150$-$50, which must contain an ordered magnetic field, is on the order of a few hundred parsecs.

\subsection{Further investigations}
Shell-like structures, such as our proposed model, should be present in other, co-located components of the ISM. This calls for further investigation of this region using additional ISM tracers. Such comparison would include neutral hydrogen \citep[\hi, such as in][]{Thomson2018} and three-dimensional dust maps~\citep[such as in][]{VanEck2017,Thomson2019}. A possible correspondence between Loop II and an \hi\ feature was commented on by \citet{Heiles1989}; however, he does not draw any strong conclusions about the relationship between this feature and Loop II. We should also note that much of the prominence of Loop II and G150$-$50 is due to the structure of the magnetic field, and not to any process that would necessarily generate enhanced emission in tracers of the neutral ISM. Additionally, the structure of Loop II is huge in angular scale, and very diffuse. This will make finding a clear corresponding feature in the ISM particularly difficult. As we discuss in Section~\ref{sec:physical}, it is also probable that emission from beyond the GMIMS-LBS polarization horizon is affecting high-frequency data we have obtained.

We note that shell-like structure is also consistent with a Faraday caustic. Faraday caustics have been considered only a few times since their introduction by \citet{Bell2011}. \citet{Bell2012} and \citet{Beck2012} discussed detecting them through processing techniques, and \citet{Ideguchi2014} found them in simulated observations. \citet{VanEck2017,VanEck2019} considered them in their analysis of LOFAR observations, but the data were unable to confirm or exclude a Faraday caustics interpretation. We are also unable to claim the definitive detection of a caustic. In fact, our model-fitting shows it is unfavourable compared to most of the other models we consider. In this case, we are limited by the potential of a rapidly changing polarization horizon as a function of $\lambda^2$.

A very useful follow-up observation would be to utilize diffuse, polarized imaging of this area from an interferometer such as the Murchison Widefield Array (MWA). The effect of the polarization horizon is highly beam-dependent. Therefore, the smaller the synthesized beam, the further the distance to the horizon. The GaLactic and Extragalactic All-sky Murchison Widefield Array (GLEAM) survey~\citep{Wayth2015} already covers the G150$-$50 region. \citet{Lenc2016} showed that large-scale, diffuse polarized emission can be successfully detected with MWA. This region does show signs of enhanced, diffuse polarized emission in GLEAM observations (Lenc, E.; Sun, X. H. private communication). The higher angular resolution would allow comparison with other ISM tracers, such as \hi. Additionally, the low frequencies would allow for very fine Faraday resolution~\citep{Riseley2018, Riseley2020}, which would be useful to compare with the GMIMS-LBS results. If the full GLEAM bandwidth could be utilized, that would provide a maximum-to-minimum frequency ratio of $~3.2$. As shown by \citet{Bell2011}, such a ratio, in combination with high Faraday resolution, can unambiguously indicate a Faraday caustic, and can even reveal details of the turbulent magnetic field along the LOS. 

Finally, the most natural extension of this work would be the incorporation of the complete GMIMS components. In particular, the mid-band surveys would allow us to test our model extrapolations and determine at what frequencies G150$-$50 no longer dominates the polarized sky. Particularly from the perspective of $QU$-fitting, the sparse $\lambda^2$ sampling from the Dwingeloo surveys is insufficient for us to be able to conclusively determine the nature of the polarized spectra towards G150$-$50. Such an effort would be an important step towards characterising, and measuring the polarization horizon for these data. This has already begun through \citet{Dickey2018}, but should be continued with the complete sky and frequency coverage GMIMS is striving towards.

\section{Conclusions}\label{sec:conclusion}
We have identified the brightest region in the Southern polarized sky at 400\,MHz in GMIMS-LBS, and we refer to it as G150$-$50. We have analysed both the morphology and spectral structure in polarization from 300\,MHz to 30\,GHz.

The region, located at $(l,b)\sim(150\degr,-50\degr)$, has two key characteristics:
\begin{enumerate}
    \item A magnetic field primarily in the plane of the sky:
    \begin{itemize}
        \item We see diffuse Stokes $I$ emission; here coming from the structure known as Loop II~\citep{Large1962}.
        \item The rotation measures from extragalactic sources are $\sim0$\,\radms.
        \item The polarized emission is \textit{not} depolarized at the low frequencies observed by GMIMS-LBS.
        \item The Faraday depths from GMIMS-LBS are also relatively low, $\phi\sim\pm3\,$\radms.
    \end{itemize}
    \item A magnetic field that is coherent and ordered:
    \begin{itemize}
        \item We find a high polarization fraction $\sim10\,$per\,cent in GMIMS-LBS.
        \item This high fraction is consistent with an ordered field at least $\sim100\,$pc deep along the LOS (to first order).
        \item We find aligned magnetic field vectors, tracing the path of Loop II.
    \end{itemize}
\end{enumerate}

In GMIMS-LBS, the G150$-$50 region appears to be separated into two regions, 1 and 2, centred on $(l,b)\sim(151\degr,-50\degr)$ and $(139\degr,-53\degr)$, respectively. Using Faraday tomography, we showed that this split is caused by beam depolarization generated by a gradient in Faraday depth across the sky, giving rise to a depolarization canal. In total intensity there is no feature that directly resembles G150$-$50, but it does appear to align with part of Loop II. With increasing frequency, Loop II appears as a polarized feature, but the prominence of G150$-$50 diminishes.

The polarized continuum and Faraday spectra from GMIMS-LBS in each region strongly indicate broad Faraday depth structure. Such a structure is caused by a Faraday dispersive medium. In particular, the Faraday spectra can be modelled as a Faraday caustic~\citep{Bell2011}. Faraday caustics are caused by a gradient in the LOS component of the Galactic magnetic field, leading to a reversal in $B_\parallel$ at some location along the line-of-sight. GMIMS-LBS is one of the first surveys capable of detecting such structures, with $\nu_{\text{max}}/\nu_{\text{min}}>1.5$, meaning that caustic structures can be resolved in Faraday depth space.

We are able to construct a physical model that matches our observations from GMIMS-LBS. We found that a simple spherical-shell magnetic field model \citep[e.g.][]{Spoelstra1972,Wolleben2007,Vidal2015}, is able to reproduce both the observed Faraday depth structure and the total intensity emission. This model indicates that Loop II and G150$-$50 are, in fact, the same structure; a shell blown into an ambient field, with a mean direction pointing away from the Galactic centre and towards the Sun.

After the application of Stokes $QU$-fitting analysis, we found that the best-fitting model requires several emitting components along the LOS, each with Faraday dispersion and helical fields. Additionally, we find that a Faraday caustic model is disfavoured in our model fitting. A multi-component model might be more reasonable in two, possibly mutual, contexts. First, in the case of an expanding shell, the shock front produces regions of relatively discrete ISM, each with different ionisation properties. Secondly, a multi-component slab, with each component modelled by uniform parameters, may mimic a single slab with non-uniform parameters.

Conversely, it is possible that the volume probed by polarization observations changes dramatically with frequency. Such a change is not accounted for by this modelling. As such, we were not able to confirm whether G150$-$50 corresponds with a Faraday caustic, as indicated by the RM synthesis results. This result is very important for consideration of future GMIMS observations, and analysis of diffuse polarized emission in general. Observations of diffuse, polarized emission that span a broad range of $\lambda^2$ may not be able to be reconciled as a whole. We must begin to constrain the distance of the polarization horizon for such analysis.

These results are key to understanding the local magneto-ionic environment. Whilst our model of the magnetic structure for G150$-$50 is relatively small on a Galactic scale, it is huge in angular scale. As such, this feature is important to characterise for studies of larger, and further Galactic structures, as well as extragalactic polarized features such as the CMB. Further investigation of both this feature, and the overall polarization horizon, can be made with the complete GMIMS survey. For our single-dish observations, beam depolarization was the key limiting factor which potentially confused our $QU$-fitting. This calls for low-frequency, high-resolution follow-up with an interferometer, such as the MWA. Faraday caustics have the ability to illuminate the turbulent component of interstellar magnetic fields~\cite{Bell2011}. If confirmed, G150$-$50 would be the first observed Faraday caustic.

\section*{Acknowledgements}

The authors thank the anonymous referee whose comments provided a substantial improvement to this work. We wish to thank Michael Bell and Torsten En{\ss}lin for their useful correspondence on this work, George Heald for insightful discussions and comments, Emil Lenc for digitizing the \citet{Mathewson1965} data, and Phil Edwards for providing a careful internal review.

AT acknowledges the support of the Australian Government Research Training Program (RTP) Scholarship and the Joan Duffield Research Award. The Dunlap Institute is funded through an endowment established by the David Dunlap family and the University of Toronto. The University of Toronto operates on the traditional land of the Huron-Wendat, the Seneca, and most recently, the Mississaugas of the Credit River; J.L.C. is grateful to have the opportunity to work on this land. J.L.C., B.M.G., and J.L.W. acknowledge the support of the Natural Sciences and Engineering Research Council of Canada (NSERC) through grant RGPIN-2015-05948, and of the Canada Research Chairs program. S. E. C. acknowledges support by the Friends of the Institute for Advanced Study Membership. C.~F.~acknowledges funding provided by the Australian Research Council (Future Fellowship FT180100495), and the Australia-Germany Joint Research Cooperation Scheme (UA-DAAD). J.L.H. is supported by the National Natural Science Foundation of China (No. 11988101 and 11833009). M.H. acknowledges funding from the European Research Council (ERC) under the European Union's Horizon 2020 research and innovation programme (grant agreement No 772663). A.S.H. is supported by an NSERC Discovery Grant. A.O is supported by the Dunlap Institute and the National Research Council Canada.

The Parkes Radio Telescope is part of the Australia Telescope National Facility which is funded by the Australian Government for operation as a National Facility managed by CSIRO. We acknowledge the Wiradjuri people as the traditional owners of the Observatory site. 

This research made use of the IPython package \citep{PER-GRA:2007}, SciPy \citep{2020SciPy-NMeth}, NumPy \citep{van2011numpy}, Matplotlib, a Python library for publication quality graphics \citep{Hunter:2007}, and Astropy,\footnote{\url{http://www.astropy.org}} a community-developed core Python package for Astronomy \citep{astropy:2013, astropy:2018}. Some of the results in this paper have been derived using the HEALPix~\citep{Gorski2005} package. We acknowledge the use of data provided by the Centre d'Analyse de Donn\'ees Etendues (CADE), a service of IRAP-UPS/CNRS \citep{Paradis2012}

\section*{Data Availability}
The data underlying this article are available publicly via the Canadian Astronomy Data Centre\footnote{\url{https://www.cadc-ccda.hia-iha.nrc-cnrc.gc.ca/en/}}, the Legacy Archive for Microwave Background Data Analysis (LAMBDA)\footnote{\url{https://lambda.gsfc.nasa.gov/}}, the Centre d'Analyse de Donn\'ees Etendues (CADE)\footnote{\url{http://cade.irap.omp.eu}}, the Max Planck Institute for Radio Astronomy (MPIfR) Survey Sampler\footnote{\url{https://www3.mpifr-bonn.mpg.de/survey.html}}, and the Max-Planck-Institut f{\"u}r Astrophysik (MPA) Information Field Theory Group\footnote{\url{https://wwwmpa.mpa-garching.mpg.de/~ensslin/research/data/data.html}}. Derived data products will be shared on reasonable request to the corresponding author.




\bibliographystyle{mnras}
\bibliography{correct} 




\appendix
\section{}
RM synthesis is a Fourier-transform-like process that maps the complex polarization ($P$) as a function of $\phi$. In Faraday tomography, this process is applied to all observed spectra over a region of sky, thereby mapping the Faraday depth across the sky. For a given set of observations of Stokes $I$, $Q$, and $U$, the complex polarization is:
\begin{equation}
	P = pI = L e^{2i\psi} = Q + iU,
	\label{eqn:pol}
\end{equation}
where $p$ is the complex fractional polarization, and $L$ is the linearly polarized intensity.

Applying Faraday tomography to such data provides the Faraday dispersion function (FDF$(\phi)$), where \cite{Burn1966} gives:
\begin{equation}
	P(\lambda^2) = \int_{-\infty}^{+\infty} \text{FDF}(\phi) e^{2i\phi\lambda^2} d\phi
	\label{eqn:rmsynth}
\end{equation}
The absolute value, $||\text{FDF}(\phi)||$, provides the linearly polarized intensity ($L=\sqrt{Q^2 +U^2}$) as a function of Faraday depth. $||\text{FDF}(\phi)||$ is spectral in nature, and here we refer to it as the `Faraday spectrum'. 

For real observations, we can only measure a finite and discrete set of $\lambda^2$ values, with $\lambda^2>0$. \citet{Brentjens2005} show that this causes the Faraday spectrum to be naturally convolved with the `RM spread function' (RMSF). Some of the effects of the RMSF can be mitigated through deconvolution techniques, such as \texttt{RM-CLEAN}~\citep{Heald2009}. 

The parameters of the Faraday spectrum are determined by the RMSF, which itself is determined by what values of $\lambda^2$ are observed. The resolution in Faraday depth ($ \delta\phi $) is defined by the FWHM of the RMSF, and is given by \citet{Dickey2018} as:
\begin{equation}
	\delta\phi \approx \frac{3.8}{\Delta\lambda^2},
	\label{eqn:faradayres}
\end{equation}
where $\Delta\lambda^2=\lambda_{\text{max}}^2 - \lambda_{\text{min}}^2$, and $\lambda_{\text{max}}^2$ and $\lambda_{\text{min}}^2$ are the maximum and minimum observed $\lambda^2$, respectively. The largest recoverable Faraday depth ($\phi_{\text{max}}$) is defined as the maximum absolute Faraday depth with more than 50\% sensitivity in the RMSF:
\begin{equation}
	\phi_{\text{max}} \approx \frac{\sqrt{3}}{\delta\lambda^2},
	\label{eqn:faradaymax}
\end{equation}
where $\delta\lambda^2$ is the size of each channel in wavelength-squared space. For most radio observations, however, the channel width is constant in frequency rather than $\lambda^2$. \cite{Pratley20} show that the channel width $\delta\lambda^2$ can vary dramatically between meter to centimetre wavelengths, which give rise to complex structures at the edge of Faraday sensitivity that are purely instrumental. This effect, however, diminishes with increasingly small frequency channels. Finally, the maximum scale observable in Faraday depth space is:
\begin{equation}
	\phi_{\text{max-scale}} \approx \frac{\pi}{\lambda_{\text{min}}^2}.
	\label{eqn:faradayscale}
\end{equation}
The `maximum scale' refers to the width ($\Delta\phi$) of a feature in the Faraday spectrum. For a given observation, with a feature having a width $\Delta\phi>\phi_{\text{max-scale}}$, the observation will only be sensitive to $<50\%$ of the emission.


\bsp	
\label{lastpage}
\end{document}